\documentclass[twocolumn]{aastex63}

\usepackage{graphicx}

\shorttitle{CARMA NAP Survey}
\shortauthors{Kong et al.}

\begin{document}

\title{High-resolution CARMA Observation of Molecular Gas in the North America and Pelican Nebulae}

\author[0000-0002-8469-2029]{Shuo Kong}
\affiliation{Steward Observatory, University of Arizona, Tucson, AZ 85719, USA}
\affiliation{Department of Astronomy, Yale University, New Haven, CT 06511, USA}

\author[0000-0001-5653-7817]{H\'ector G. Arce}
\affiliation{Department of Astronomy, Yale University, New Haven, CT 06511, USA}

\author[0000-0003-2251-0602]{John M. Carpenter}
\affiliation{Joint ALMA Observatory, Avenida Alonso de C\'ordova 3107, Vitacura, Santiago, Chile}

\author[0000-0001-8135-6612]{John Bally}
\affiliation{Department of Astrophysical and Planetary Sciences, University of Colorado, Boulder, Colorado, USA}

\author[0000-0002-8351-3877]{Volker Ossenkopf-Okada}
\affiliation{I.~Physikalisches Institut, Universit\"at zu K\"oln,
              Z\"ulpicher Str. 77, D-50937 K\"oln, Germany}

\author[0000-0002-3078-9482]{\'Alvaro S\'anchez-Monge}
\affiliation{I.~Physikalisches Institut, Universit\"at zu K\"oln,
              Z\"ulpicher Str. 77, D-50937 K\"oln, Germany}

\author[0000-0002-4633-5098]{Anneila I. Sargent}
\affiliation{Division of Physics, Mathematics and Astronomy, California Institute of Technology 249-17, Pasadena, CA 91125, USA}

\author[0000-0003-0412-8522]{S\"umeyye Suri}
\affiliation{Max Planck Institute for Astronomy, K\"onigstuhl 17, 69117 Heidelberg, Germany}

\author[0000-0003-0948-6716]{Peregrine McGehee}
\affiliation{Department of Earth and Space Sciences, College of the Canyons, Santa Clarita, CA 91355}

\author[0000-0002-0500-4700]{Dariusz C. Lis}
\affiliation{Jet Propulsion Laboratory, California Institute of Technology, 4800 Oak Grove Drive, Pasadena, CA 91109, USA}

\author[0000-0002-0560-3172]{Ralf Klessen}
\affiliation{Universit\"{a}t Heidelberg, Zentrum f\"{u}r Astronomie, Albert-Ueberle-Str. 2, 69120 Heidelberg, Germany}
\affiliation{Universit\"{a}t Heidelberg, Interdisziplin\"{a}res Zentrum f\"{u}r Wissenschaftliches Rechnen, INF 205, 69120 Heidelberg, Germany}

\author[0000-0002-6956-0730]{Steve Mairs}
\affiliation{East Asian Observatory, 660 N. A`oh\={o}k\={u} Place, Hilo, Hawai`i, 96720, USA}

\author[0000-0002-2250-730X]{Catherine Zucker}
\affiliation{Harvard Astronomy, Harvard-Smithsonian Center for Astrophysics, 60 Garden St., Cambridge, MA 02138, USA}

\author[0000-0002-0820-1814]{Rowan J. Smith}
\affiliation{Jodrell Bank Centre for Astrophysics, Department of Physics and Astronomy, University of Manchester, Oxford Road, Manchester M13 9PL, UK}

\author[0000-0001-5431-2294]{Fumitaka Nakamura}
\affiliation{National Astronomical Observatory of Japan, 2-21-1 Osawa, Mitaka, Tokyo 181-8588, Japan}

\author[0000-0002-0786-7307]{Thushara G.S. Pillai}
\affiliation{Institute for Astrophysical Research, Boston University, Boston, MA, USA.}

\author[0000-0002-5094-6393]{Jens Kauffmann}
\affiliation{Haystack Observatory, Massachusetts Institute of Technology, 99 Millstone Road, Westford, MA 01886, USA}

\author[0000-0003-2549-7247]{Shaobo Zhang}
\affiliation{Purple Mountain Observatory, \& Key Laboratory for Radio Astronomy, Chinese Academy of Sciences, Nanjing 210023, China}

\begin{abstract}

We present the first results from a CARMA 
high-resolution $^{12}$CO(1-0), $^{13}$CO(1-0), 
and C$^{18}$O(1-0) molecular line survey of the 
North America and Pelican (NAP) Nebulae. 
CARMA observations have been combined with single-dish 
data from the Purple Mountain 13.7m telescope to add
short spacings and produce 
high-dynamic-range images. We find
that the molecular gas is predominantly shaped by the 
W80 HII bubble that is driven by an O star. 
Several bright rims are probably remnant molecular clouds heated and
stripped by the massive star.
Matching these rims in molecular lines and optical images,
we construct a model 
of the three-dimensional structure of the NAP complex.
Two groups of molecular clumps/filaments are on the 
near side of the bubble, one being pushed toward us, 
whereas the other is moving toward the bubble.
Another group is on
the far side of the bubble and moving away.
The young stellar objects in the Gulf region reside in three different clusters,
each hosted by a cloud from one of the three molecular clump groups. 
Although all gas content in the NAP is impacted by feedback
from the central O star, some regions
show no signs of star formation, while other areas 
clearly exhibit star formation activity.
Other molecular gas being carved by feedback includes
the cometary structures in the Pelican Head region and
the boomerang features at the boundary of the Gulf region.
The results show that the NAP complex is an ideal place
for the study of feedback effects on star formation. 

\end{abstract}


\section{Introduction}\label{sec:intro}

Feedback from young massive stars is crucial for the evolution of the interstellar medium of galaxies. In particular,  bubbles driven by young massive stars  are common in our Milky Way \citep{2006ApJ...649..759C}. They carry momentum and energy  and significantly impact their environments. However, details of these regulatory processes are not well understood, largely because the structure of the gas is often complicated by the effects of multiple massive stars \citep{2015MNRAS.450.1199D}. Studies of the effects of a single massive star on its natal molecular environment are clearly needed to address this issue.

In this regard, the North America and Pelican (NAP) Nebulae provide a favorable target for investigating the response of a star-forming molecular cloud to a single massive star and its feedback. Within this complex, an HII region, W80, appears as an ionized bubble at radio wavelengths \citep{1958BAN....14..215W}.
The ionizing source is an O3.5 star, 2MASS J20555125+4352246 \citep{2005A&A...430..541C}, also known as the Bajamar Star \citep[with the possibility of being a spectroscopic binary,][]{2016ApJS..224....4M}. 

\begin{figure*}[htbp]
\epsscale{1.15}
\plotone{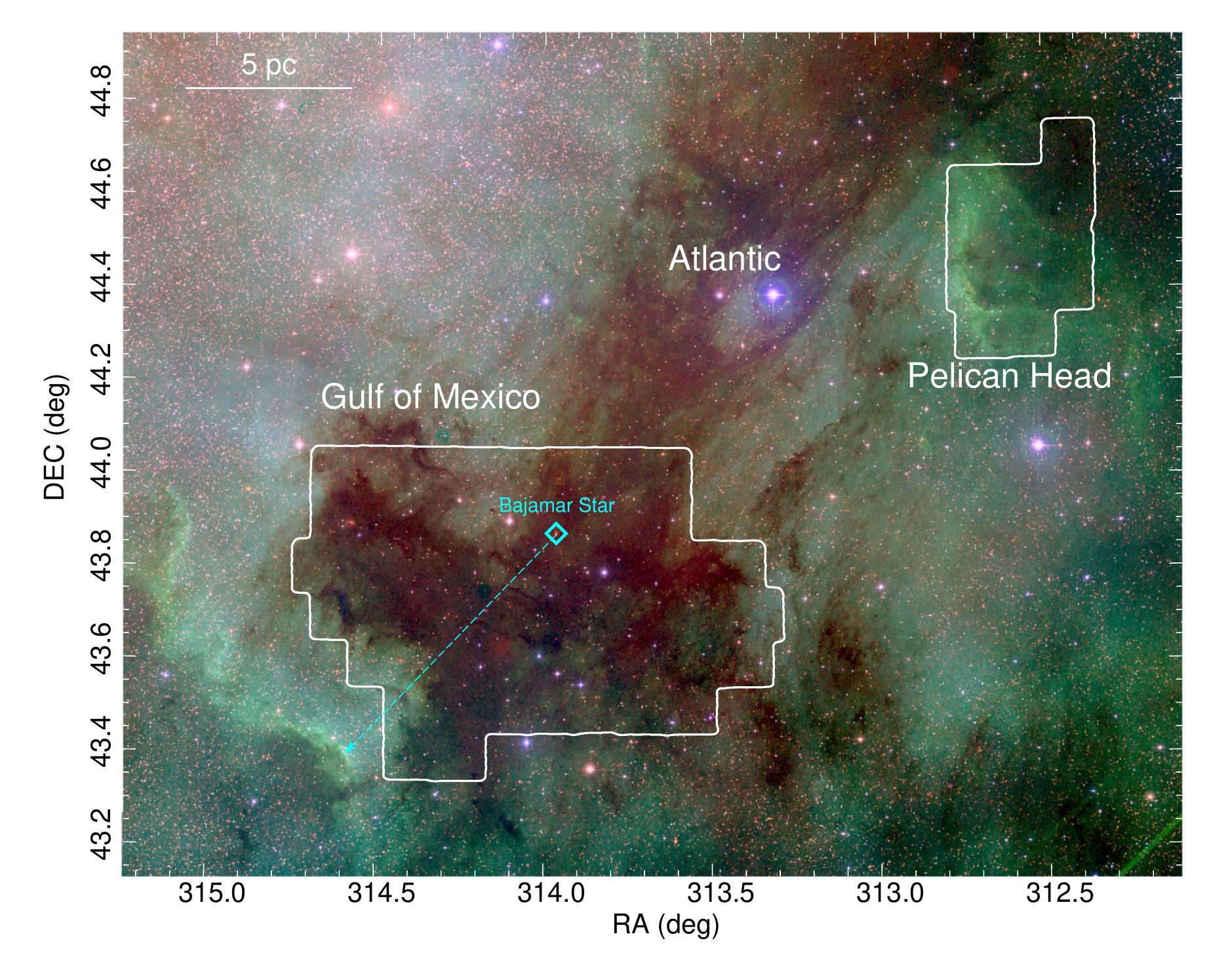}
\vspace{-0.5cm}
\caption{
The footprint of the CARMA NAP map (white
boundaries) overlaid on a POSS-II false
color RGB image using the 
0.9$\mu$m (Red), 
0.7$\mu$m (Green), and
0.43$\mu$m (Blue) bands. 
The observed regions include the Gulf of Mexico
(the dark area to the southeast) and the Pelican
Head (the area to the northwest).
The cyan diamond marks the position of the
O star confirmed by \citet{2005A&A...430..541C}.
The cyan arrow points from the 
O star to the two boomerang features at
the south-east corner. 
The scale bar is derived assuming a 800 pc
distance to the NAP complex (\S\ref{subsec:gaia}).
\label{fig:DSScoverage}}
\end{figure*}

An optical/near infrared image of the NAP is shown in Figure \ref{fig:DSScoverage}. A dark cloud, L935, is seen against the bright W80 bubble. The position of the Bajamar Star is indicated by a cyan diamond. The HII region has been produced by a single massive ionizing source still embedded in an actively star-forming molecular cloud \citep[][hereafter B14]{2014AJ....148..120B}. This is an ideal area for a detailed study of the effects of massive star feedback on the surrounding environment. 

Based on the recent Gaia data, we have learned that the NAP region is about 800 pc away \citep[][also see \S\ref{subsec:gaia}]{2020A&A...633A..51Z}. At this distance, a physical scale of 0.1 pc translates to an angular scale of 25\arcsec. The scale corresponds to the typical size of star-forming cores and filaments, which are most relevant to star formation studies. To reasonably resolve these structures in the NAP cloud, we have used the Combined Array for Research in Millimeter-wave Astronomy (CARMA) to map emission from common molecular gas tracers (\S\ref{subsec:obs}). The NAP complex is vast, so we limit our maps to regions with mostly dense gas, i.e., the Gulf of Mexico and the Pelican Head. These regions are outlined with white boundaries in Figure \ref{fig:DSScoverage}.


CARMA high-resolution 
molecular line observations are critical for examining  the molecular gas/HII region interactions. However, the analysis of the feedback mechanism also requires an understanding of the effects of a massive star on the overall star formation process. We have
therefore complemented our molecular line imaging program with studies of the relationship between the molecular gas and YSOs, using Rebull's comprehensive YSO catalog    \citep[][hereafter R11]{2011ApJS..193...25R} .

In this paper, we aim to create a 3-dimensional view of the NAP complex through a comparison
of multi-wavelength images. By pinpointing the 
interaction between the molecular gas and the
HII region, especially elevated dust emission associated with molecular gas, we aim to distinguish the line-of-sight locations of the molecular gas clouds relative to the HII region. We also study the number and distribution of YSOs in the associated molecular gas in order to shed light on how star formation is impacted
by the feedback. 

\section{Observations and Data Combination}\label{sec:obs}

As already noted, the NAP region is an example of a relatively simple configuration of a  star-forming molecular cloud, young massive star, and associated HII region. The areas of the Gulf of Mexico and Pelican Head mapped by CARMA observations are shown in Figure \ref{fig:DSScoverage}. These were selected to highlight the most dense gas.

\subsection{Observations}\label{subsec:obs}
CARMA observations were carried out in 2017 during the last season of array operations. The instrument set-up was the same as that adopted for the CARMA-NRO Orion Survey \citep[][hereafter K18]{2018ApJS..236...25K}.
Briefly, the 15-element array of six 10-m diameter and nine 6-m antennae was used to map the area of interest. The total mosaicked area of about 1 deg$^2$ is made up of 126 pointings of 6'x 6' subfields. The `D'' and ``E'' configurations of the array led to
uv-coverage between 2.5-40 k$\lambda$, where
$\lambda$ is 2.6 mm for $^{12}$CO(1-0), resulting in an angular resolution of about 7\arcsec\ (see Table \ref{tab:observations}).

As for the Orion Survey, the spectral line setup focused on $^{12}$CO(1-0), $^{13}$CO(1-0),
C$^{18}$O(1-0) but also included the
 CN N=1-0,J=3/2-1/2, 
SO 2(3)-1(2), and CS(2-1) lines. Here we report only on the $^{12}$CO(1-0), $^{13}$CO(1-0),
and C$^{18}$O(1-0) observations. 
For $^{12}$CO the correlator  
set-up provided a bandwidth of 31 MHz (81 $\rm
km~s^{-1}$) and a spectral resolution of
98 kHz ($\sim$ 0.25 $\rm km~s^{-1}$).
For the other spectral lines, a bandwidth of
8 MHz ($\sim 21$ $\rm km~s^{-1}$) with 
spectral resolution 24 kHz ($\sim 0.067$
$\rm km~s^{-1}$) was used. 

To compensate for the fact that the interferometer  observations did not include 
baselines shorter than 2.5 k$\lambda$, effectively filtering out emission from structures larger than 100\arcsec, the CARMA observations were combined with single-dish maps of the same regions. These were obtained at 
the Purple Mountain Observatory Delingha 13.7m telescope
(hereafter DLH14)
\citep[][hereafter Z14]{2014AJ....147...46Z}, and are part of the Milky Way Imaging Scroll 
Painting project \citep{2019ApJS..240....9S}.

\begin{deluxetable}{ccccc}
\tablecolumns{5}
\tablewidth{0pt}
\tablecaption{Final sensitivity \label{tab:observations}}
\tablehead{
\colhead{Transition} &
\colhead{Beam} &
\colhead{PA} &
\colhead{$\Delta_{V}$} &
\colhead{$\sigma_{K}$}\\
\colhead{} & 
\colhead{} & 
\colhead{(deg)} & 
\colhead{$(\rm km~s^{-1})$} & 
\colhead{(K)}}
\startdata
$^{12}$CO(1-0) & $7\arcsec\times6\arcsec$ & 75 & 0.25 & 1.1 \\  
$^{13}$CO(1-0) & $7\arcsec\times6\arcsec$ & 75 & 0.16 & 0.8 \\    
C$^{18}$O(1-0) & $7\arcsec\times6\arcsec$ & 75 & 0.16 & 0.7 \\
\enddata
\end{deluxetable}

\subsection{Data Combination}

\begin{figure}[htbp]
\epsscale{1.1}
\plotone{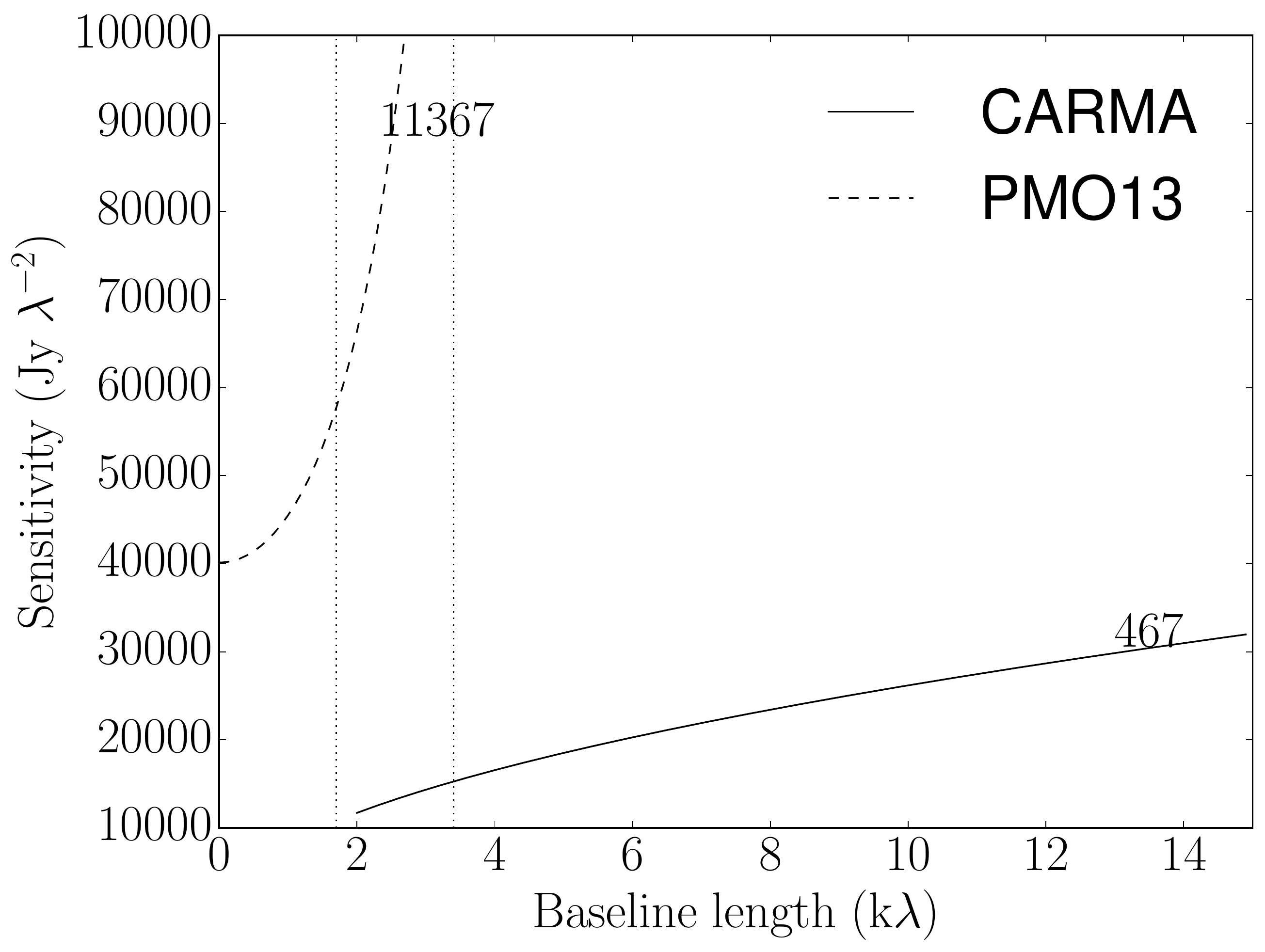}
\caption{
Visibility sensitivity as a function of uv-distance for both
DLH14 (dashed curve) and CARMA (solid curve). The imaging sensitivity is labeled next to the
curves in units of 
mJy as in \citet{2011ApJS..193...19K}. 
The vertical dotted lines mark the overlapping region between the two datasets.
\label{fig:uvsens}}
\end{figure}

The CARMA and DLH14 data were combined in the uv plane, 
following the procedures outlined in 
\citet{2011ApJS..193...19K} and \citet{2018ApJS..236...25K}.
The resulting images are based on observations that range from zero-spacing fluxes to measurements on 
the maximum baselines provided by CARMA. 
Figure \ref{fig:uvsens} shows a comparison of CARMA and DLH14 uv sensitivities. The uv-plane pixel sensitivity as a function of baseline length was calculated following  appendix C in \citet{2011ApJS..193...19K}. It is
a function of the image-plane sensitivity and 
the dish size or, for an array, baseline length. 

In general, to ensure good resulting imaging quality when combining observations from two instruments, it is desirable that the sensitivities match where the baselines overlap. Figure \ref{fig:uvsens}
shows this is not the case here; the sensitivity of the DLH14 data is much lower that that of CARMA. The sensitivity mismatch also affects  
flux scale determination adversely. As described in \citet{2018ApJS..236...25K}, the relative fluxes 
of the single-dish data and the
CARMA can be calibrated based on the flux ratio within
the overlapping baselines. Here, the 
DLH14 data is sufficiently noisy beyond baselines of 2 k$\lambda$, that no meaningful flux scale factor can be established (so we adopt the factor of 1). 
Nevertheless, as we demonstrate below, the DLH14 observations add a useful contribution to our final images despite their quantitative issues.
Table \ref{tab:observations} summarizes the final sensitivity for the combined data.
To match the single-dish observations,
a spectral resolution of 0.16 km s$^{-1}$ was adopted for the combined  
$^{13}$CO and C$^{18}$O data set.


\begin{figure*}[htbp]
\epsscale{0.42}
\centering
\hspace{-18pt}
\plotone{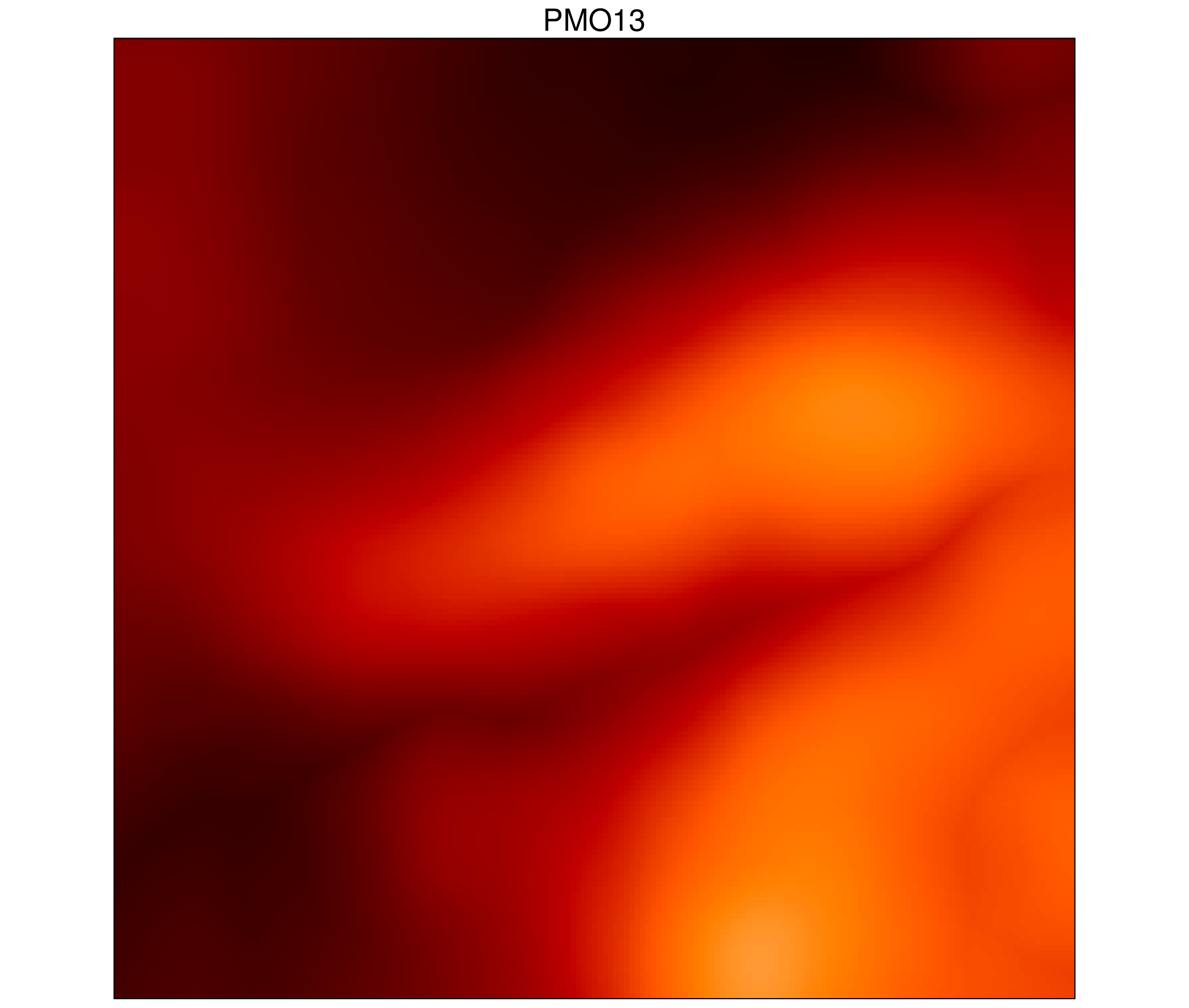}
\hspace{-18pt}
\plotone{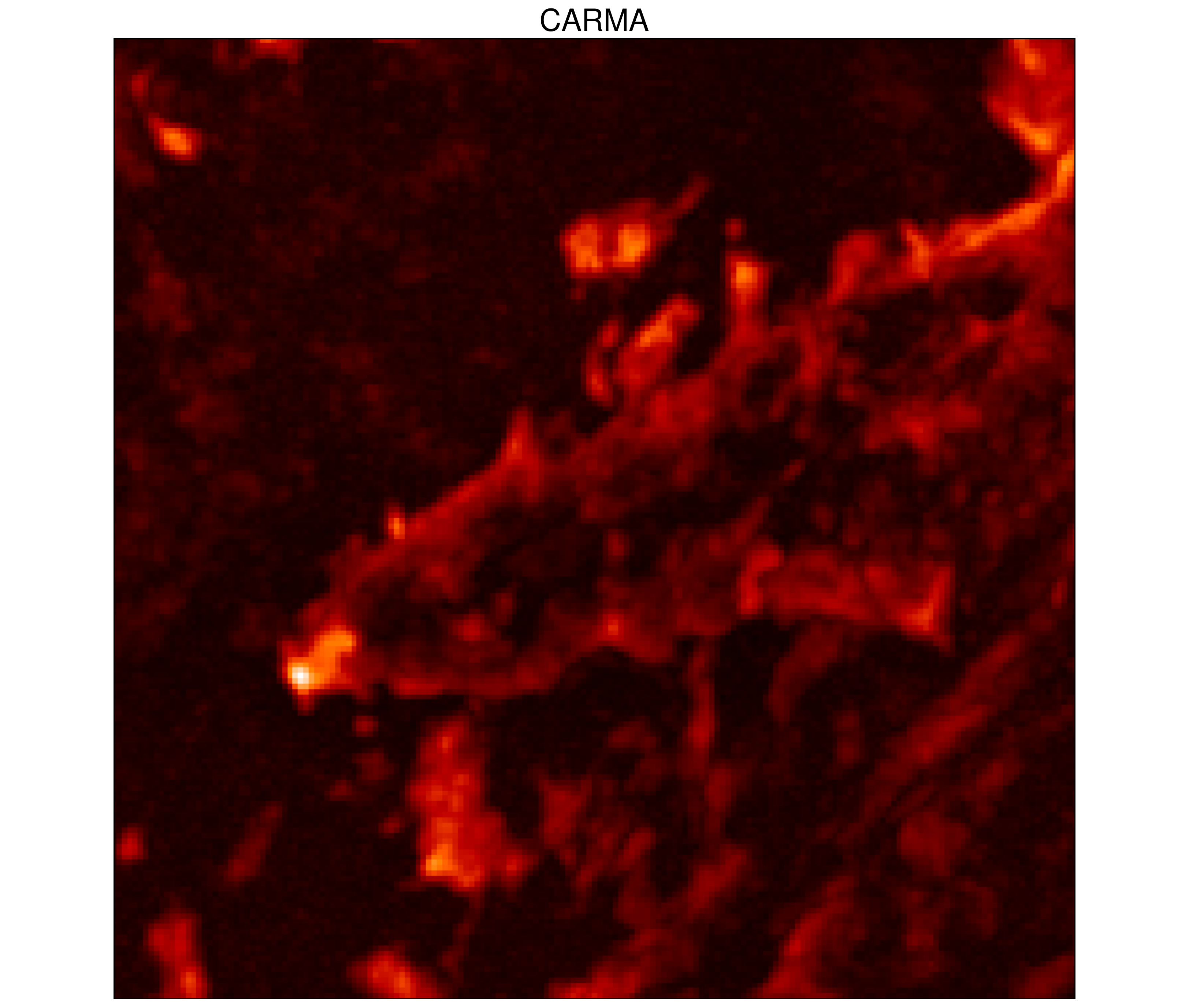}
\hspace{-18pt}
\plotone{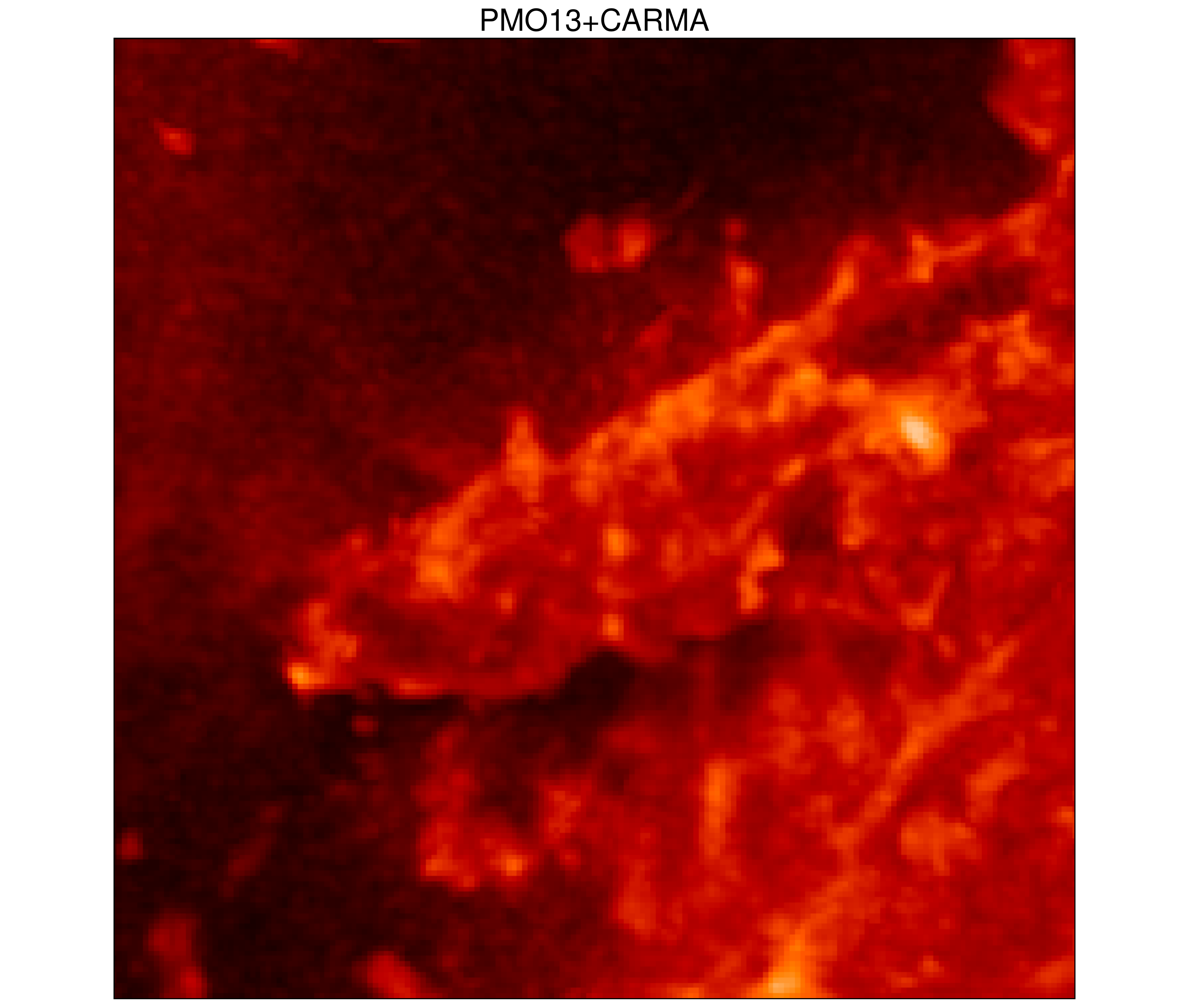}
\hspace{-18pt}
\caption{Comparison between DLH14,
CARMA and combined images.
{\it Left:} DLH14 peak intensity image.
{\it Middle:} CARMA image. {\it Right:} Combined 
CARMA+DLH14 images. All panels show the same
6\arcmin\ by 6\arcmin\ 
(1.4 pc by 1.4 pc) area centered around Comet1.
\label{fig:combine}}
\end{figure*}

\subsection{Image Quality}\label{sec:3panel}

To allay concerns about the reliability of the NAP images that result from combining CARMA and DLH14 observations, we compare $^{12}$CO integrated intensity  images of the Pelican Head region from DLH14, CARMA, and combined CARMA+DHL14  data. The Pelican Head area was selected since it appears to encompass multi-scale structures (see Figure \ref{fig:combine}). 
The DLH14 map (left panel) shows smooth, extended emission with very little structure, while the CARMA map (middle panel) is irregular and prominently  shows the (sharp) edges of the 
cometary-like cloud in this region.
The combined image (right panel) clearly displays the usefulness of adding short spacing flux information; considerably more structure appears to be present, both in the edges and body of the cloud.

\begin{figure*}
\centering
\vspace{-1.5cm}
\includegraphics[width=1.\columnwidth]{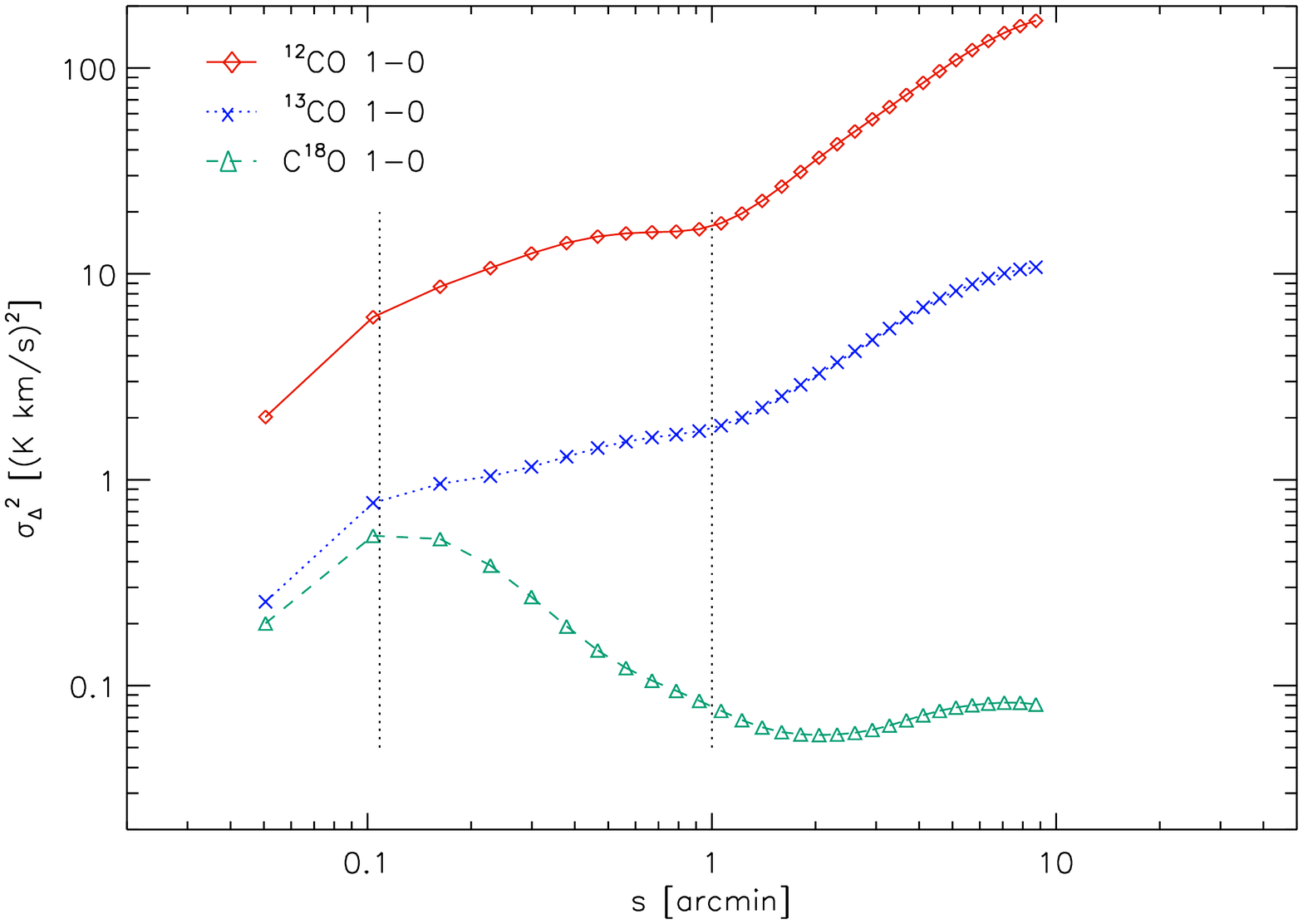}
\includegraphics[width=1.\columnwidth]{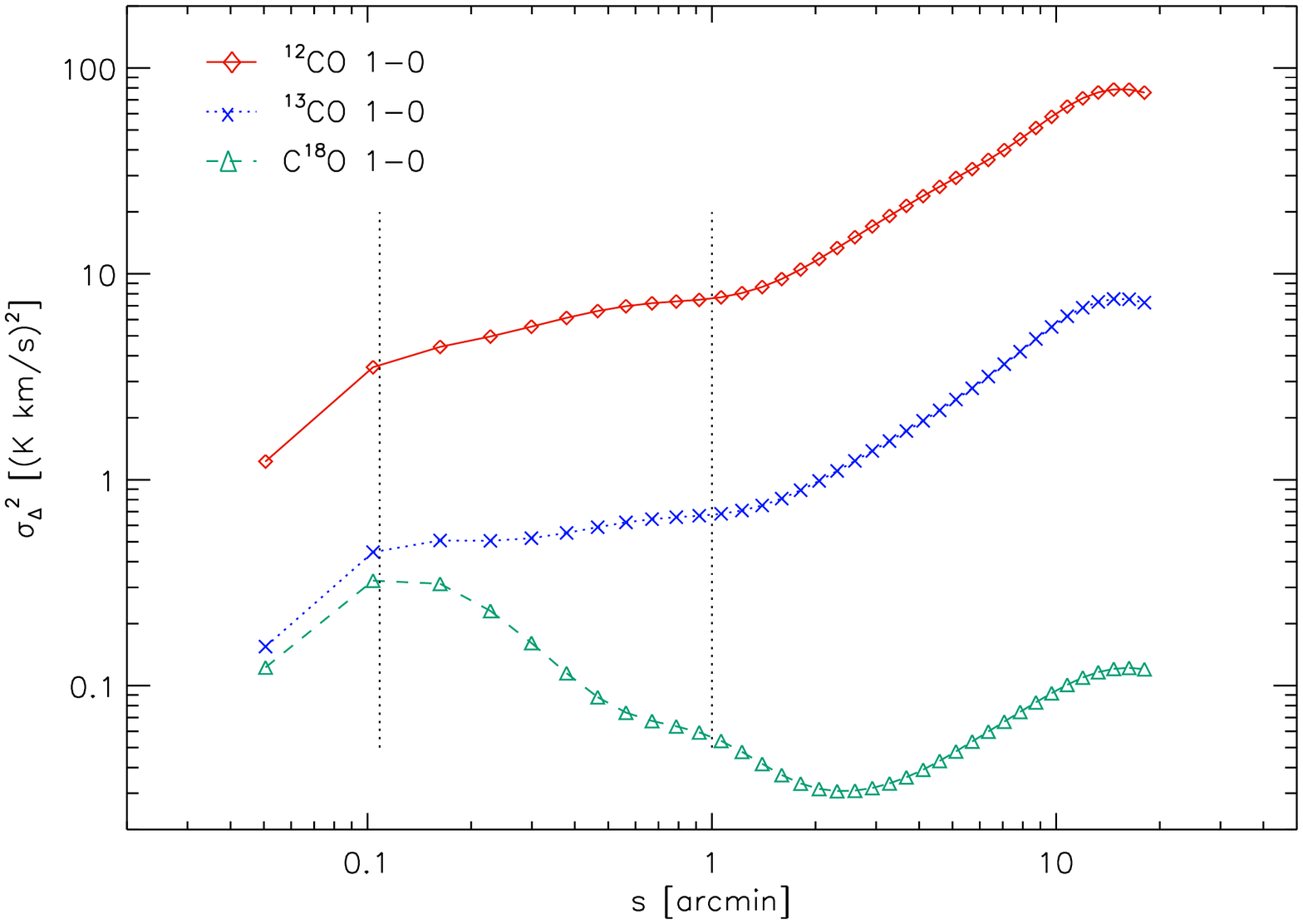}
\caption{$\Delta$-variance spectra for the
$^{12}$CO(1-0), $^{13}$CO(1-0), and C$^{18}$O(1-0)
integrated intensity maps from the 
combined (CARMA+DLH14) dataset.
The x-axis shows the scale measured by the wavelet size. 
The y-axis displays the variance of the structural 
fluctuations in the maps at that scale in units 
of the square of the measured map. The left figure 
shows the results for the Pelican Head region, the
right figure for the Gulf of Mexico. The vertical dotted lines show the scales of the CARMA beam ($\sim$6\arcsec) and the DLH14 beam ($\sim$60\arcsec).}
\label{fig:Dvar1}
\end{figure*}

Following the CARMA-NRO Orion Survey (K18), 
we carry out a $\Delta$-variance analysis
\citep{Stutzki1998,Ossenkopf2008}
for the channel maps and integrated intensity maps
to examine the quality of the data combination. 
All main features are visible in the integrated intensity maps.
Figure~\ref{fig:Dvar1} shows the $\Delta$-variance spectra for the 
Pelican Head (left panel) and the Gulf of Mexico (right panel) regions. 
In both regions and for all tracers, 
there is a structural break at a scale of 
1\arcmin\ - 2\arcmin, 
corresponding to the gap in our spatial 
sensitivity at 1.7 - 3.4k$\lambda$  due to the lack of sensitivity at these scales (see Figure \ref{fig:uvsens}). 
The dip in the $\Delta$-variance at that
scale also reflects the mismatch of sensitivity in the overlapping baseline lengths.

\section{Results}\label{sec:results}

\subsection{Feedback Features in IR and the Ionizing Source}\label{subsec:midir}

\begin{figure*}[htbp]
\epsscale{1.15}
\plotone{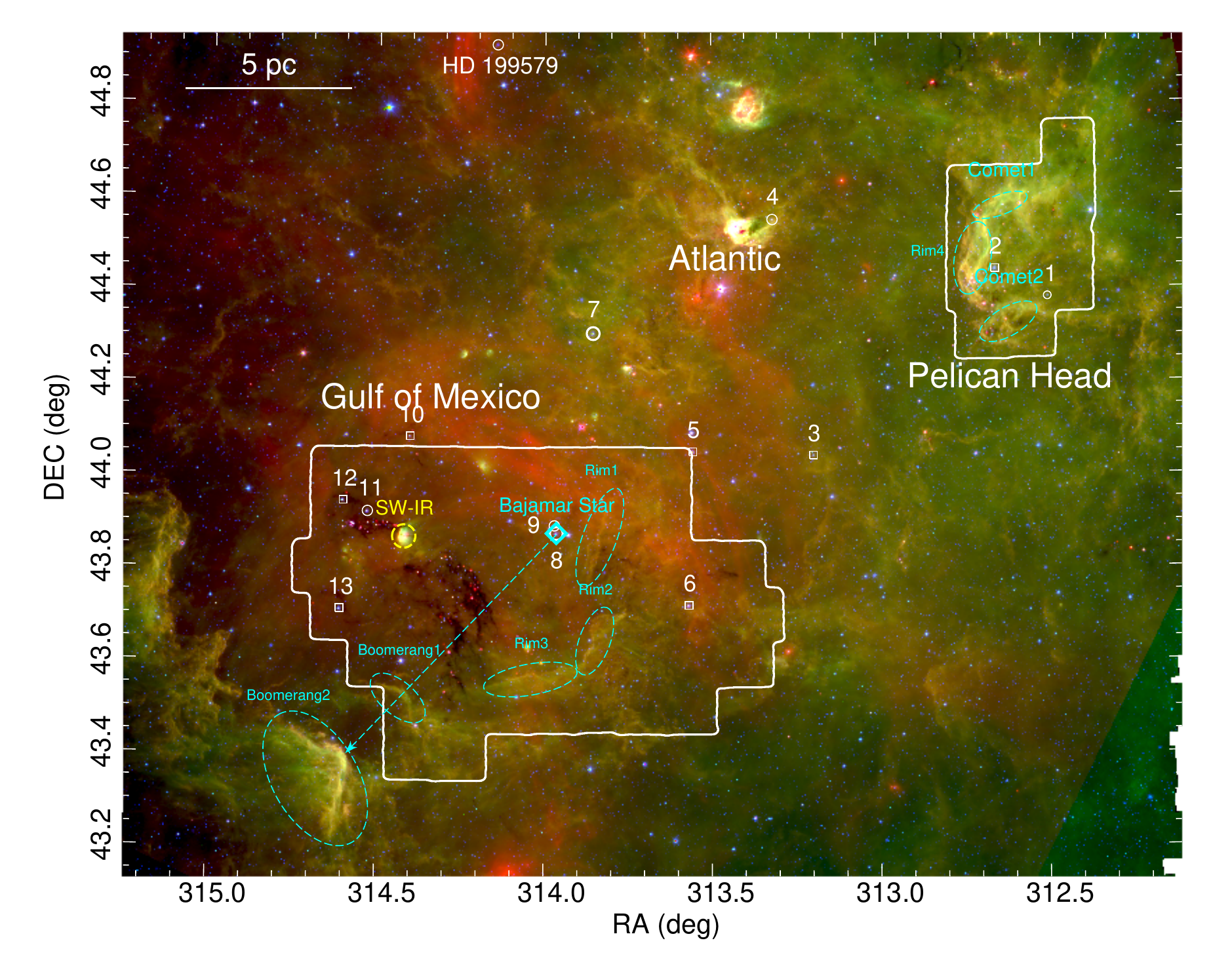}
\vspace{-0.5cm}
\caption{
Spitzer false color RGB image using the MIPS
24$\mu$m band (Red), and the IRAC 
8$\mu$m (Green), and
4.5$\mu$m (Blue) bands. 
The images are results from Spitzer program 20015 and program 462 (PI: L. Rebull, see R11).
The cyan diamond marks the position of the
Bajamar Star. The cyan arrow points south-east from the 
Bajamar Star to the bright-rim feature with a boomerang-like morphology.
The cyan dashed ellipses mark several feedback
features discussed in the text 
(and seen in 
molecular gas in Figures \ref{fig:peakg} and \ref{fig:peakp}). 
The white boundaries show 
the footprint of the CARMA NAP map.
The stars with numbers 1-13 are from SL08 (as shown in B14).
Circles are stars classified as spectral type OB. 
Squares are possible AGB stars (see SL08 and B14).
The Bajamar Star is \#8. 
\label{fig:spitzercover}}
\end{figure*}

Figure \ref{fig:spitzercover} shows a composite  image of the Gulf of Mexico and Pelican regions
(same area shown in Figure \ref{fig:DSScoverage})
made from Spitzer data
at 24$\mu$m (red), 8$\mu$m  (green), and 
4.5$\mu$m (blue) . 
Figure \ref{fig:wisecover}
shows a WISE three-color composite image of the same region in which we assigned red for the 
22$\mu$m emission, green for the 12$\mu$m emission, and blue for
the 4.6$\mu$m emission.
As in Figure 1, We show the extent of the CARMA maps with white boundaries. 

In both Spitzer and WISE images, the Gulf region 
shows a few relatively bright-rim structures. 
These bright rims trace hot dust in the area.
For instance, the WISE W4 band (22 $\mu$m) 
traces the warm dust thermal radiation.
Meanwhile, the WISE W3 band (12 $\mu$m) includes many
emission features from PAH (at 7.7 $\mu$m, 
8.5 $\mu$m, 10 $\mu$m, and 11.3 $\mu$m)
and fine-structure lines from [Ne II] 
(12.8 $\mu$m) and [Ne III] (15.6 $\mu$m).
These lines are likely pumped by UV emission 
\citep{1989ApJS...71..733A} and are believed to
be good indicators of star formation and stellar 
feedback \citep[e.g.,][]{2007ApJ...658..314H}.
In any case, the hot dust should be heated
by a nearby ionizing source.

Here we argue that the ionizing source
responsible for the bright rims seen in these
IR images is the Bajamar Star
simply because of the proximity of this high-mass,
UV-irradiating young star to these features. 
As will be shown in \S\ref{subsec:gas},
the labeled bright rims that are inside the
CARMA map have clear molecular line 
counterparts. They provide the key connection
between the molecular gas and the massive star
feedback. Hereafter, we name the bright-rim 
filamentary structures west and south of the
Bajamar Star as  Rim1, Rim2, and
Rim3. We denote the bright-rim structures with a
boomerang-like morphology to the south-east of the
Bajamar Star as Boomerang1 and Boomerang2. 
There are a few more  bright-rim structures 
in the infrared images which we
do not label as they are less prominent or 
do not have  molecular line  counterparts.

The morphology of the bright-rim structures can 
also be used to deduce that the Bajamar Star is
the most likely source of feedback in this region. 
A line that passes through the centers of the two
boomerang structures and extends to the north-west
passes through the position of the Bajamar Star
(see the cyan arrow in Figures 
\ref{fig:DSScoverage} and \ref{fig:spitzercover}).
There are two more arc-like features to the
northeast of Boomerang2. These features, 
which are not as bright as Boomerang2, 
point towards the general direction of the
Bajamar Star (see Figures \ref{fig:DSScoverage} 
and \ref{fig:spitzercover}). 

As stated earlier, 
there is evidence that the Bajamar Star 
is the main ionizing source in the Gulf region.
\citet[hereafter SL08]{2008BaltA..17..143S} reported
12 additional massive star candidates in the NAP.
We determined the distances to these massive
stars (including the Bajamar Star) from the Gaia 
DR2 results \citep{2018AJ....156...58B}. 
The Bajamar Star (star \#8 in SL08)
has a distance of $668^{+39}_{-35}$ 
pc. However, updated results based on Gaia EDR3 find a distance to the Bajamar's star of $d=785\pm 16$ pc, in excellent agreement with the 3D dust and YSO-based results which we discuss in \S \ref{subsec:gaia}.
The IRS4 source (star \#2 in SL08)
in the Pelican region \citep{1980ApJ...239..121B}
has a distance of 799$_{-157}^{+255}$ pc.
However, SL08 suspected that IRS4 is a carbon star.
All other candidates are beyond 1500 pc, and there
are two sources with no Gaia distances 
(star \#11 in SL08, a.k.a, 2MASS J20580673+4355141; 
and star \#12 in SL08,  a.k.a.
2MASS J20582424+4356386). As shown in 
Figure 1 of B14, these two sources are far 
to the east of the Bajamar Star.
Given the alignment between the Bajamar Star and the
bright-rim features discussed above, 
the Bajamar Star is likely the only massive star
responsible for these feedback features in the Gulf region. 

We further argue that the Bajamar Star is 
the main ionizing star impacting the 
Pelican region as well, as suggested by 
\citet{2017A&A...602A.115D}.
The Pelican Head region is clearly being 
impacted by UV radiation. In
Figures \ref{fig:spitzercover} and \ref{fig:wisecover}
we notice that the Pelican Head region has 
very bright infrared emission (mostly notable
at $\lambda\ga5\mu$m, i.e., visible in
IRAC band 3 at 5.7$\mu$m but barely seen
in WISE band 2 at 4.6$\mu$m). The same
region also has plenty of H$_2$ fluorescent
emission caused by UV emission 
(as reported by B14), which
matches well with the infrared filaments here. 
In addition, as discussed in \S\ref{subsec:gas}, 
there are two bright emission features with a
comet-like (i.e., head-tail) morphology in the
Pelican Head region which were very likely caused 
by feedback from a massive ionizing star and 
point in the general direction of the Bajamar Star.
Hereafter, we therefore assume all the
feedback features are the result of
feedback from the Bajamar Star.

\subsection{Gas Distribution}\label{subsec:gas}

\begin{figure*}[htbp]
\epsscale{0.55}
\plotone{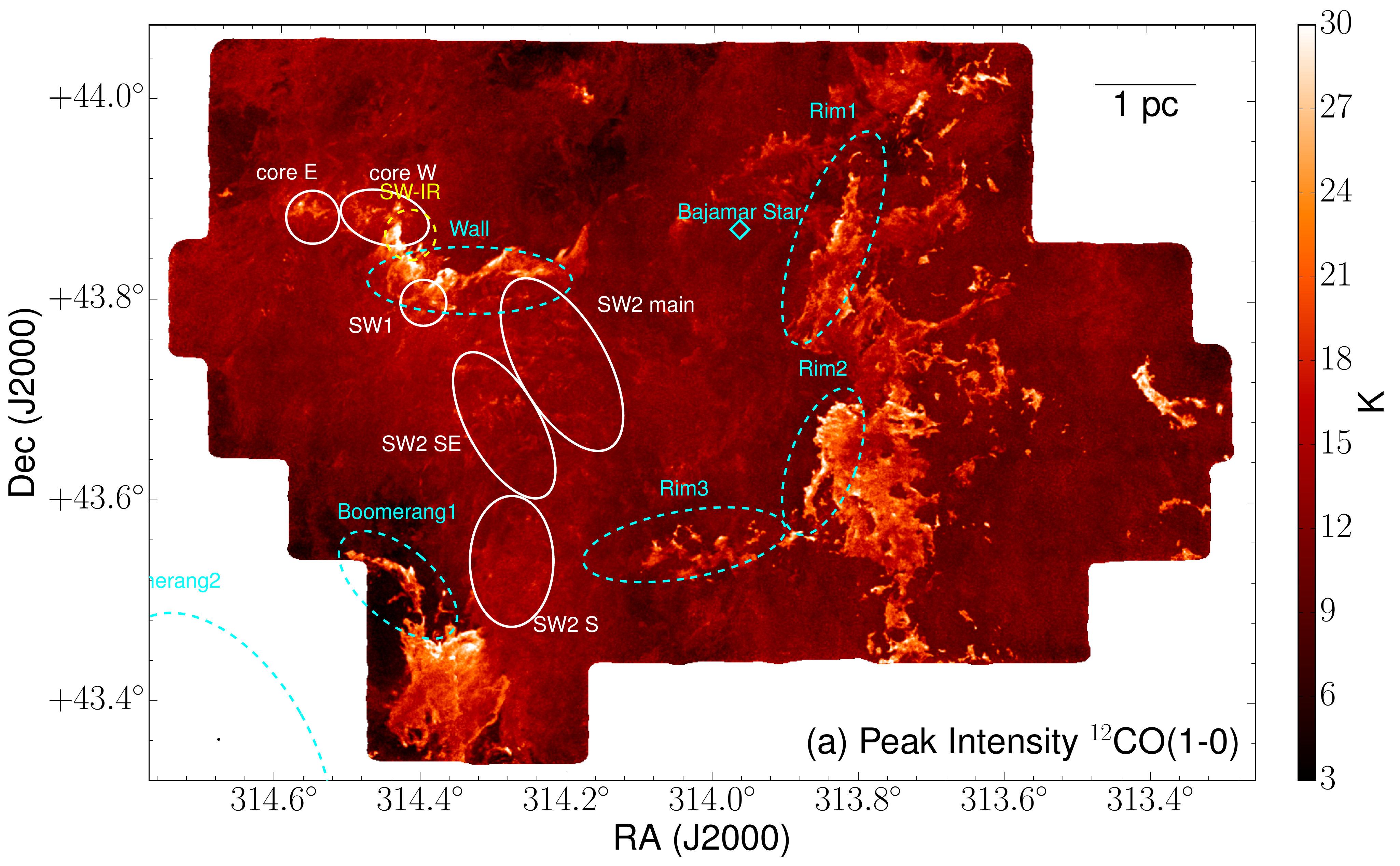}
\plotone{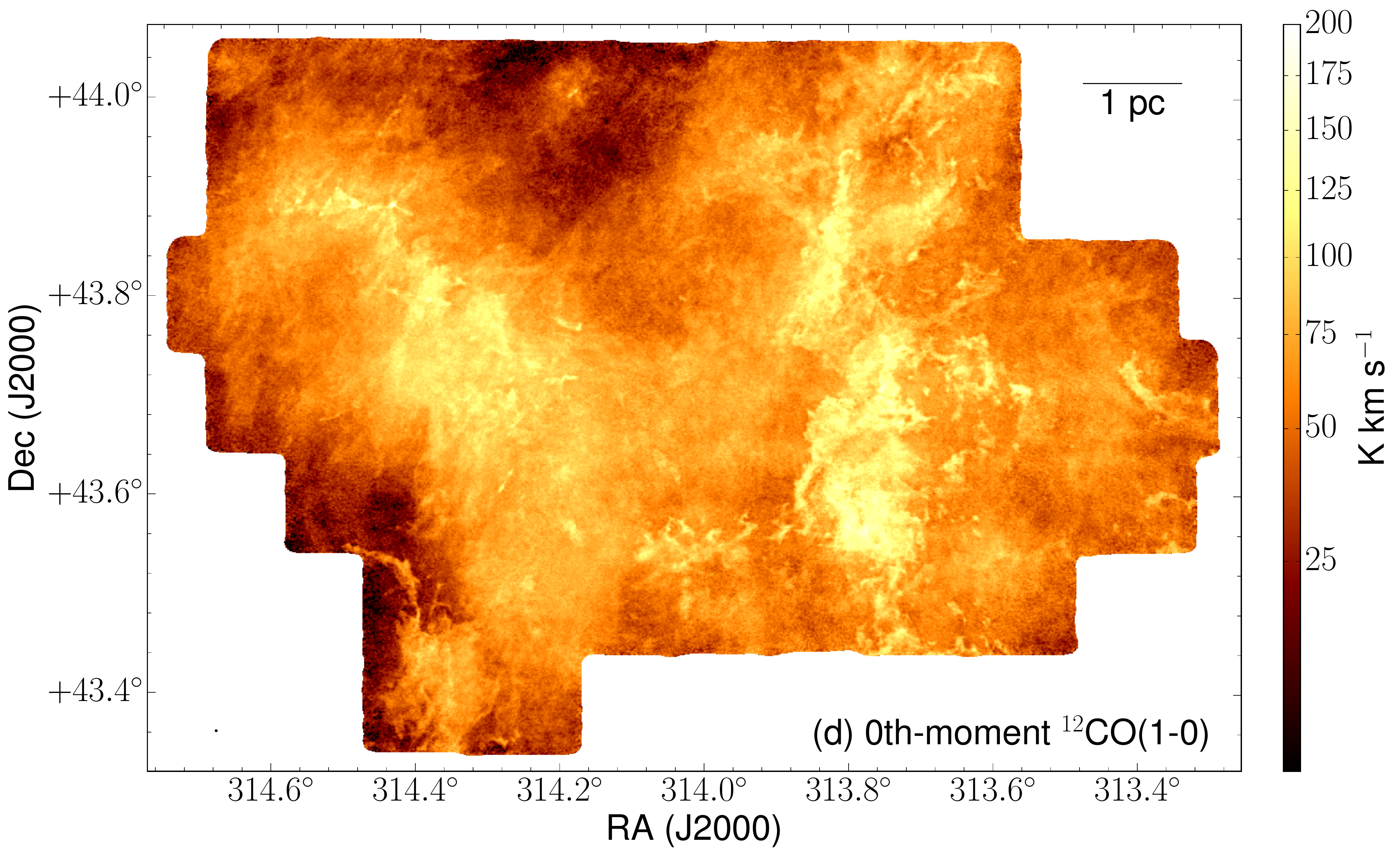}\\
\plotone{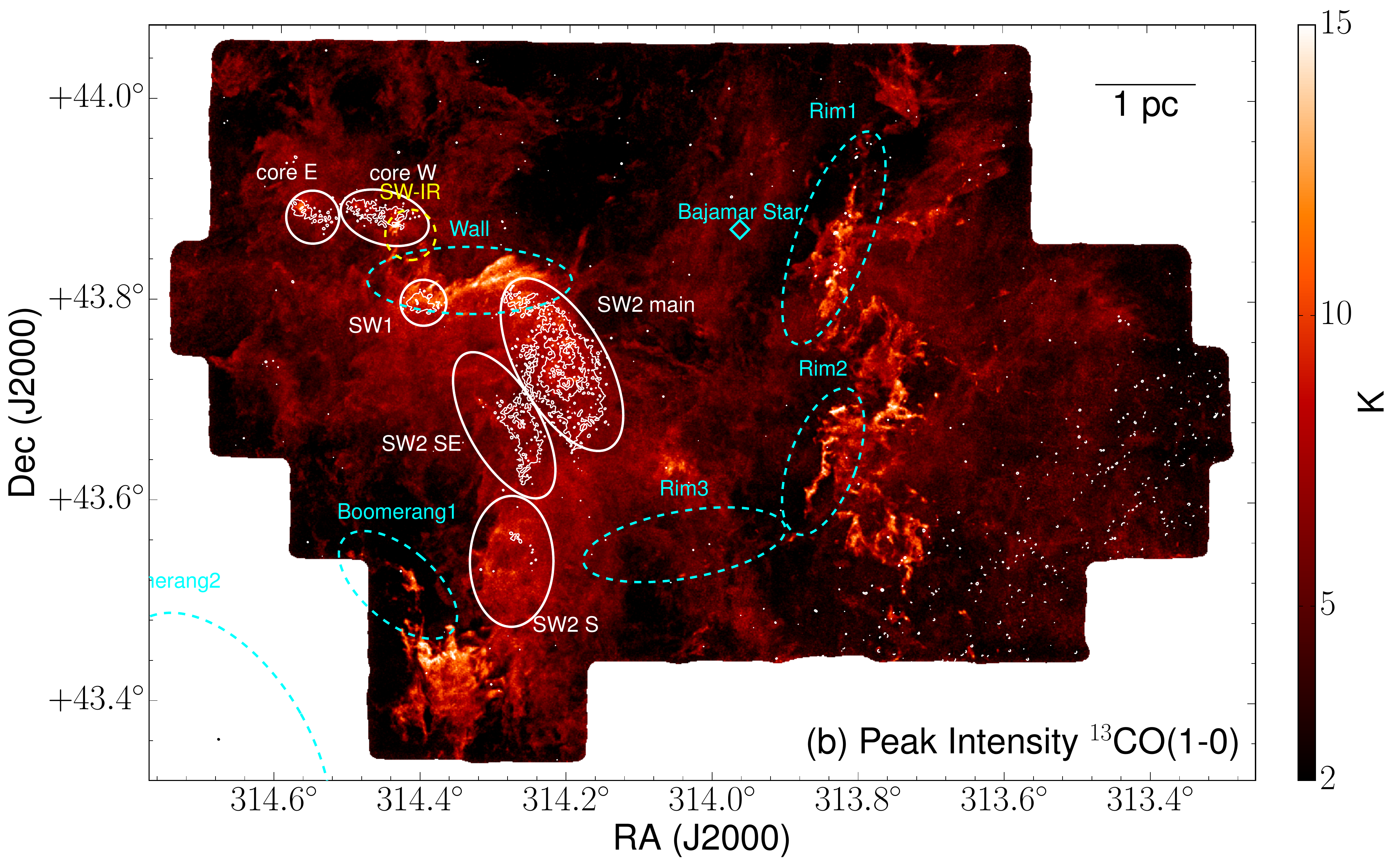}
\plotone{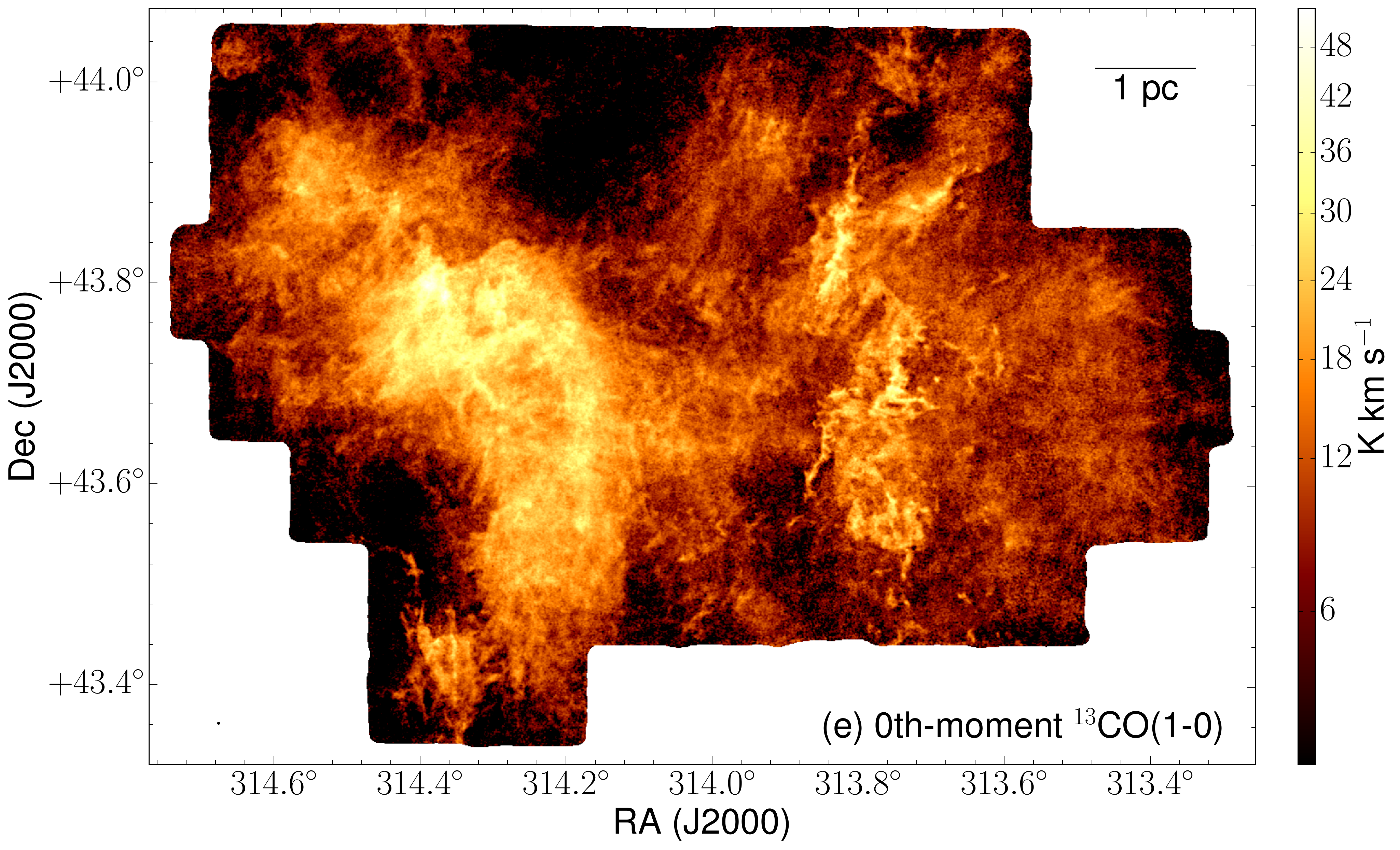}\\
\plotone{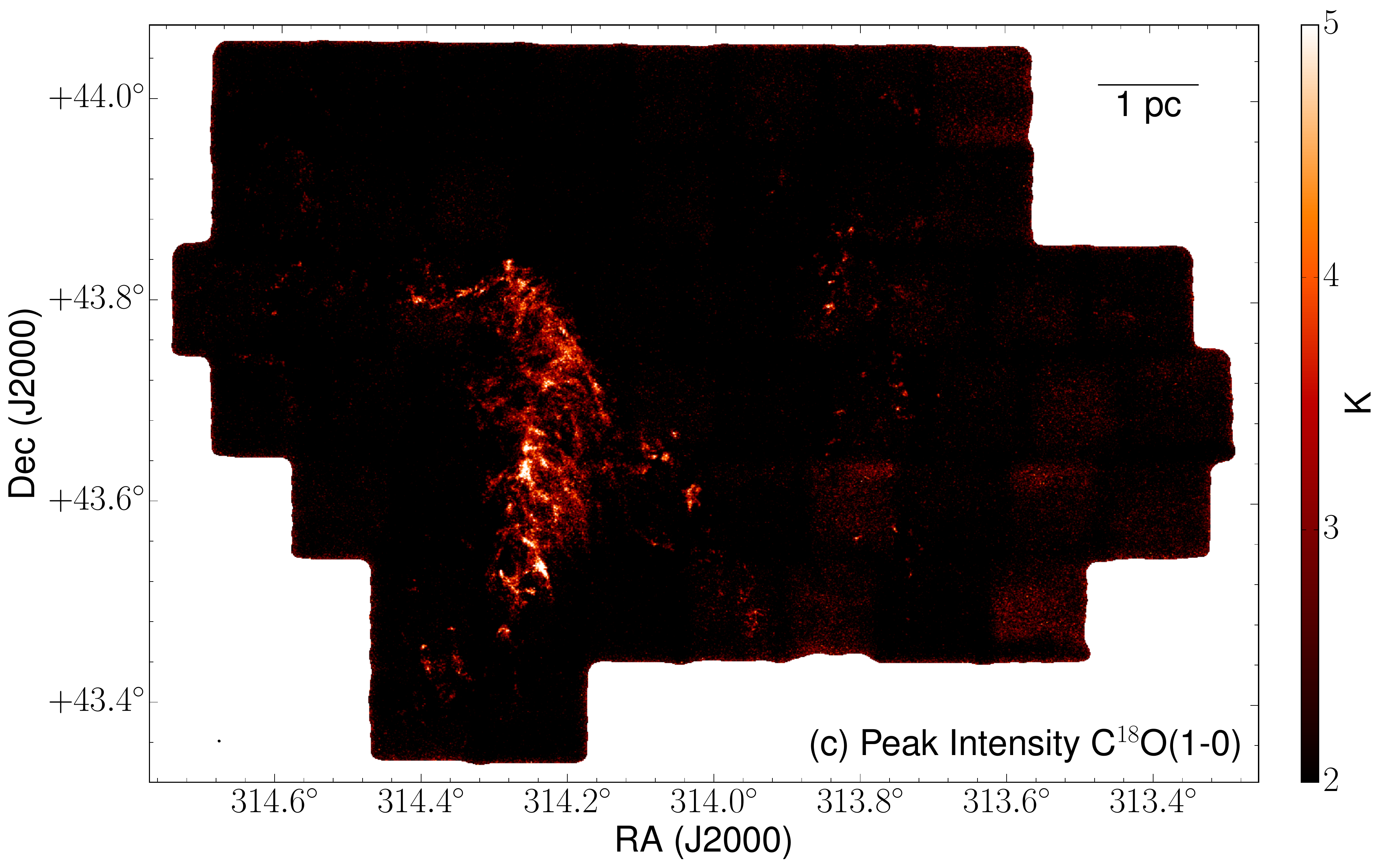}
\plotone{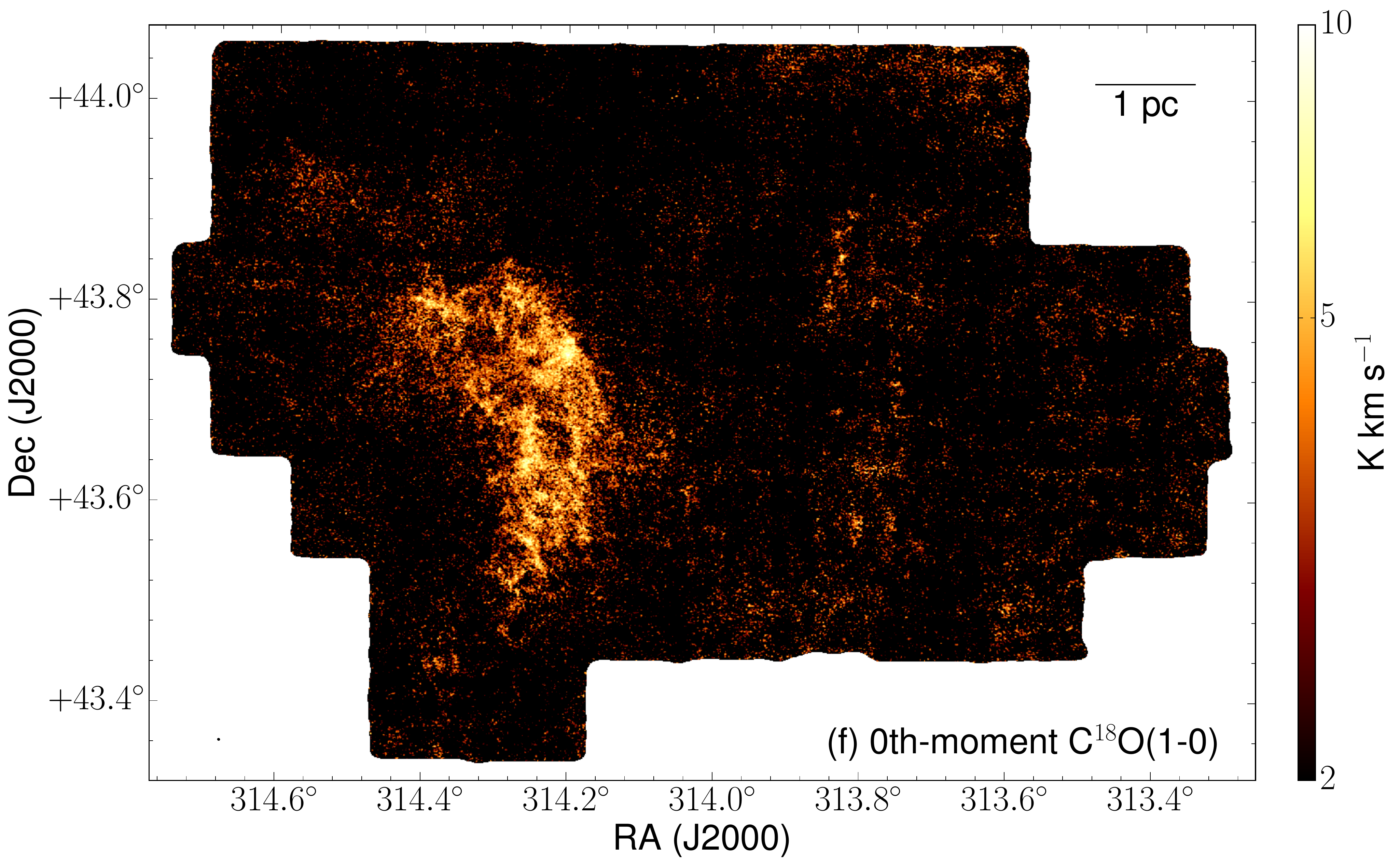}
\caption{
{\bf (a):} $^{12}$CO peak intensity map for the Gulf region. The synthesized beam is at the lower-left corner.
The white ellipses are the BGPS cores defined in B14.
The cyan ellipses are the same as in Figure \ref{fig:spitzercover}.
{\bf (b):} $^{13}$CO peak intensity map for the Gulf region. The white contours show the BGPS 1.1 mm continuum,
starting from 0.4 Jy beam$^{-1}$ and increasing with
 steps of 0.5 Jy beam$^{-1}$.
{\bf (c):} C$^{18}$O peak intensity map for the Gulf region.
{\bf (d):} $^{12}$CO 0th-moment map for the Gulf region. The synthesized beam is at the lower-left corner.
{\bf (e):} $^{13}$CO 0th-moment map for the Gulf region. 
{\bf (f):} C$^{18}$O 0th-moment map for the Gulf region.
}\label{fig:peakg}
\end{figure*}

\begin{figure*}[htbp]
\epsscale{1.15}
\plotone{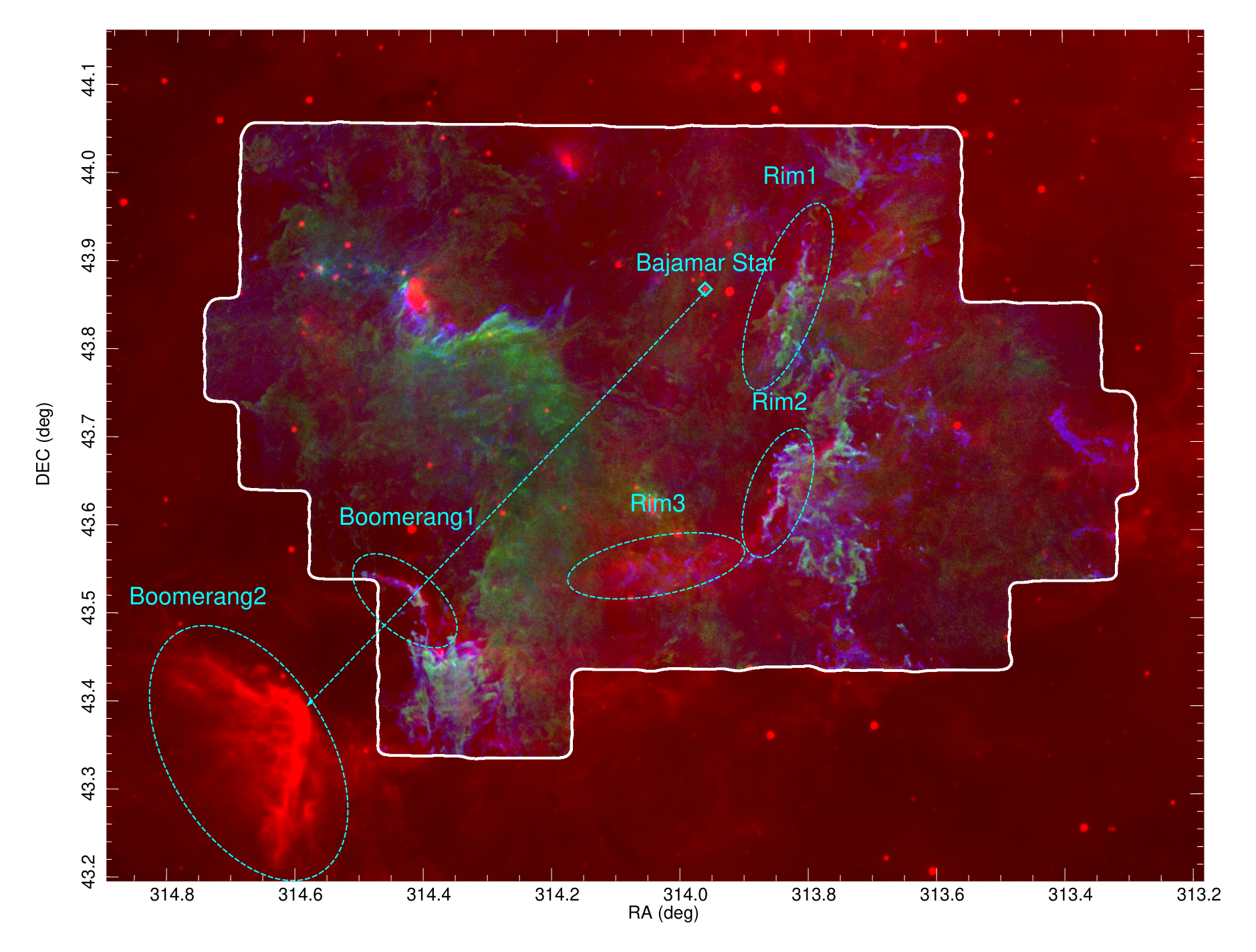}
\vspace{-0.5cm}
\caption{
RGB false-color image including the WISE 12 $\mu$m
band (red), the peak intensity map of $^{13}$CO(1-0)
(green), and the peak intensity map of $^{12}$CO(1-0)
(blue). The cyan ellipses mark the locations of the 
feedback features in Figure \ref{fig:spitzercover}.
They match well with the thin CO filaments in the region covered by our CARMA observations.
\label{fig:wise1213}}
\end{figure*}

\begin{figure*}[htbp]
\epsscale{0.37}
\plotone{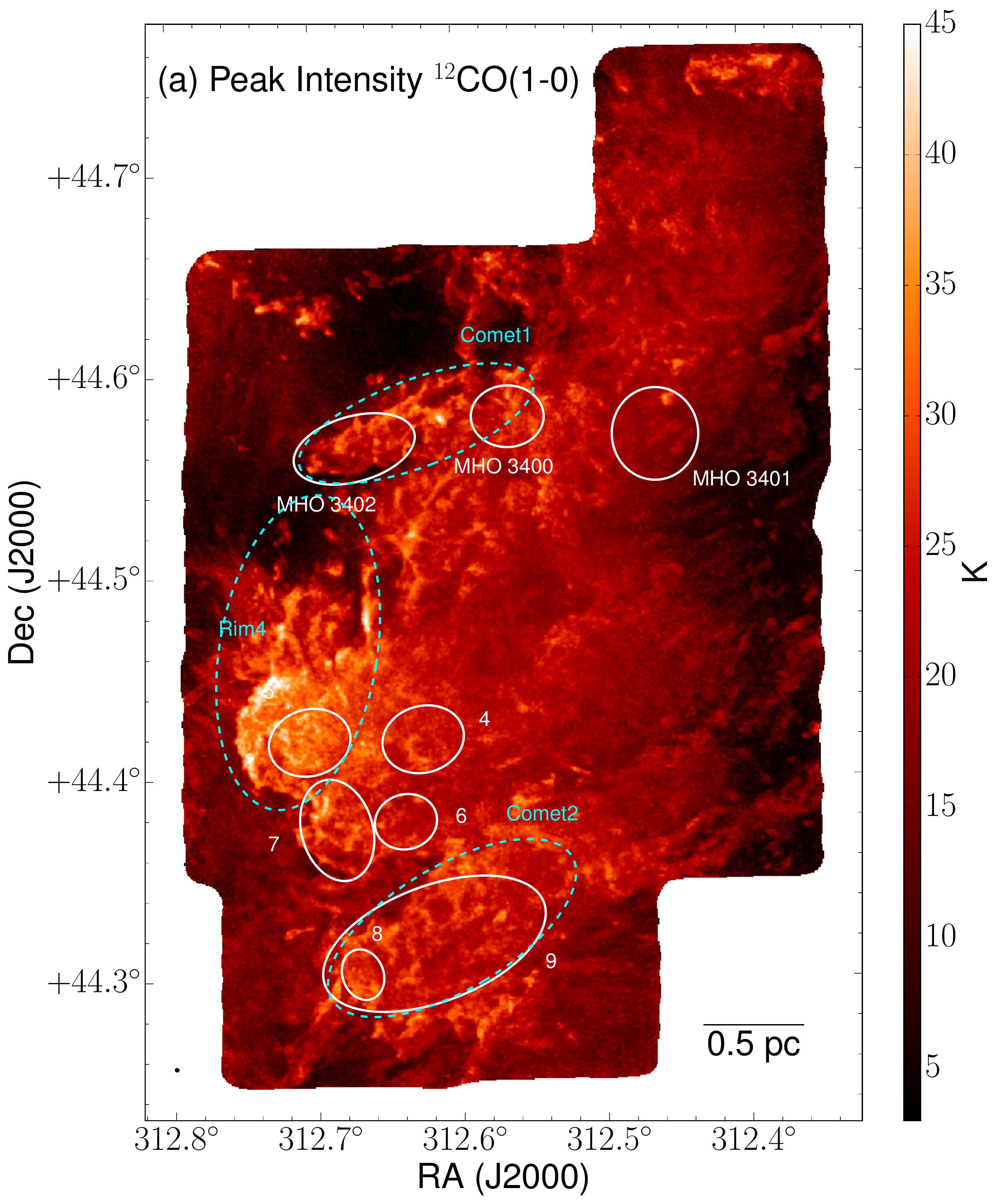}
\plotone{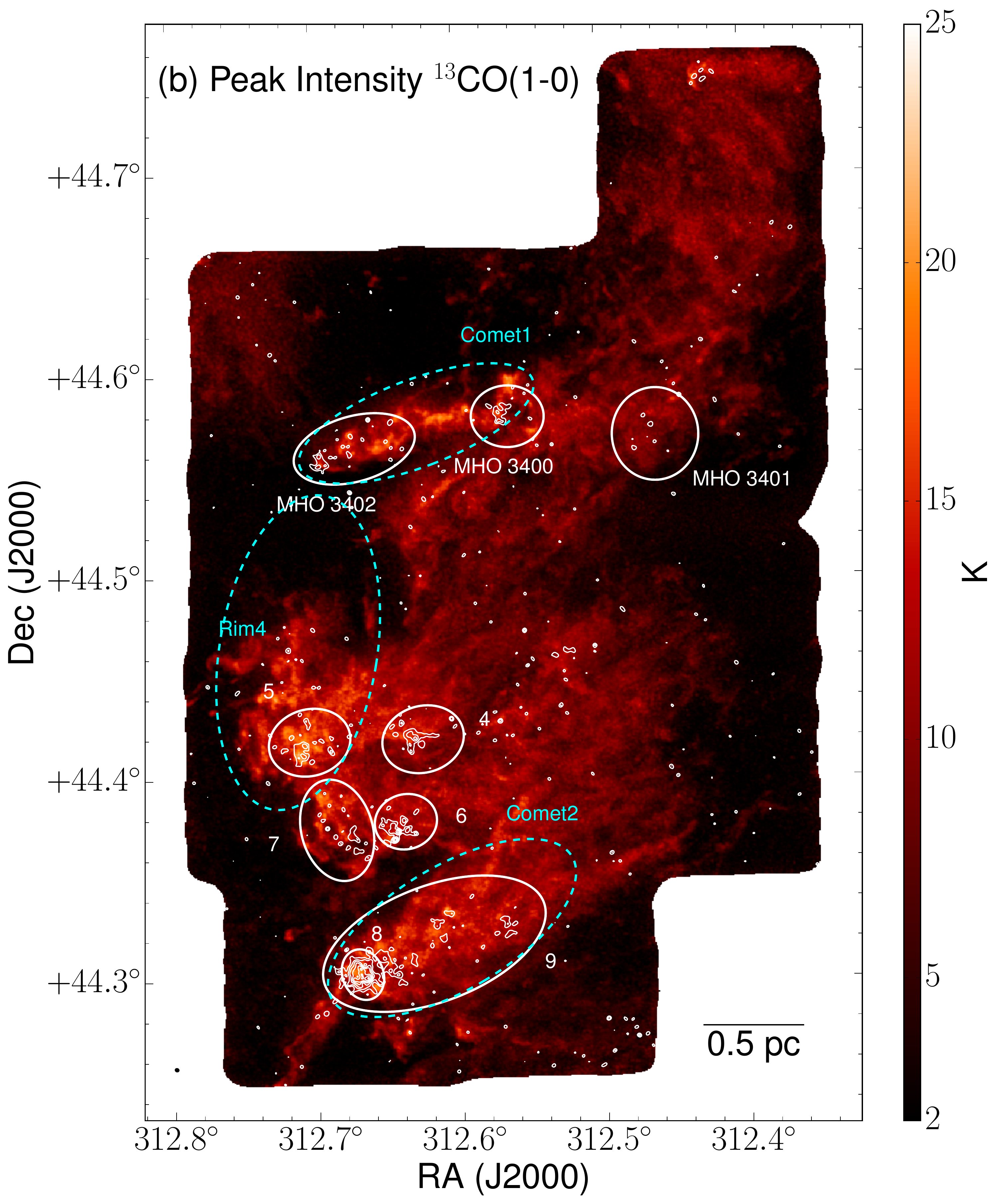}
\plotone{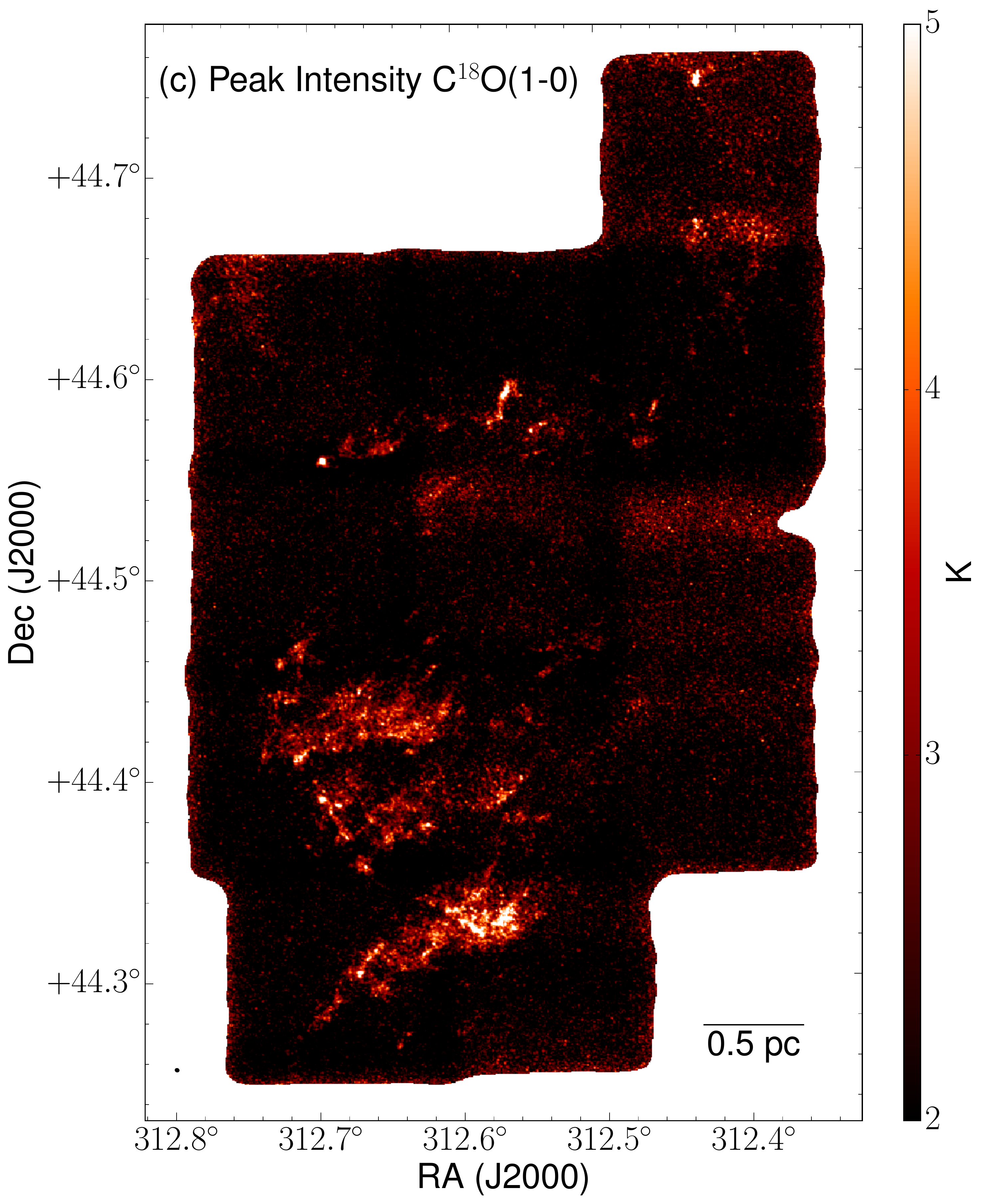}\\
\plotone{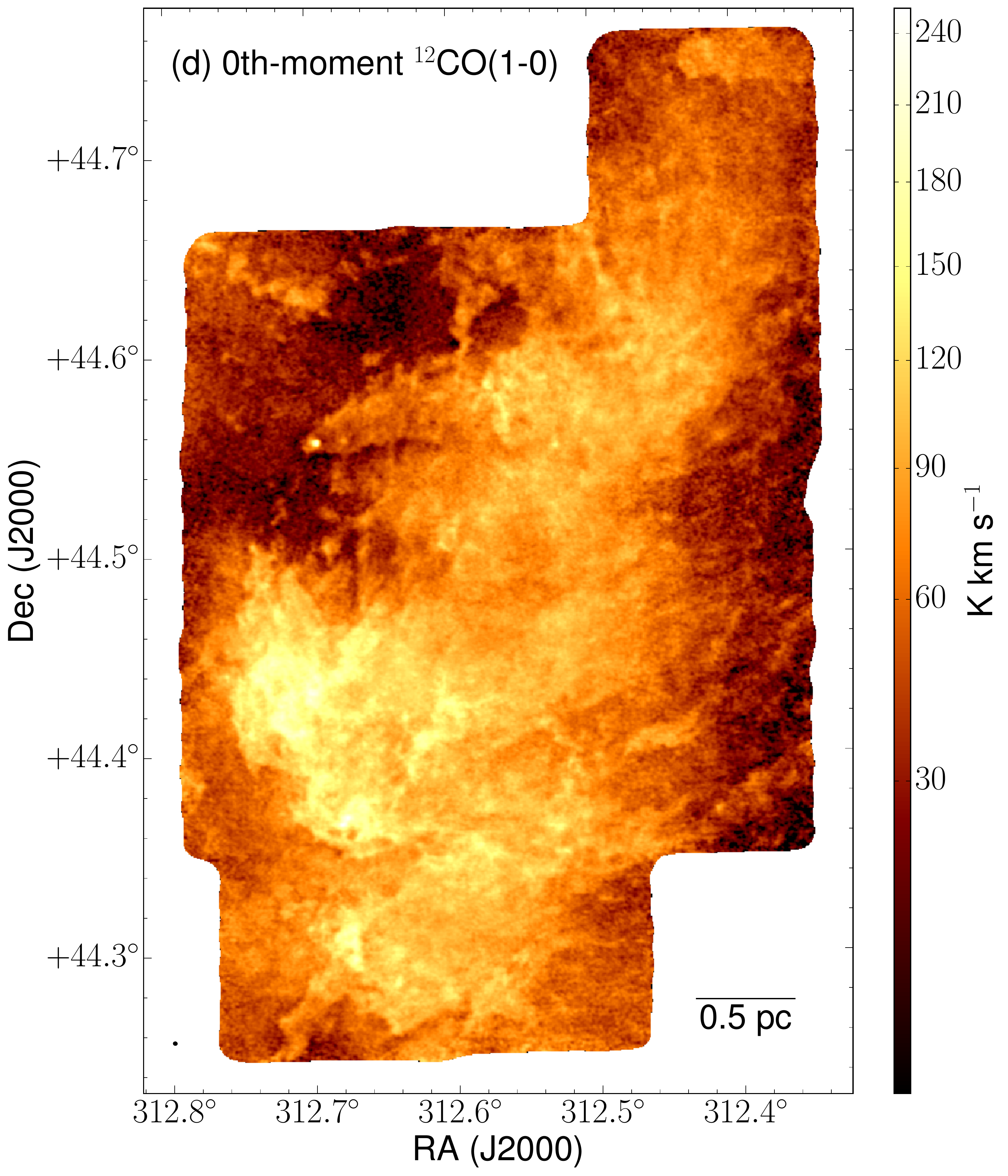}
\plotone{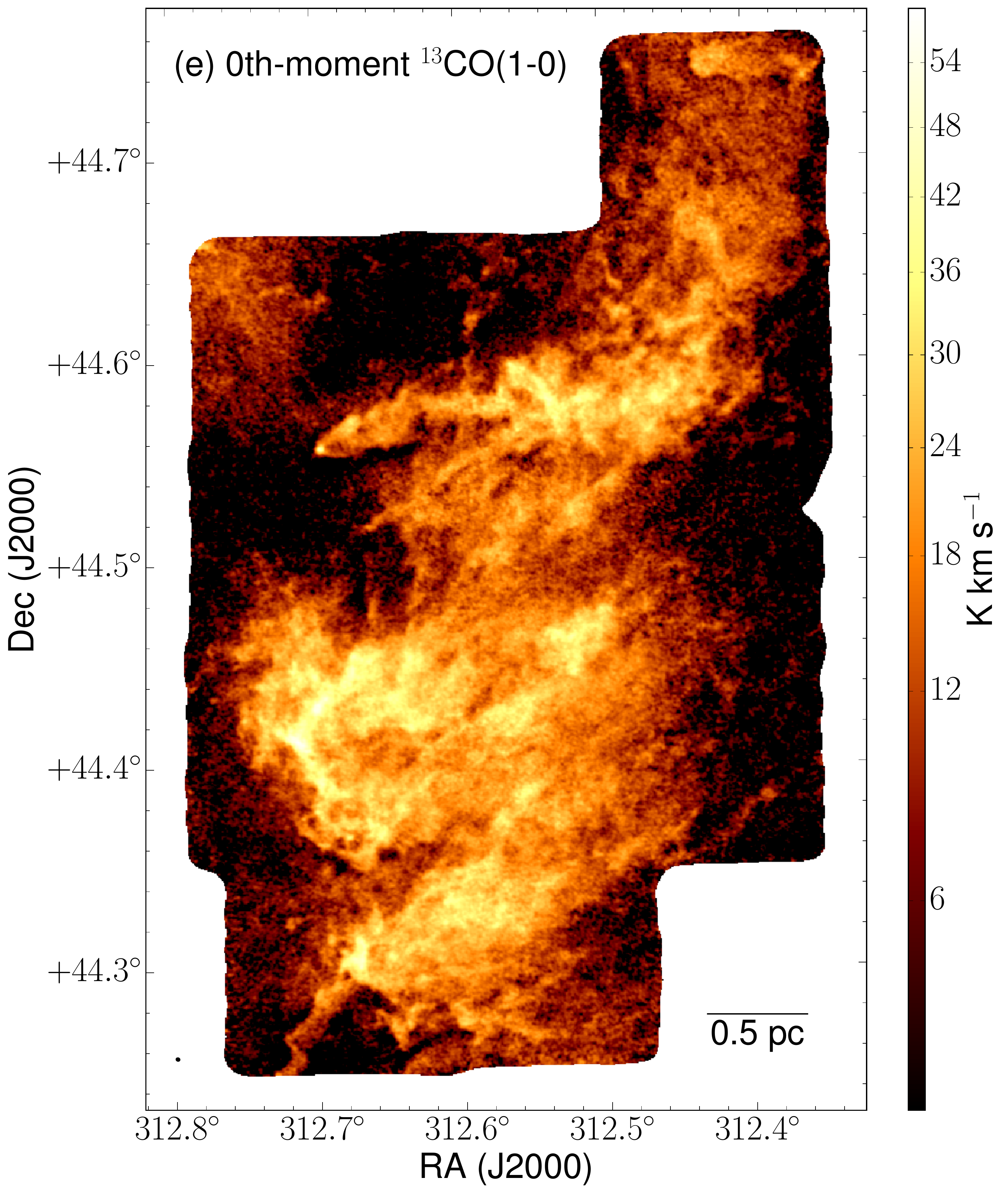}
\plotone{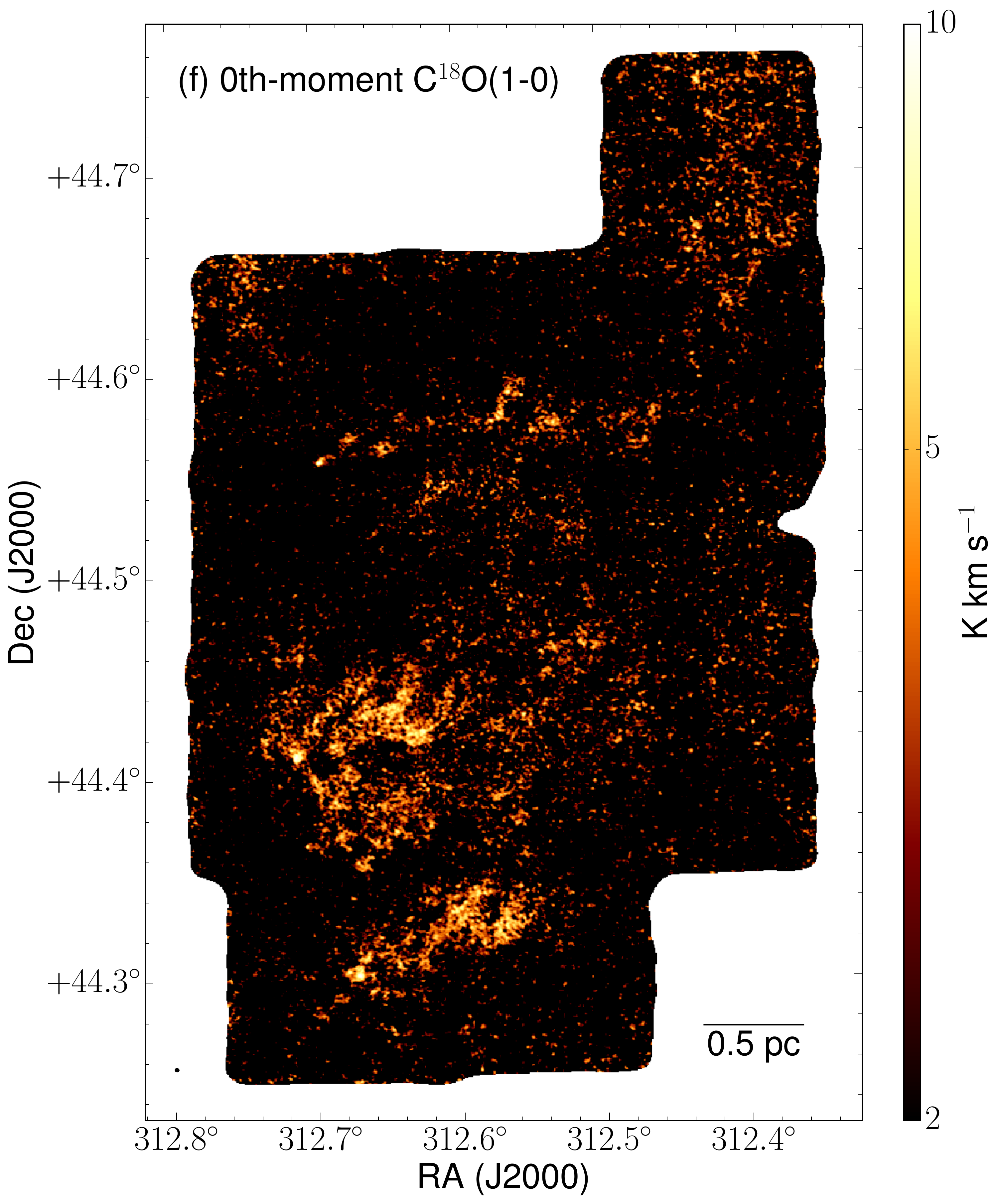}
\caption{
{\bf (a):} $^{12}$CO peak intensity map for the Pelican region. The synthesized beam is at the lower-left corner.
The white ellipses are the BGPS cores defined in B14.
The cyan ellipses are the same as in Figure \ref{fig:spitzercover}.
{\bf (b):} $^{13}$CO peak intensity map for the Pelican region. The white contours show the BGPS 1.1 mm continuum,
starting from 0.4 Jy beam$^{-1}$ and increasing with
steps of 0.5 Jy beam$^{-1}$.
{\bf (c):} C$^{18}$O peak intensity map for the Pelican region.
{\bf (d):} $^{12}$CO 0th-moment map for the Pelican region. The synthesized beam is at the lower-left corner.
{\bf (e):} $^{13}$CO 0th-moment map for the Pelican region. 
{\bf (f):} C$^{18}$O 0th-moment map for the Pelican region. 
\label{fig:peakp}}
\end{figure*}

As can be seen in the $^{12}$CO and $^{13}$CO
moment 0 maps of the Gulf region (Figures
\ref{fig:peakg}(d) and \ref{fig:peakg}(e)),
there is molecular gas practically over 
the entire area. There is only notable 
decrease in emission towards the southeast
corner (near the Boomerang1 structure) 
and towards the northern-central edge of the map.
There are also a number of bright prominent
regions throughout the maps. These are seen
better in the peak intensity maps shown in 
Figures \ref{fig:peakg}(a) and \ref{fig:peakg}(b) 
and discussed below. In contrast, the C$^{18}$O
emission is concentrated in the central-eastern
part of the map (Figures \ref{fig:peakg}(c) and
\ref{fig:peakg}(f)). This emission, as discussed
below, coincides with the denser parts of the
cloud where 1.1 mm dust continuum emission 
has been detected. 

Figures \ref{fig:peakg}(a)-\ref{fig:peakg}(c)
show the peak intensity  maps of $^{12}$CO(1-0),
$^{13}$CO(1-0), and C$^{18}$O(1-0) for the
Gulf region. We overlay the Bolocam Galactic
Plane Survey (BGPS) 1.1 mm continuum emission
\citep{2013ApJS..208...14G}, shown in white contours.
In the Gulf region, the figures show there is 
a group of six
prominent continuum sources toward the east side of the maps.
They are all part of the group of cores traced in 1.1 mm continuum emission labeled as ``Gulf total'' 
by B14. This includes  
the two cores named ``Core E'' and ``Core W'' in the northeast part of the map. South of Core W lies the ``SW1'' core, and towards the southwest of SW1 
lie (from north to south) 
the ``SW2 main'', ``SW2 SE'', ``SW2 S''
cores (B14). These last three (the SW2 cores) 
have prominent C$^{18}$O counterparts 
(Figure \ref{fig:peakg}(c)).

There are various prominent features of bright molecular line emission in the Gulf  region. Between continuum cores W and SW2 main lies a region of high $^{12}$CO and $^{13}$CO emission that mostly extends from east to west which we call the ``Wall'' (see Figure
 \ref{fig:peakg}(a)
and  \ref{fig:peakg}(b)). It is likely that the Wall, which is at least twice as bright as the average emission in the rest of the map,
traces the warm gas heated by a  young star that lies just north of this structure and is responsible for a bright IR nebulosity (hereafter SW-IR) seen in Figures \ref{fig:spitzercover} and \ref{fig:wisecover}.

Additional prominent line emission regions can be seen in the south-east corner and the mid-west part of the map. In the south-east,
the bright emission is coincident with the 
Boomerang1 feature
discussed above (see \S\ref{subsec:midir} and Figure  \ref{fig:spitzercover}). 
In the central-western region of the map, the bright emission is associated with the IR features labeled Rim1, Rim2 and Rim3 (see Figure \ref{fig:spitzercover} and \S\ref{subsec:midir}).
These, in general, show a sharp increase in intensity towards the edge of the emission structure facing the Bajamar Star
(aligning with the IR bright rims defined in
\S\ref{subsec:midir} and Figure \ref{fig:spitzercover})
and  have a chaotic morphology in $^{12}$CO 
(Figure \ref{fig:peakg}(a)) and $^{13}$CO 
(Figure \ref{fig:peakg}(b)) away from these bright rims. As discussed later, these structures are part of the filament F-1 defined in the study by Z14.

We show a three-color image in Figure
\ref{fig:wise1213} comparing the WISE 
12 $\mu$m band, the peak intensity map of
$^{13}$CO, and the peak intensity map of $^{12}$CO.
It is very clear that the bright-rim features in
the IR image are associated with the bright
filaments in molecular line emission. 
Rim3 overlaps with ripple-like features in CO
emission. Moreover, the entire molecular gas
structure  that connects with the rims appear as
shredded pieces. A plausible explanation is that
the expanding bubble has reached (or is very close to) the molecular gas associated with these structures,
and the gas has this chaotic morphology because it is being dispersed by the UV photons
from the massive star.

In the Pelican Head region, we detect $^{12}$CO 
(Figure \ref{fig:peakp}(a)(d))
throughout most of the mapped area which roughly
follows a morphology similar to the structure
delineated by the yellow-colored features in the 
Spitzer and WISE three-color composite images.
The $^{12}$CO, $^{13}$CO and C$^{18}$O integrated
intensity maps (Figures \ref{fig:peakp}(d) to 
\ref{fig:peakp}(f)) show three separate regions of
relatively high-intensity emission: 
in the north (around decl.~44.58\arcdeg), 
center (around decl.~44.43\arcdeg),
and south (around decl.~44.3\arcdeg).
These structures are also clearly seen in the peak
intensity maps (Figures \ref{fig:peakp}(a) to \ref{fig:peakp}(c)). 
The northern and southern high-intensity regions 
(which we name Comet1 and Comet2, respectively)
have comet-like structures with bright narrow heads 
and with tails extending to the northwest.
The central high-intensity structure 
(which we name Rim4) has a wide arc-like edge and
extended (bright) emission to the west, 
somewhat similar to the Rim2 
structure in the Gulf region. 
All these features roughly 
point toward the Bajamar Star, 
and thus very likely caused by feedback 
from the massive star.

Comet1 is shown in detail in Figure \ref{fig:combine}.
The CARMA high-resolution data shows some ripples at the 
boundaries of Comet1 (especially the northern boundary).
Interestingly, both cometary clouds
have BGPS sources at their head location,
reminiscent of triggered star formation by radiatively
driven implosion \citep{1989ApJ...346..735B}.
The northern one corresponds to the ``Pelican
MHO 3402'' core in B14 (Figure 3, Table 4),
while the southern one corresponds to the
``Pelican 8'' core in B14.

\subsection{Gas kinematics}\label{subsec:kin}



\begin{figure*}[htbp]
\epsscale{0.55}
\plotone{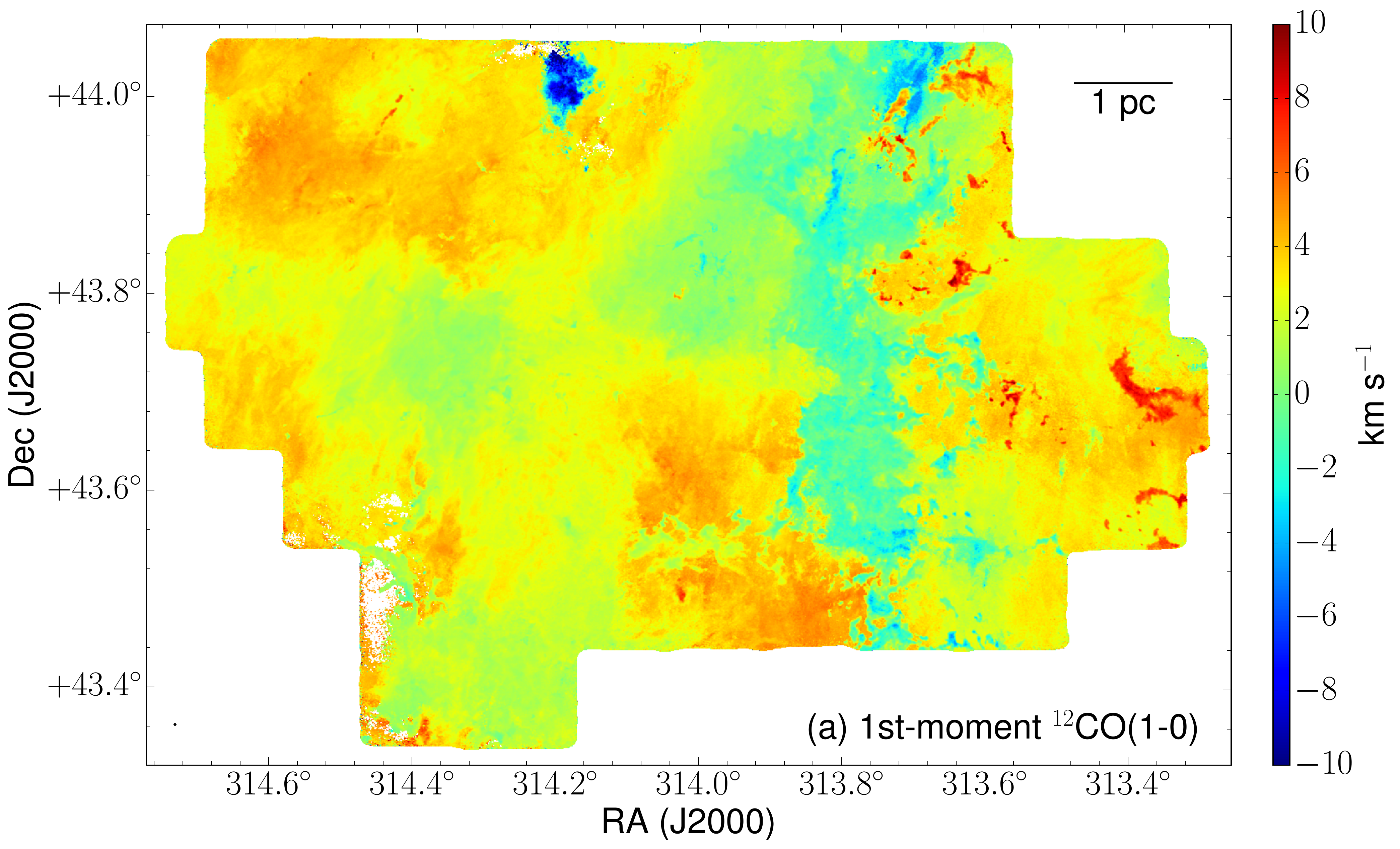}
\plotone{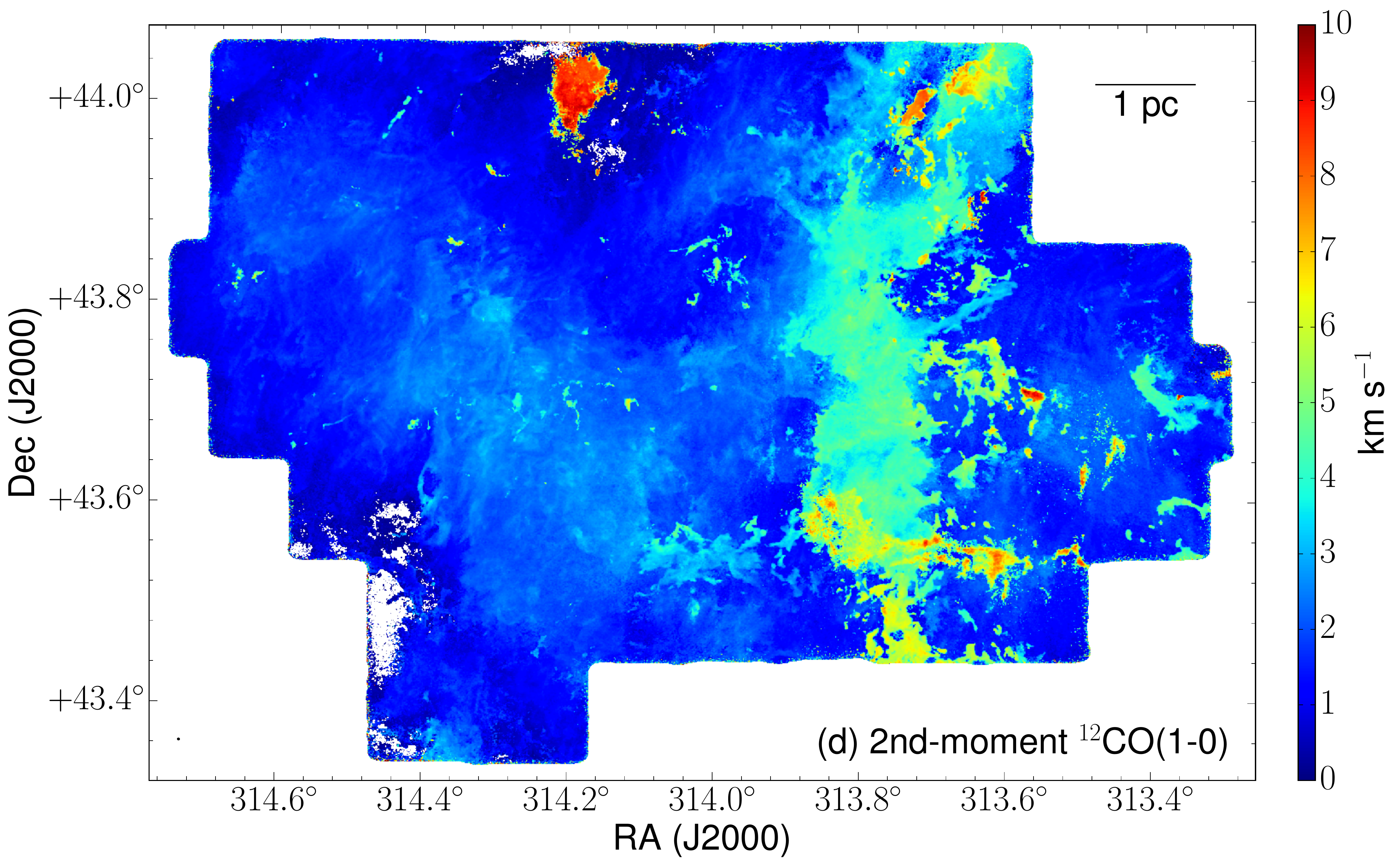}\\
\plotone{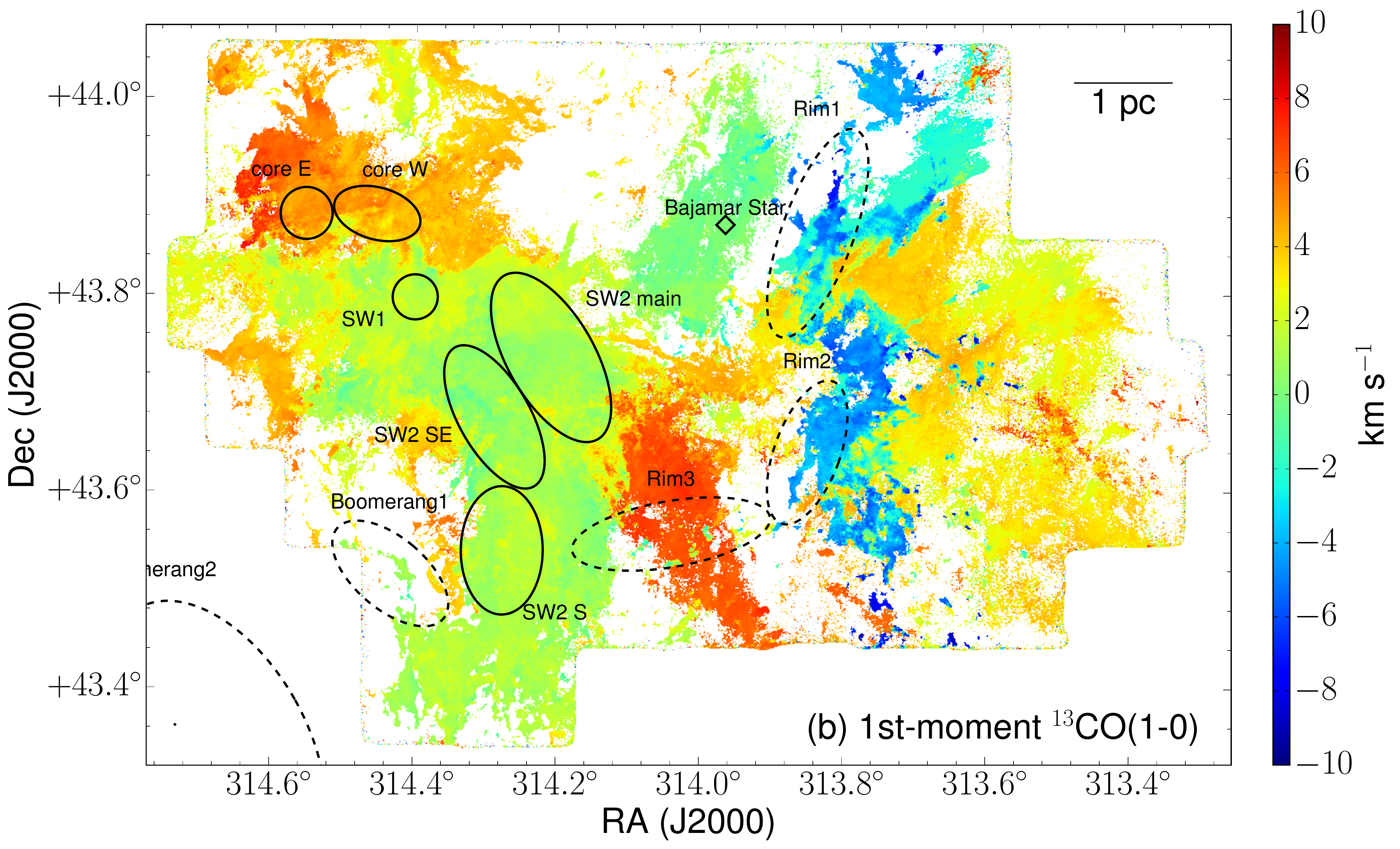}
\plotone{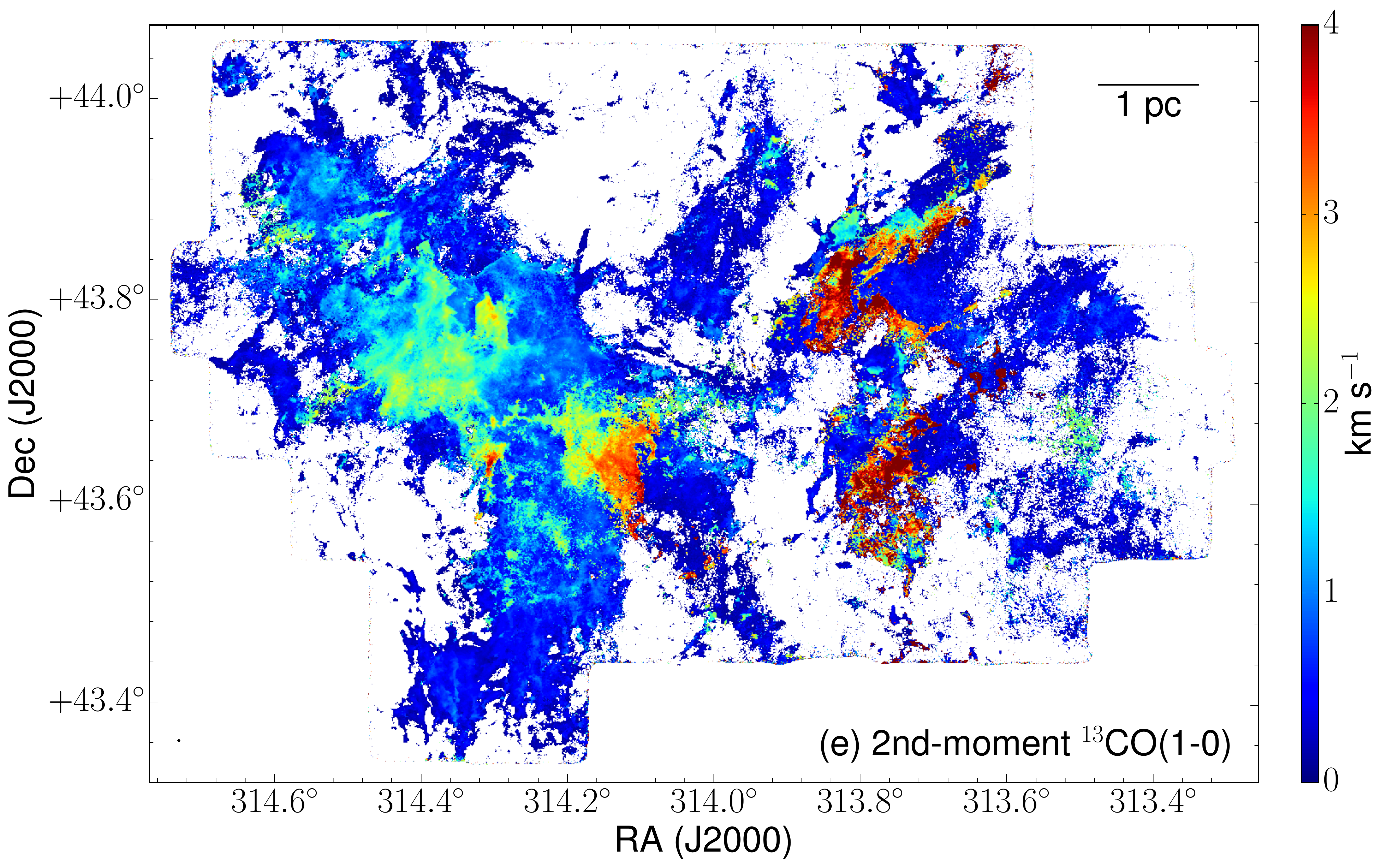}\\
\plotone{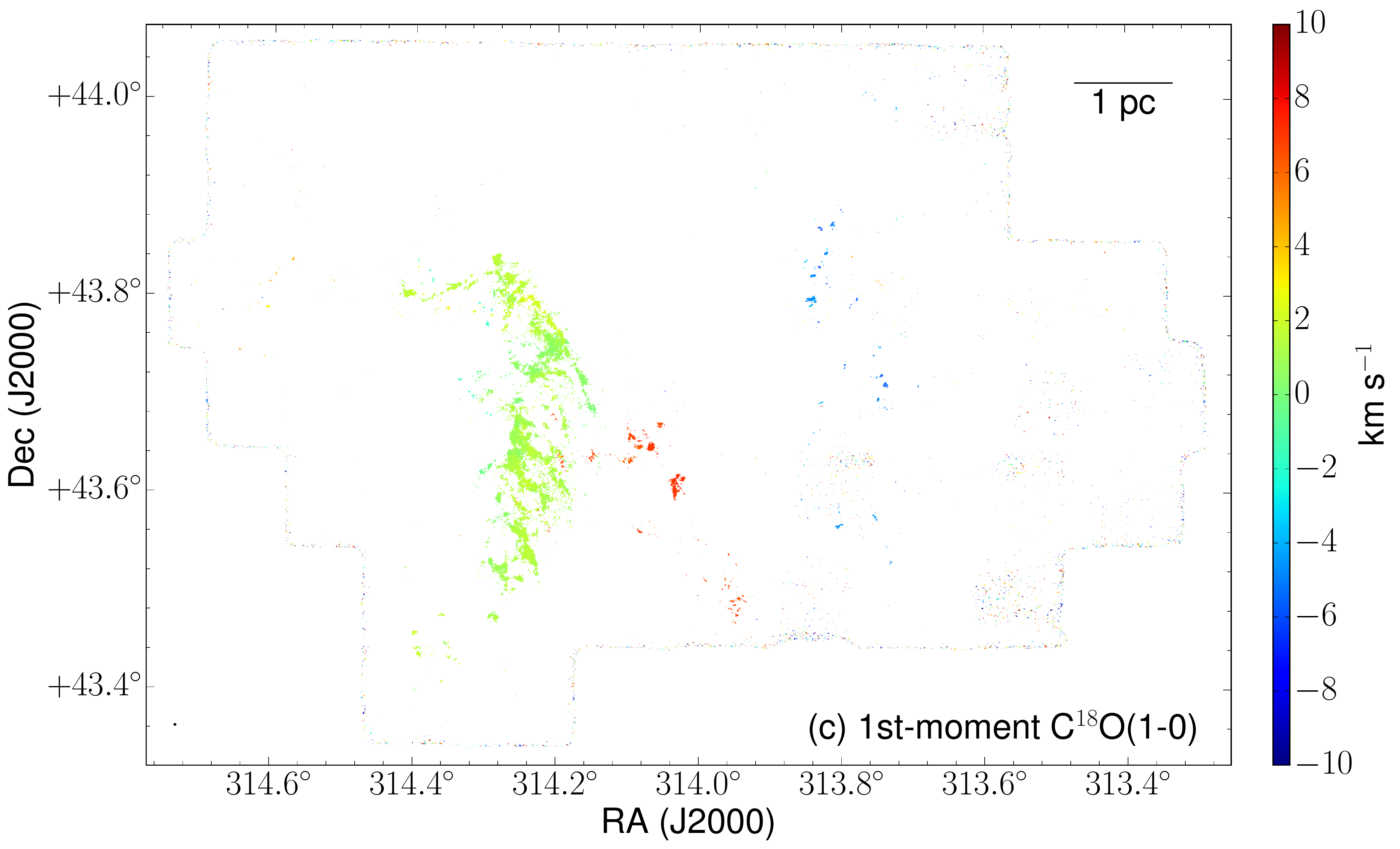}
\plotone{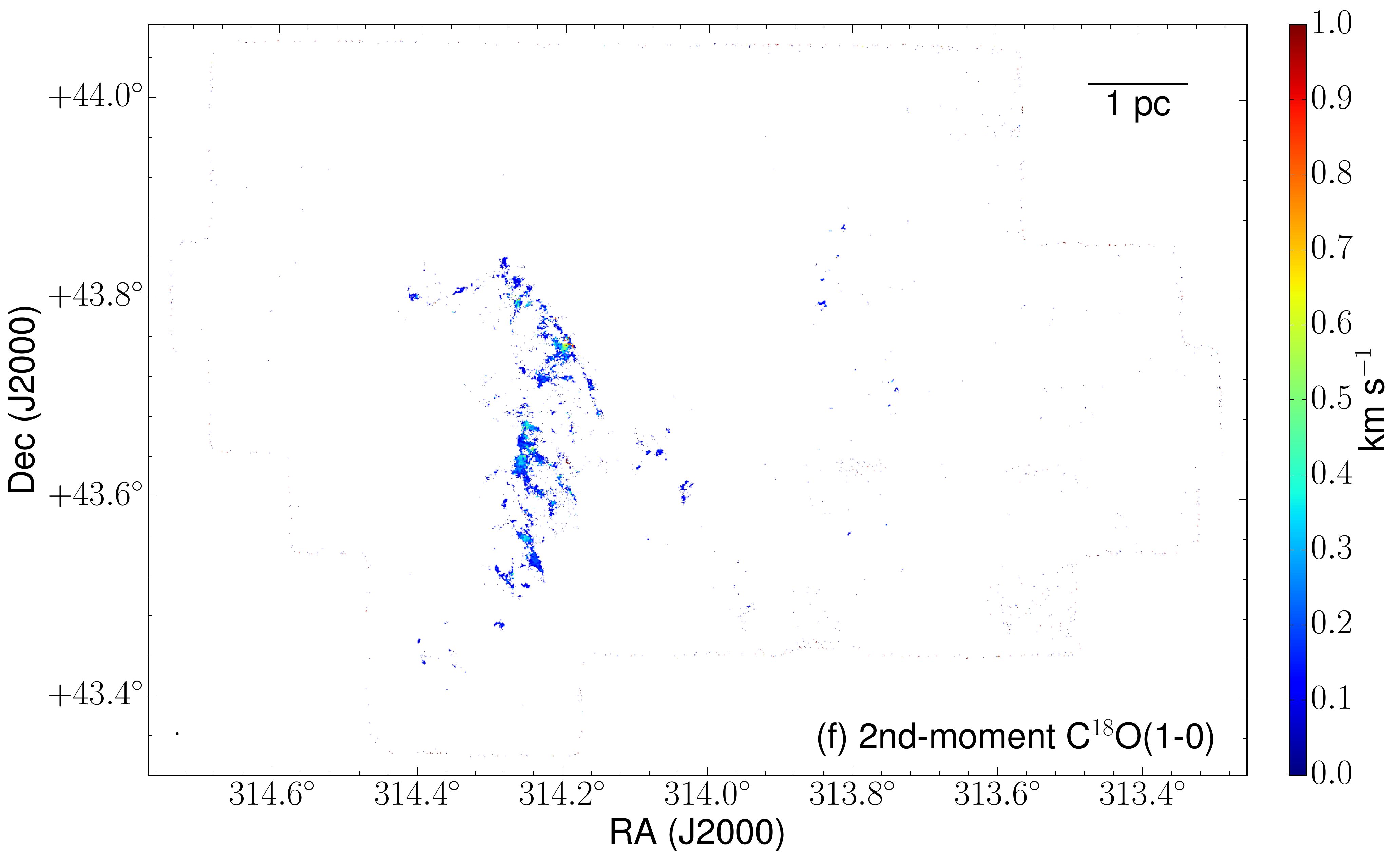}
\caption{
{\bf (a):} $^{12}$CO 1st-moment map for the Gulf region.
The synthesized beam is at the lower-left corner.
{\bf (b):} $^{13}$CO 1st-moment map for the Gulf region.
The  ellipses with black solid lines are the BGPS cores defined in B14.
The dashed ellipses are the same as the cyan ones in Figure \ref{fig:spitzercover}.
{\bf (c):} C$^{18}$O 1st-moment map for the Gulf region.
{\bf (d):} $^{12}$CO 2nd-moment map for the Gulf region.
The synthesized beam is at the lower-left corner.
{\bf (e):} $^{13}$CO 2nd-moment map for the Gulf region.
{\bf (f):} C$^{18}$O 2nd-moment map for the Gulf region.
}\label{fig:king}
\end{figure*}

\begin{figure*}[htbp]
\epsscale{0.37}
\plotone{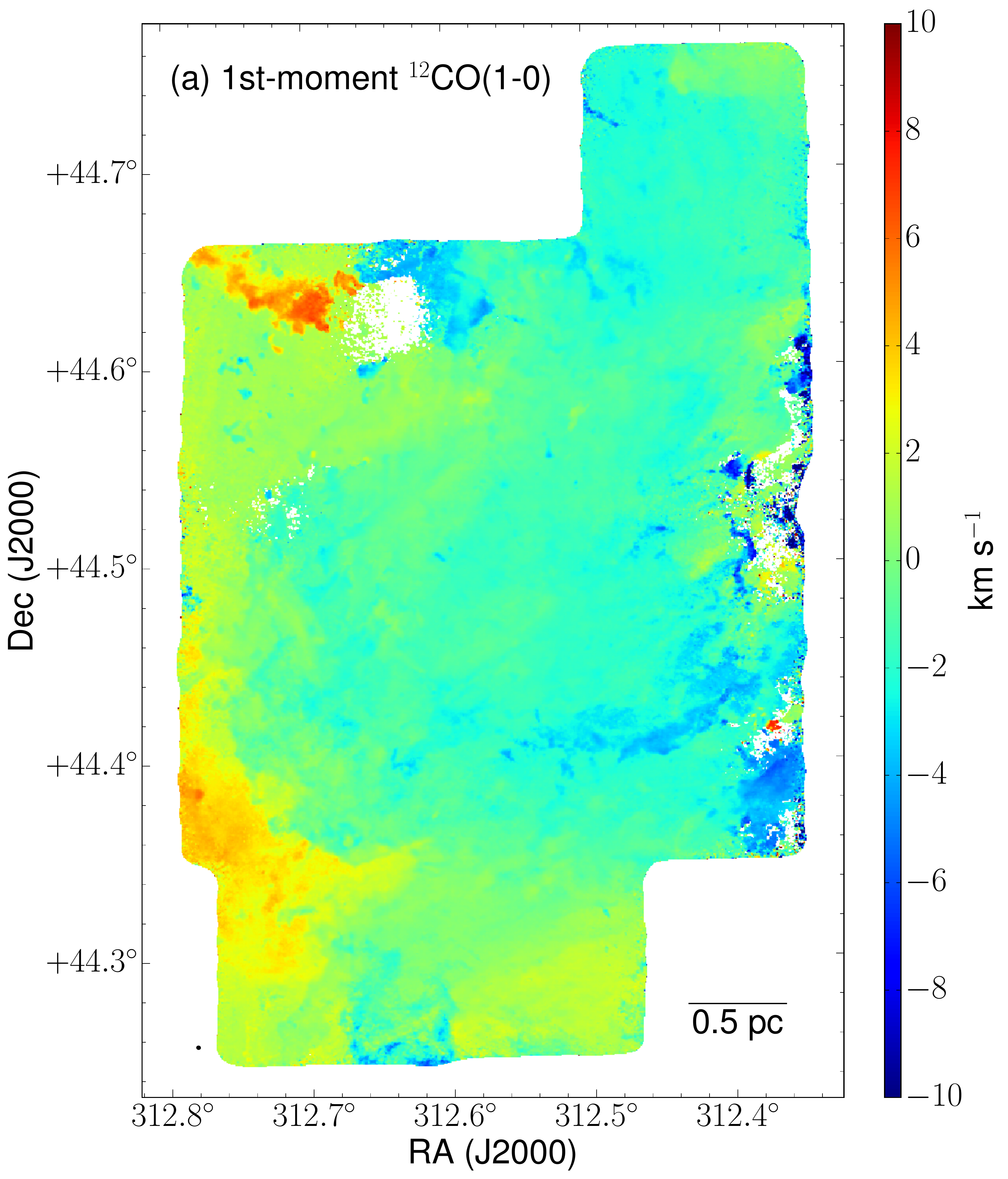}
\plotone{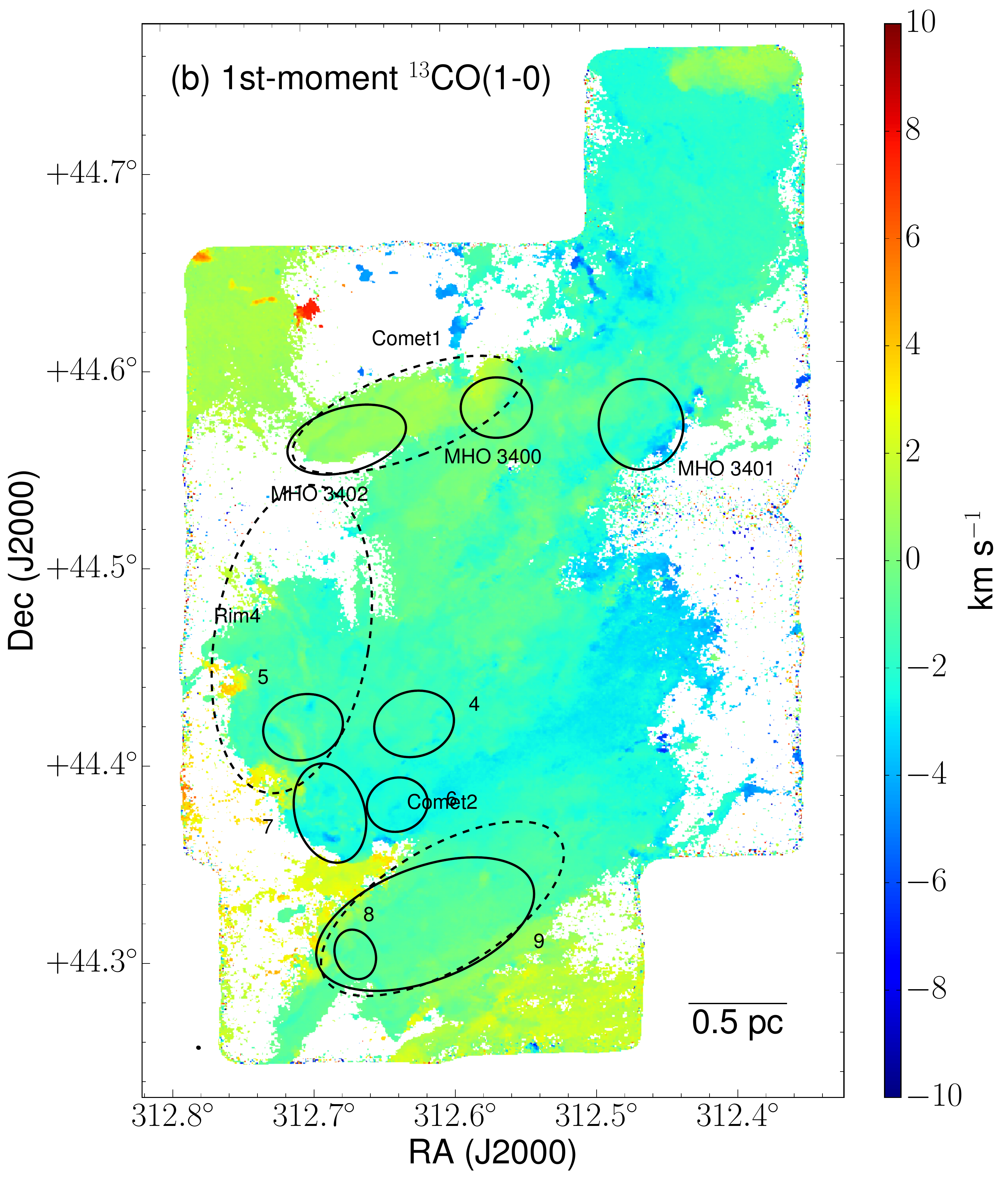}
\plotone{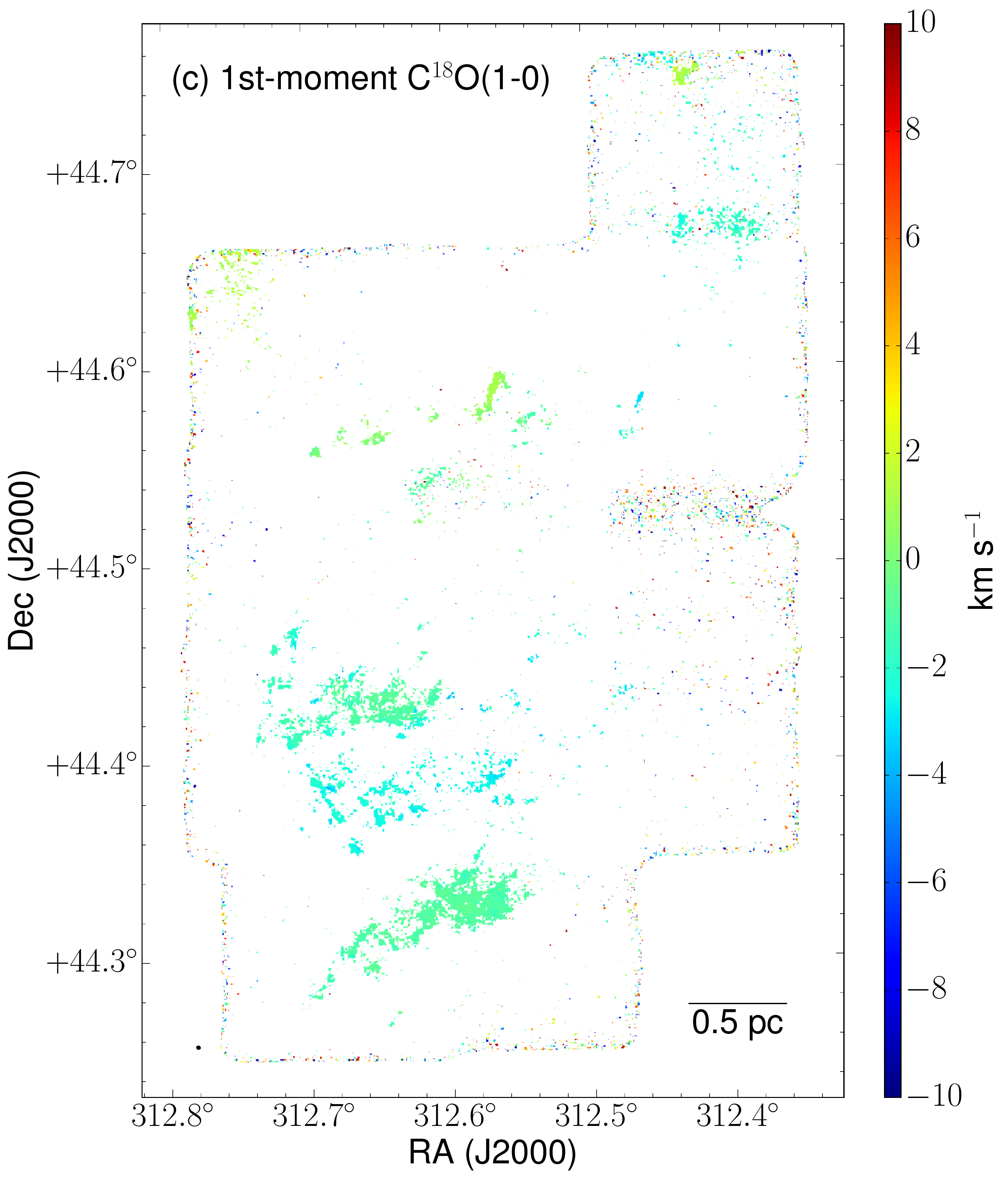}\\
\plotone{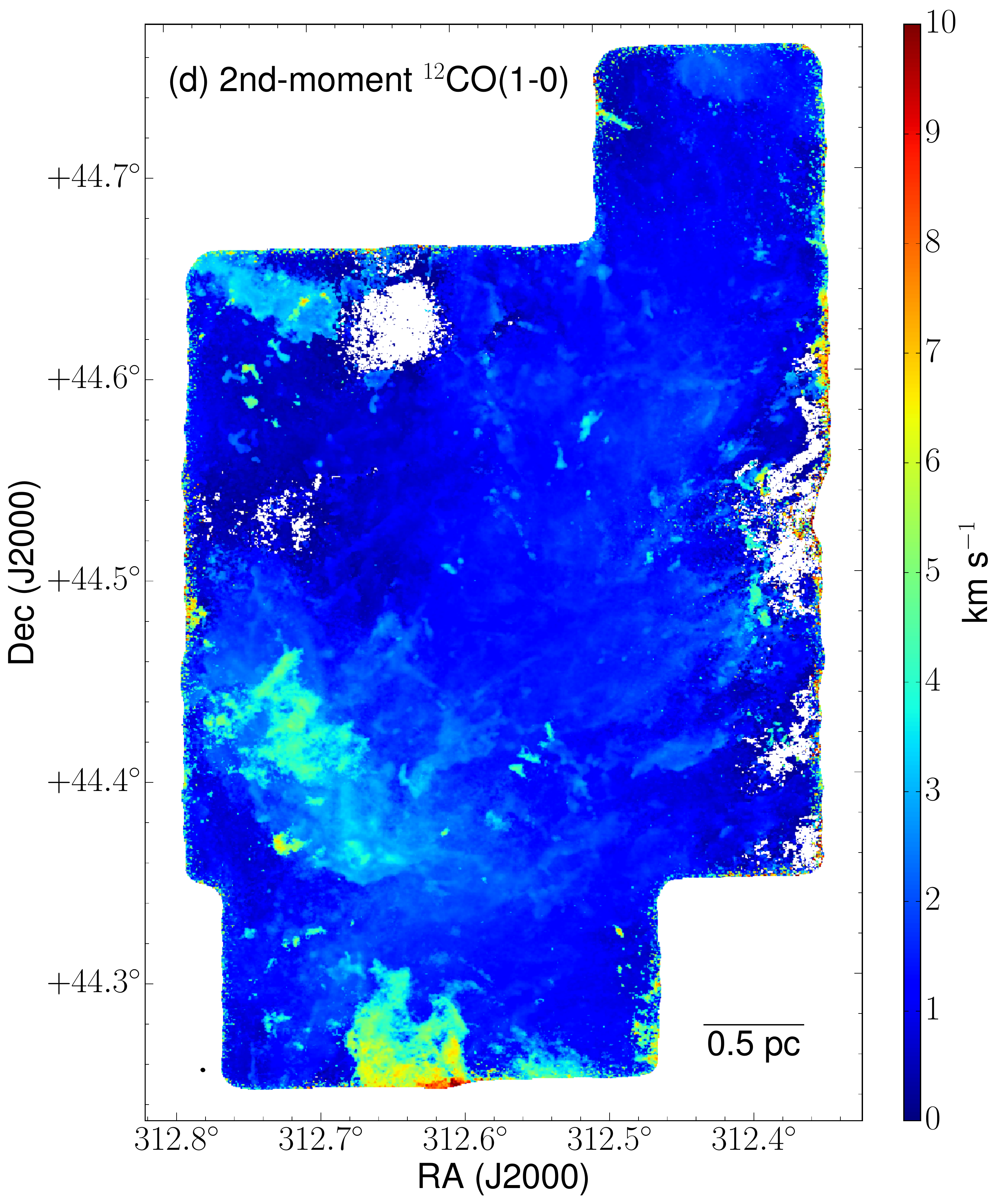}
\plotone{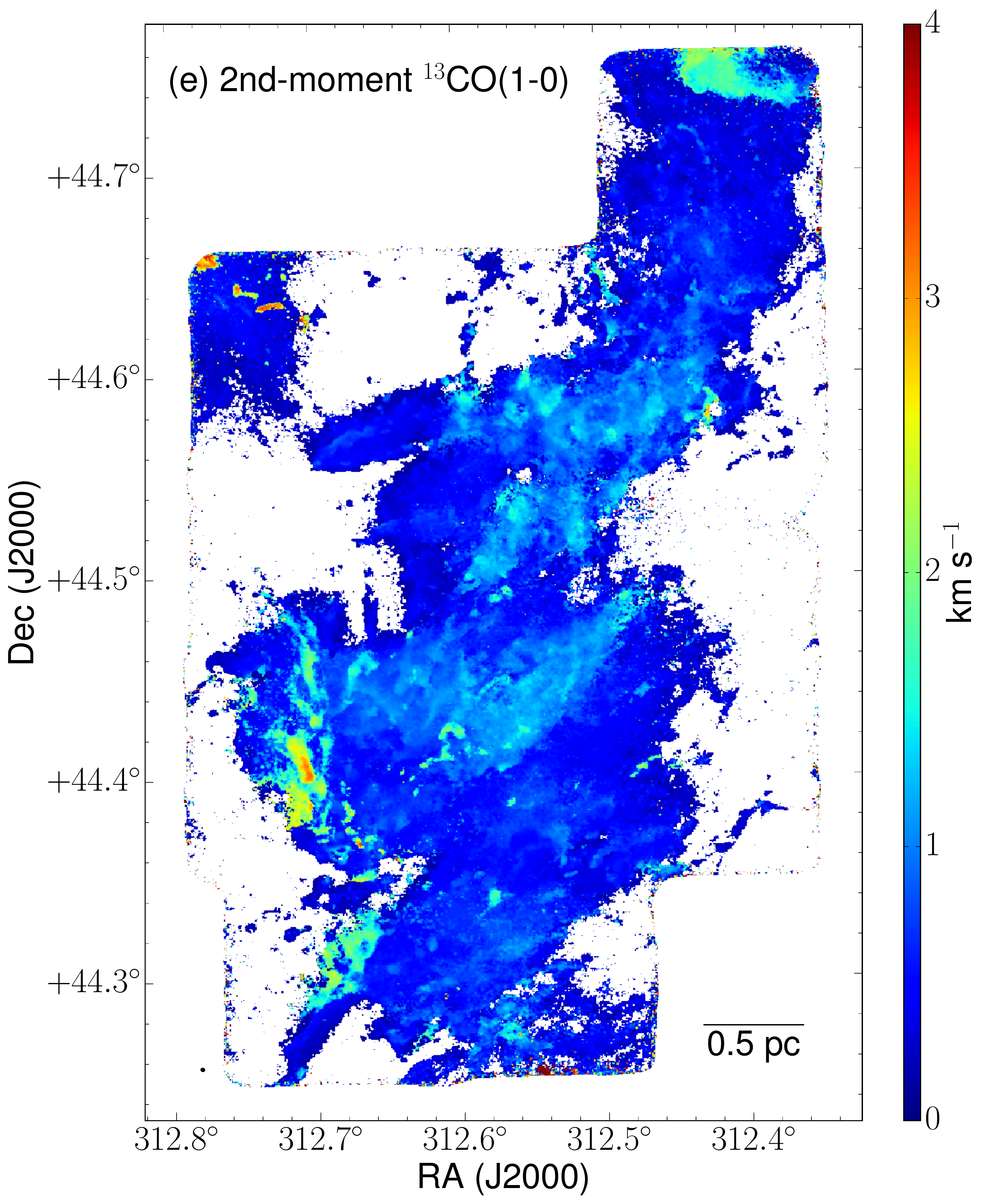}
\plotone{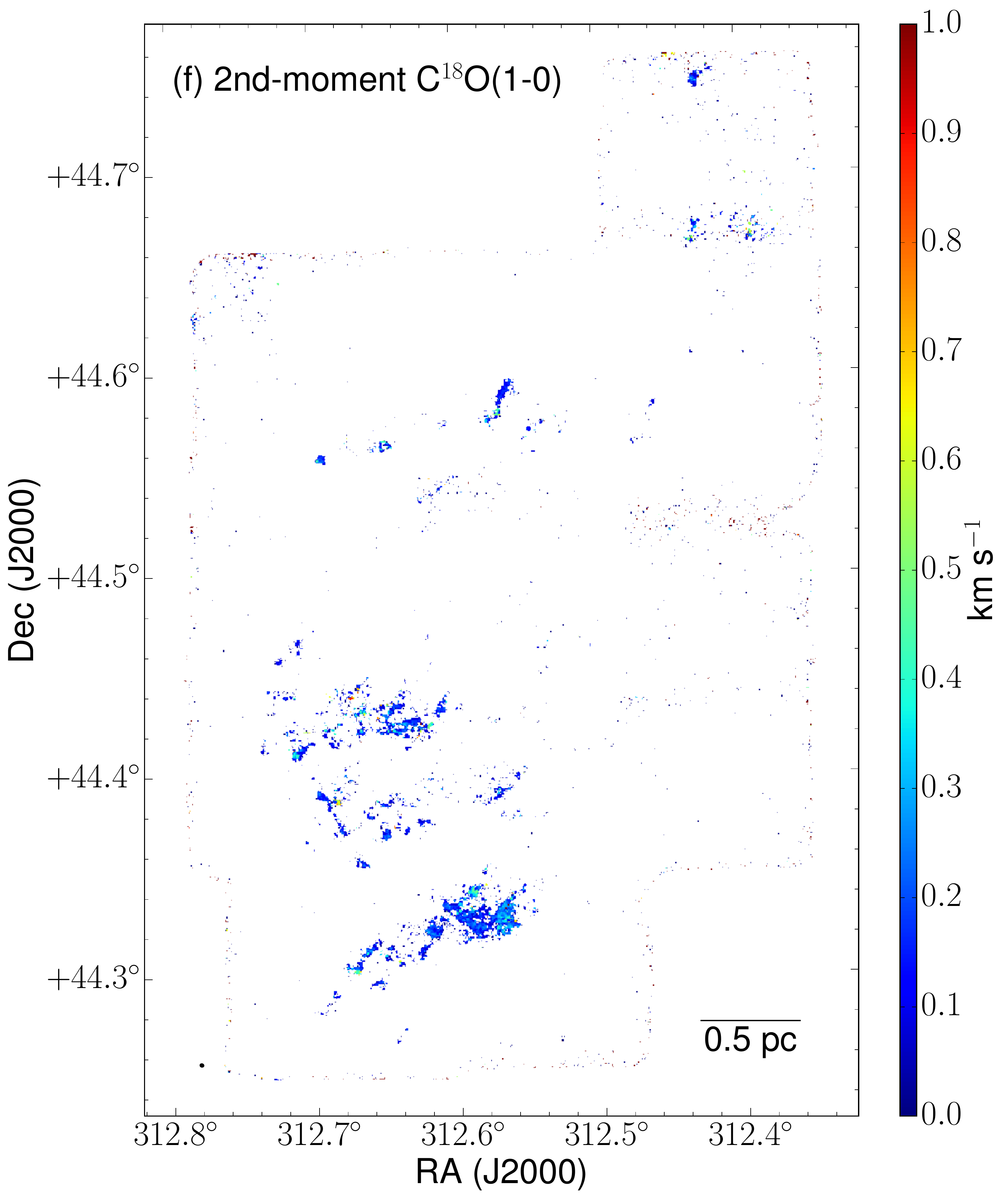}
\caption{
{\bf (a):} $^{12}$CO 1st-moment map for the Pelican region.
The synthesized beam is at the lower-left corner.
{\bf (b):} $^{13}$CO 1st-moment map for the Pelican region.
The  ellipses with black solid lines are the BGPS cores defined in B14.
The dashed ellipses are the same as the cyan ones in Figure \ref{fig:spitzercover}.
{\bf (c):} C$^{18}$O 1st-moment map for the Pelican region.
{\bf (d):} $^{12}$CO 2nd-moment map for the Pelican region.
The synthesized beam is at the lower-left corner.
{\bf (e):} $^{13}$CO 2nd-moment map for the Pelican region.
{\bf (f):} C$^{18}$O 2nd-moment map for the Pelican region.
}\label{fig:kinp}
\end{figure*}

Figures \ref{fig:king}(a) to \ref{fig:king}(c)
show the intensity-weighted average velocity  
(1st-moment) maps of the Gulf region. They
show a general picture of the complex kinematics
in this region. 
In the $^{12}$CO and $^{13}$CO 1st-moment maps
(Figures \ref{fig:king}(a) and 
\ref{fig:king}(b)), the molecular gas associated  
with the IR-bright rims (i.e., Rim1, Rim2)
can be seen as gas with average velocities 
of about -5 km~s$^{-1}$, in sharp contrast with 
adjacent gas at more redshifted velocities. 

In contrast to the Gulf region, the Pelican Head 
region (Figures \ref{fig:kinp}(a)
to \ref{fig:kinp}(c))
shows a significantly narrower range of
intensity-weighted average velocities. 
Here most of the emission is concentrated at 
velocities between -2 and 2 km~s$^{-1}$. 
In the $^{13}$CO 1st-moment map
(Figures \ref{fig:kinp}(b)),
Comet1 is discernible as an oval region with  
velocities around 2 km~s$^{-1}$. Comet2 and Rim4
structures are not as conspicuous as Comet1,
but are similar to the other feedback features in 
the NAP, in that the gas beyond the sharp edge is 
at redshifted velocities compared to the feedback structure.

Figures \ref{fig:king}(d) to \ref{fig:king}(f) 
show the 2nd-moment maps of the Gulf region.
In the $^{12}$CO and $^{13}$CO maps
(Figures \ref{fig:king}(d) and 
\ref{fig:king}(e)), the gas associated with
the bright rims shows large velocity dispersion.
On the other hand, the C$^{18}$O gas (Figure \ref{fig:king}(f)) has narrow velocity widths,
in sharp contrast with the 
$^{12}$CO and $^{13}$CO gas. 
Figures \ref{fig:kinp}(d) to \ref{fig:kinp}(f)
show the 2nd-moment maps of the Pelican region.
The gas facing the Bajamar Star shows larger 
velocity dispersion. The gas behind this front
is more quiescent. Interestingly, Comet1 does
not show large dispersion in either $^{12}$CO
or $^{13}$CO maps.

\subsection{Gas Structures Around W80}\label{subsec:loc}

The moment maps by themselves are not enough to
differentiate among the different cloud components. 
The distribution of dark patches in the optical 
image of the region in concert with the location 
and kinematics of the molecular gas can be used
to derive an approximate three-dimensional 
distribution of the gaseous structures observed 
in our map. 

The dark patches in the optical image
of the NAP in Figure \ref{fig:DSScoverage},
particularly the Gulf and Atlantic regions,
have been generally thought to be caused by molecular
clouds in front of the HII region, mostly at the same
distance (B14, Z14). However, a closer look at 
Figure \ref{fig:DSScoverage} suggests a more complicated
story. The dark clouds in this image show at 
least two shades of darkness.
One is a semi-transparent gray silhouette that 
runs from the top-right (northwest) edge of the
figure throughout the Atlantic area, 
extending south of the Bajamar Star. 

A large area
southeast of the Bajamar Star also shows this 
semi-transparent grayish hue. 
We call areas with this shade  as ``gray regions''. 
In Figure~\ref{fig:DSScoverage},
we also see a collection of patchy
clumps that are much more opaque than the gray regions,
which traverse from east to
west across the entire Gulf region. We name
these ``dark regions''. The two different shades,
combined with the information from images at other
wavelengths (and the gas kinematics), 
imply that there  may be more than one group of
molecular gas (possibly at different distances)
along the line of sight, consistent with our
implications based on the kinematics information.
Determining the relation between these two 
groups of clouds is important for identifying the 
molecular gas responsible for the star formation
that is taking place in this region and for 
understanding the impact of feedback on the
molecular clouds in the NAP.

\begin{figure*}[htbp]
\epsscale{1.2}
\plotone{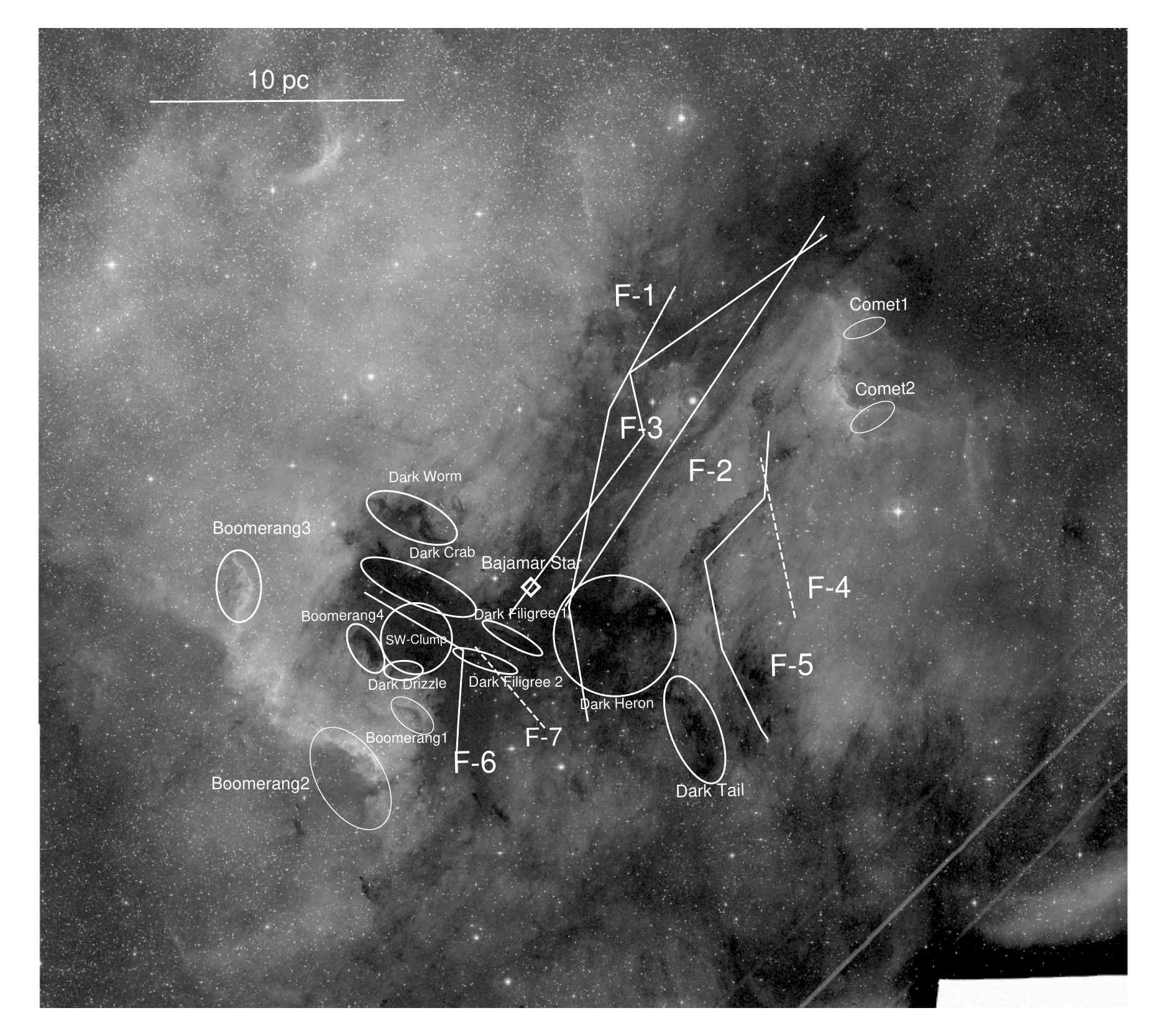}
\caption{
Definition of molecular gas filaments and clumps 
by comparing the POSS-II red color (0.7 $\mu$m)
and the $^{13}$CO(1-0) cubes (\S\ref{subsec:loc}, 
\S\ref{app:struc}, Table \ref{tab:name}).
{\bf Background:} POSS-II red color plate.
The regions mark the filaments and clumps defined
in \S\ref{subsec:loc} and \S\ref{app:struc}.
The filaments with solid lines are in the foreground.
Those with dashed lines are behind the bubble.
Filaments F-1, F-2, F-3, F-4 were defined in Z14.
The green boundaries show the CARMA mosaic footprints.
\label{fig:compare}}
\end{figure*}

In \S\ref{app:struc} and Figures \ref{fig:compare1}
to \ref{fig:compare7}, we  
identify distinctive, coherent features in the
$^{13}$CO position-position-velocity (PPV) cube, 
and compare the gas distribution with
different features seen in the POSS-II
red color plate.
Figure \ref{fig:compare} summarizes the major
structures defined based on this comparison.

\begin{deluxetable}{ccccc}
\tablecolumns{5}
\tablewidth{0pt}
\tablecaption{Cloud Naming \label{tab:name}}
\tablehead{
\colhead{Name} &
\colhead{RA} &
\colhead{DEC} &
\colhead{$v_{\rm lsr}$} &
\colhead{Figure}\\
\colhead{} & 
\colhead{(deg)} & 
\colhead{(deg)} & 
\colhead{$(\rm km~s^{-1})$} & 
\colhead{}}
\startdata
F-1 & 313.64623 & 44.09655 & -4.9 & \ref{fig:compare1} \\  
F-2 & 313.31135 & 44.36088 & -2.7 & \ref{fig:compare2} \\  
F-3 & 313.47317 & 44.35101 & 0 & \ref{fig:compare3} \\
F-4 & 312.99136 & 43.98752 & 3.5 & \ref{fig:compare5} \\  
F-5 & 313.11716 & 43.81911 & 0 & \ref{fig:compare3} \\
F-6 & 313.32268 & 43.60290 & 0 & \ref{fig:compare3} \\  
F-7 & 314.04069 & 43.57983 & 6.5 & \ref{fig:compare7} \\  
Boomerang1 & 314.43230 & 43.51767 & 1.5 & \ref{fig:compare4} \\  
Boomerang2 & 314.67420 & 43.34278 & 0 & \ref{fig:compare3} \\  
Boomerang3 & 315.10378 & 43.88507 & -2.7 & \ref{fig:compare2} \\  
Boomerang4 & 314.61085 & 43.70758 & 4.9 & \ref{fig:compare6} \\  
Comet1 & 312.62712 & 44.57655 & 2 & \ref{fig:combine},\ref{fig:kinp} \\  
Comet2 & 312.60718 & 44.32575 & 0 & \ref{fig:kinp} \\  
Dark Crab & 314.40155 & 43.87826 & 4.9 & \ref{fig:compare6} \\  
Dark Drizzle & 314.46471 & 43.64457 & 4.9 & \ref{fig:compare6} \\  
Dark Filigree 1 & 314.03919 & 43.73130 & 3.5 & \ref{fig:compare5} \\  
Dark Filigree 2 & 314.14630 & 43.66815 & 4.9 & \ref{fig:compare6} \\  
Dark Heron & 313.64063 & 43.73207 & 3.5 & \ref{fig:compare5} \\  
Dark Tail & 313.33844 & 43.46264 & 3.5 & \ref{fig:compare5} \\  
Dark Worm & 314.42482 & 44.07318 & 3.5 & \ref{fig:compare5} \\  
SW-Clump & 314.41301 & 43.73359 & 3.5 & \ref{fig:compare5} \\  
\enddata
\tablecomments{F-1, F-2, F-3, F-4 were defined in Z14. Figure \ref{fig:compare} shows a summary of the features overlaid on the POSS-II red color image.}
\end{deluxetable}

\section{Three-dimensional Structure}

The most useful information from the small-scale structures (captured by CARMA) is the morphology matching with the bright infrared rims. The matching between F-1 and the rims confirms that the rims are part of the molecular filament. This finding shows that the filament is being heated and dispersed by the massive star. The morphology match between the molecular filament and the gray regions in the POSS-II image thus reveal that the gray regions in the optical images likely trace  foreground molecular gas that is being destroyed by the Bajamar Star, and therefore
very close to the expanding bubble.
As a result of this, the three-dimensional picture of the entire NAP becomes clearer.

\subsection{Three-dimensional structure of the NAP complex}\label{subsec:3d}

\begin{figure*}[htbp]
\epsscale{1.1}
\plotone{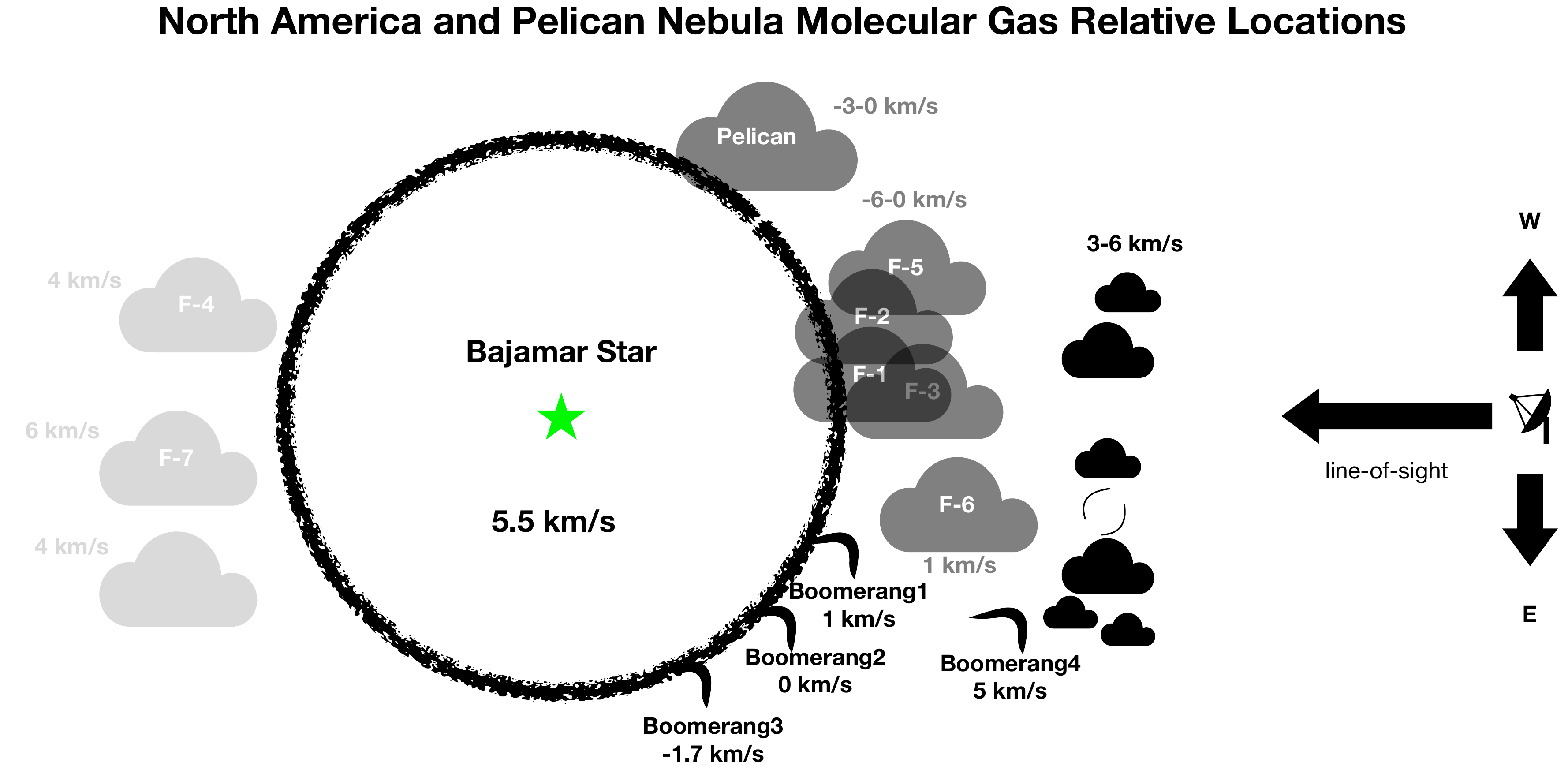}
\caption{
The line-of-sight locations of the filaments
and clumps relative to the W80 bubble (black circle).
The Bajamar Star is at the center
of the bubble. Three groups of clouds are shown,
including the far side clouds (F-4, F-7, and another
clump to the northeast of the Gulf region, see 
\S\ref{subsec:loc} and Figure \ref{fig:compare5}),
the foreground contacting gray-clouds, 
and the foreground dark clumps
and filigrees (closest to the observer). 
The LSR velocities of the clouds are also included.
We assign a velocity of 5.5 km s$^{-1}$ for W80
based on the radio recombination line observation
by \citet{1989ApJS...71..469L}.
\label{fig:3d}}
\end{figure*}

Figure \ref{fig:3d} shows a schematic of the 
line-of-sight structure of the NAP complex.
The locations are in relative sense and
are based on the findings from \S\ref{subsec:loc}
and \S\ref{app:struc}.
We include major structures that
are defined in \S\ref{app:struc}. 
The structures are mainly in
the Gulf and Atlantic regions, with a few in the
Pelican region and behind the W80 bubble.

There are three major groups of clumps and 
filaments based on their line-of-sight distance.
The closest to us
is a group of dark clumps with varying morphologies
(see Figure \ref{fig:compare} and the rightmost dark
structures in Figure \ref{fig:3d}).
They are in front of the W80 bubble and appear
as dark shadows in the POSS-II images.
The group members are small and clumpy, 
with many $\sim$ 10000 AU globules. 
Their $V_{lsr}$ is similar to that of the bubble. 

Between the dark group and the W80 bubble
is the group of gray clouds. 
F-1 is very close to the bubble being part of this group; it
is lit in the mid-to-far infrared and shows
bright rims (\S\ref{subsec:loc} and Figures 
\ref{fig:spitzercover}, \ref{fig:wisecover}, \ref{fig:peakg}, \ref{fig:compare1}).
F-2, F-3, F-5 are also close to the bubble, 
especially in the Atlantic region,
based on their similar extinction and velocity
as the F-1 filament. F-6 may not be so close to
the bubble as F-1, but it is 
between the bubble and the Dark Crab.
Boomerang1 and Boomerang4 
appear to be dark in the optical image. 
They are on the near side of the bubble. 
Boomerang2 and Boomerang3 are probably
on the edge of the expanding bubble.

Behind the bubble, there are several background clumps,
including F-4 and F-7. They do not show gray/dark
counterparts in the optical images.
They are moving away at $\sim$ 5 km s$^{-1}$.
It is not clear whether they are close to the bubble.
We suspect that F-7 (covered by our CARMA observations) is not very close to the HII region as it does not
exhibit the ``shredded'' and chaotic structure seen in F-1. 

\subsection{Correlation between Gas and Extinction}\label{subsec:corr}

\begin{figure*}[htbp]
\epsscale{1.15}
\plotone{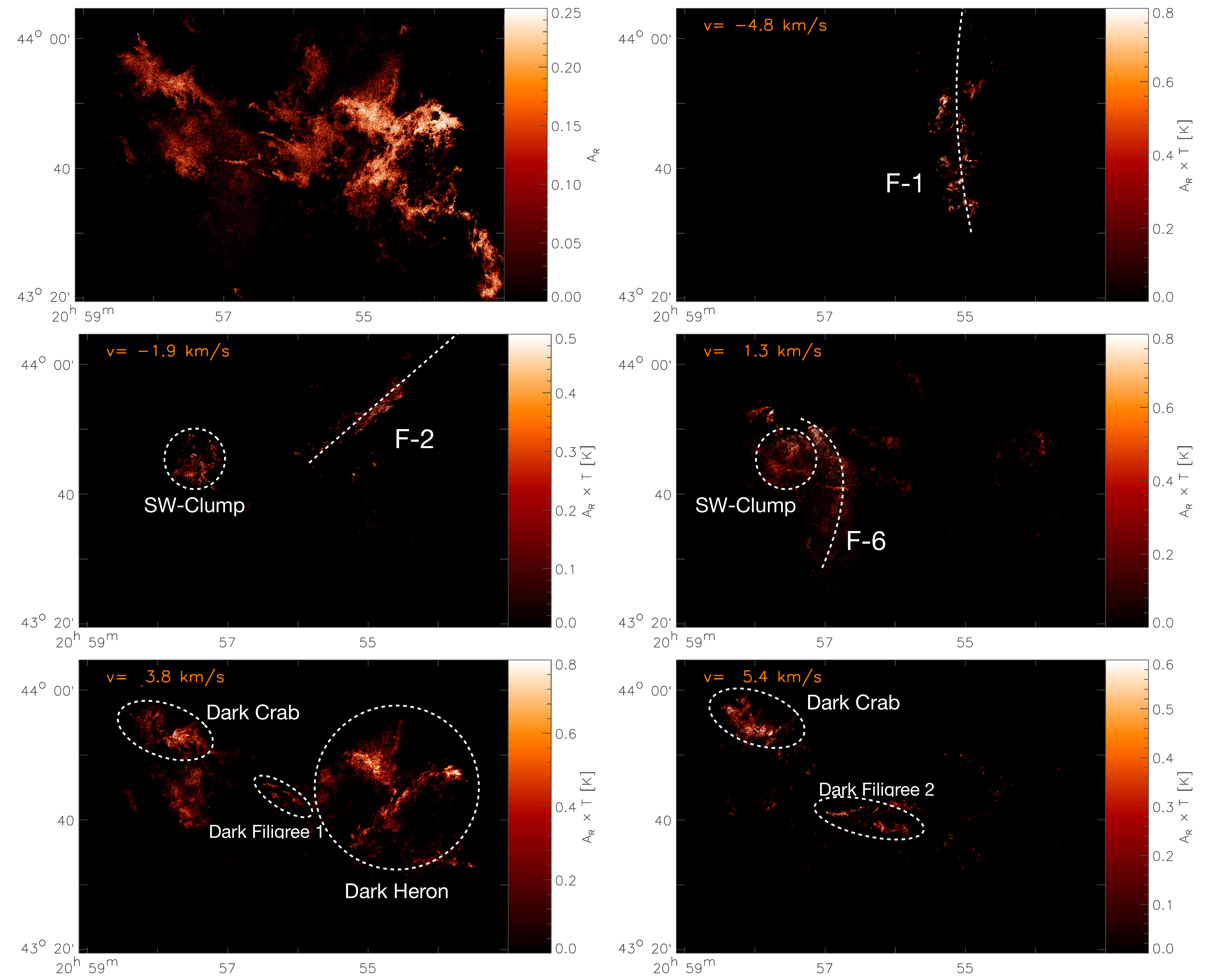}
\caption{
Maps showing the correlation between optical extinction and $^{13}$CO emission. {\bf Top Left:} Map of foreground extinction based on the POSS-II red image in Figure \ref{fig:compare}.1. {\bf Other Panels:} Covariance maps 
of the foreground with selected $^{13}$CO velocity channels. The velocity is labeled at top left corners.
\label{fig:covariance_maps}}
\end{figure*}

As an independent approach to associate the different
velocity components to structures seen in the optical
image as foreground layers we correlate the $^{13}$CO 
data with the foreground extinction. We focus on the 
combined data in the Gulf region. Instead of using 
existing extinction maps that trace the whole column 
of material \citep[e.g.][]{2002AJ....123.2559C}, 
we obtain an approximate foreground extinction map
from the POSS-II red image in Figure \ref{fig:compare}
by assuming that the highest intensity in the 
region represents the typical radiation from the 
bubble and translating all lower intensities into 
an an effective red extinction. The resulting foreground
extinction map is shown in the top left panel of Figure \ref{fig:covariance_maps}.

We investigate the correlation between $^{13}$CO gas
and the extinction by computing the covariance 
between the optical image and $^{13}$CO channel 
maps\footnote{In contrast to the mathematically 
correct covariance, we do not subtract the mean 
values before the multiplication of both maps to 
focus on emission only.}. For the $^{13}$CO channel 
covariance maps (shown in \ref{fig:covariance_maps}), 
only contributions above $4\sigma$ of the 
noise were considered. 

We find strong correlation around -5 km s$^{-1}$
(the channel including F-1, Figure \ref{fig:compare1}),
-2 km s$^{-1}$ (including F-2, Figure \ref{fig:compare2}), 
and 4 km s$^{-1}$ (including SW-Clump, Dark Filigree 1,
Dark Heron, Figure \ref{fig:compare5}). 
At 1.3 km s$^{-1}$ (including F-6, Figure \ref{fig:compare4}) 
and 6.4 km s$^{-1}$ (including Dark Crab,
F-7, Figure \ref{fig:compare7}),
we actually find a significantly weaker
correlation because of additional emission 
at those velocities stemming from the background.
Most importantly, the calculation suggests
that the Dark Crab is a distinct structure from
the F-6 filament. Although it appears at the same velocity as the 
F-7 gas both the dark appearance at near-infrared wavelengths and
the well resolved structure indicate that it must represent material
in front of the bubble, closer towards us.
This is important for being able to  match
the known YSO clusters in the NAP region with their  host molecular gas  (see \S\ref{subsec:sf}). 
Of particular interest is the  SW-Clump, which seems to have  counterparts at
-2 and 4 km s$^{-1}$. In the molecular line 
cubes of $^{12}$CO and $^{13}$CO, we find in fact that this
clump spans a wide velocity range, encompassing
the velocity of the F-6 filament. The average spectrum
of SW-Clump shows two peaks about
-2 and 4 km s$^{-1}$. A possible explanation is that the clump is
elongated along the line of sight, and its far
side is being pushed by the bubble to produce
the negative velocity component. The SW-Clump is likely
connected or associated with the F-6 filament.

\subsection{Distance to the NAP Region}\label{subsec:gaia}

In this section, we discuss the distance to the NAP based on YSOs, 3D dust mapping, and the presumed ionizing star (Bajamar's star).

\begin{figure*}[htbp]
\epsscale{1.2}
\plotone{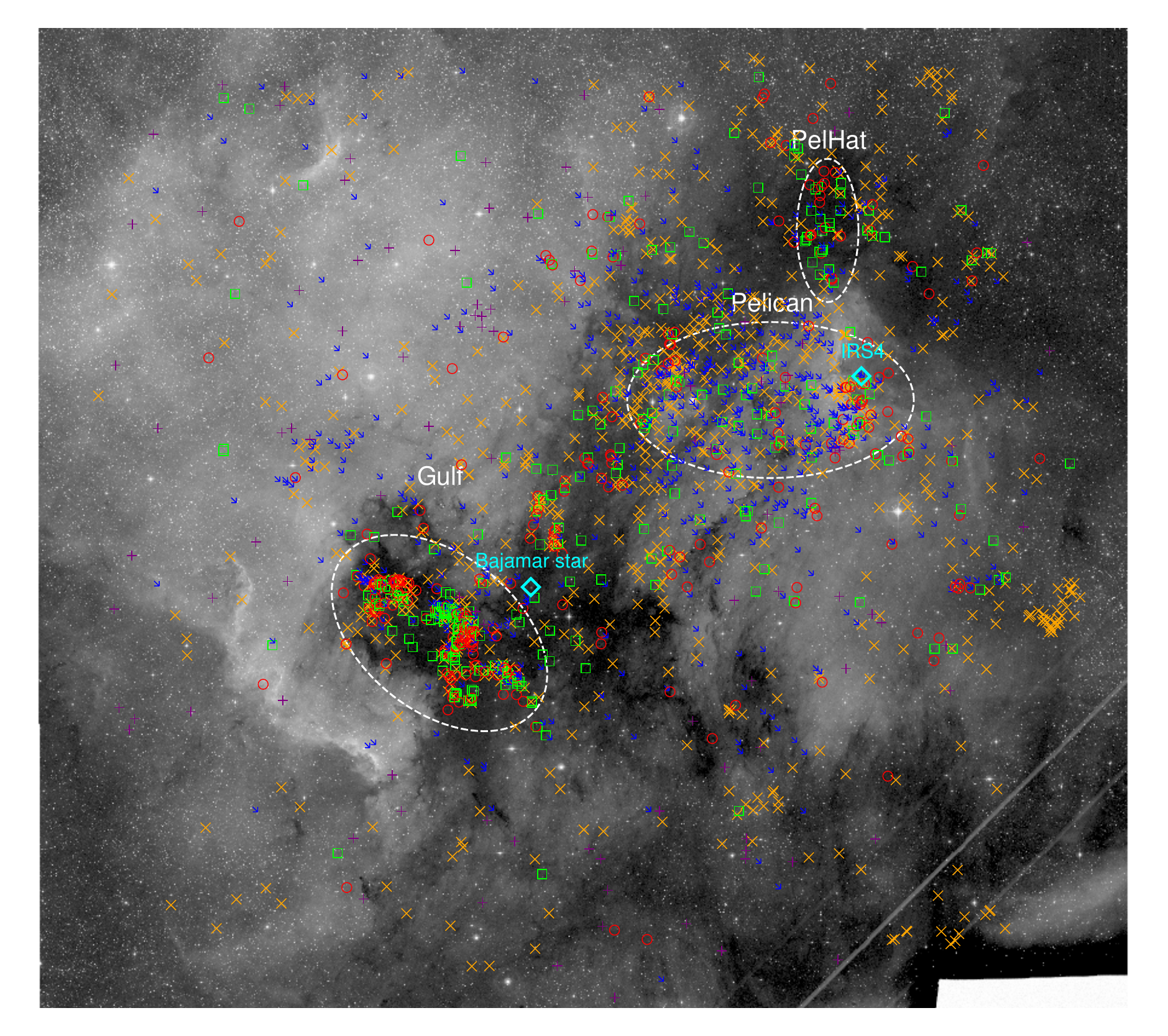}
\caption{
Distribution of young stellar objects in the NAP region.
Grayscale background:
POSS-II red band image from Figure \ref{fig:DSScoverage}.
YSOs from R11 are shown
as different symbols: red circle - Class I, green box
- Class flat, blue arrow - Class II, purple cross 
- Class III, orange ``x'' - unknown.
The three white dashed circles mark the locations
of the clusters (Gulf, Pelican, PelHat) 
defined by R11.
\label{fig:rebull}}
\end{figure*}

First, we start by using  the Gaia distances 
\citep{2018AJ....156...58B} of YSOs associated to dark clouds in this region to evaluate
the distance to the NAP region. Figure \ref{fig:rebull} shows an overview of 
the YSOs in the NAP region from R11. The YSO classification was based on the
near- to mid-IR spectral slope (from 2 to 24 
$\mu$m, see \S5.2 of R11)\footnote{See \citet{2020ApJ...904..146F} for 
the latest spectroscopic study of the young stars in NAP.}.
We cross-matched each of the more than 2000 sources in the catalog of NAP YSOs produced by R11
with the Gaia distance catalog \citep{2018AJ....156...58B}
using a search radius of 2\arcsec, and found that  
a little more than 1000 sources have Gaia measurements. 

\begin{figure*}[htbp]
\epsscale{1.15}
\plotone{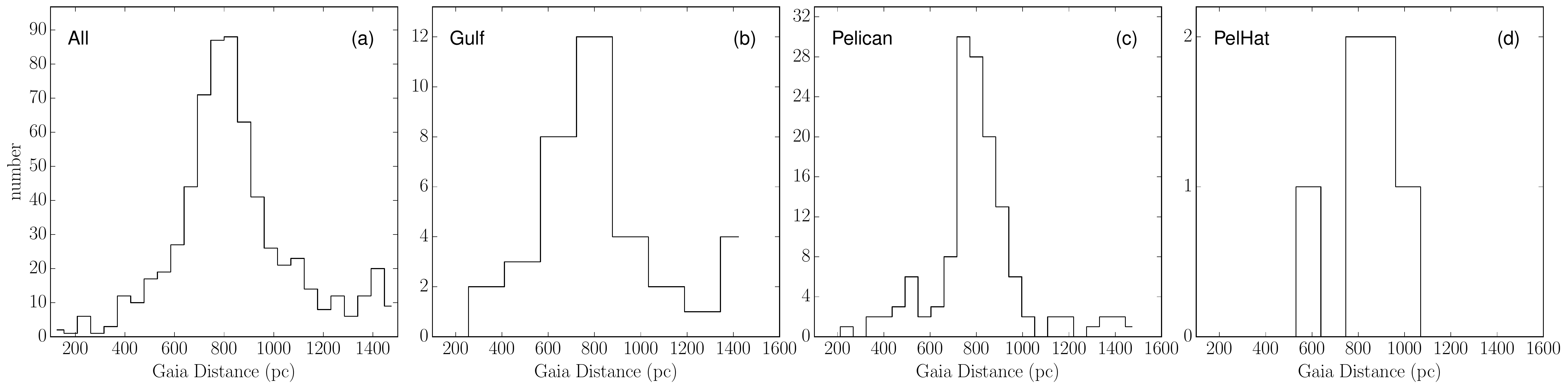}
\caption{
Distribution of (available) Gaia distances to the YSOs identified by R11.
Panel (a) shows the 
histogram for all YSOs in the region. 
Panels (b) to (d) show the distribution of Gaia
distance for the three different 
clusters in the region. The cluster is identified on the top left corner of each panel.
The y-axis shows the number 
of YSOs in each distance bin.
Note that we limit the x-axis to distances between
0 to 1500 pc. YSOs beyond 1500 pc
are likely not associated with the NAP complex.
\label{fig:hist}}
\end{figure*}

In Figure \ref{fig:hist}, we show a histogram 
of Gaia distances for all R11 YSOs that 
are available in the Gaia catalog. We zoom in
to the range of distances between 0 and 1500 pc, as sources beyond
1500 pc are highly unlikely 
to be associated with the NAP 
complex. As shown in the histogram, there is
clearly a peak around 800 pc (there is a tail
extending from 1500 pc to about 7000 pc, but 
no other peaks). Assuming the YSOs are close
to or associated with the molecular gas in the NAP complex,
then the histogram implies these clouds are likely
at a distance of $\sim$800 pc. Recently, 
\citet{2020A&A...633A..51Z} reported a similar 
distance toward the NAP region based on 3D dust mapping. 
\citet{2020ApJ...899..128K} also reached 
a similar conclusion based on a more detailed
study of stellar components of the NAP region.

The estimated distance of 800 pc using the YSOs is at odds with the original distance of  $668^{+39}_{-35}$ pc that we derived to Bajamar's star based on Gaia DR2 parallax measurements.
Even if one considers the upper bound of the uncertainty, the deduced distance would be  about 100 pc 
closer than the distance derived from YSOs or 3D dust mapping. Fortunately, the higher precision astrometry and improved treatment of systematic effects available in Gaia EDR3 has resolved the discrepancy in the previous DR2 parallax measurement, and \citet{Kuhn_2020_RNAAS} report a new distance to the Bajamar's star of $785 \pm 16$ pc.


In panels (b) to (d)
of Figure \ref{fig:hist}, we separately 
plot the distance histogram for the three
clusters defined in R11.  Their names are Gulf, Pelican, and PelHat, and 
we mark their locations with white dashed
ellipses in Figure \ref{fig:rebull}.
One can see from Figure \ref{fig:hist}
that all of the three YSO clusters peak
at $\sim$800 pc. The PelHat cluster has
a limited number of members so its distance
is uncertain. However, the Gulf cluster,
which is associated with the foreground
gray/dark features (Dark Crab and F-6, see
\S\ref{subsec:sf}), also has a peak distance 
of $\sim$800 pc. Since the W80 region is
behind the gray/dark features, the distance to
the W80 region must be farther than the
Gulf cluster. While this contradicts the Gaia DR2
distance measurement for the Bajamar Star, it is entirely consistent with the more accurate Gaia EDR3 measurement \citep[see][]{Kuhn_2020_RNAAS}.

\begin{figure}[htbp]
\epsscale{1.2}
\plotone{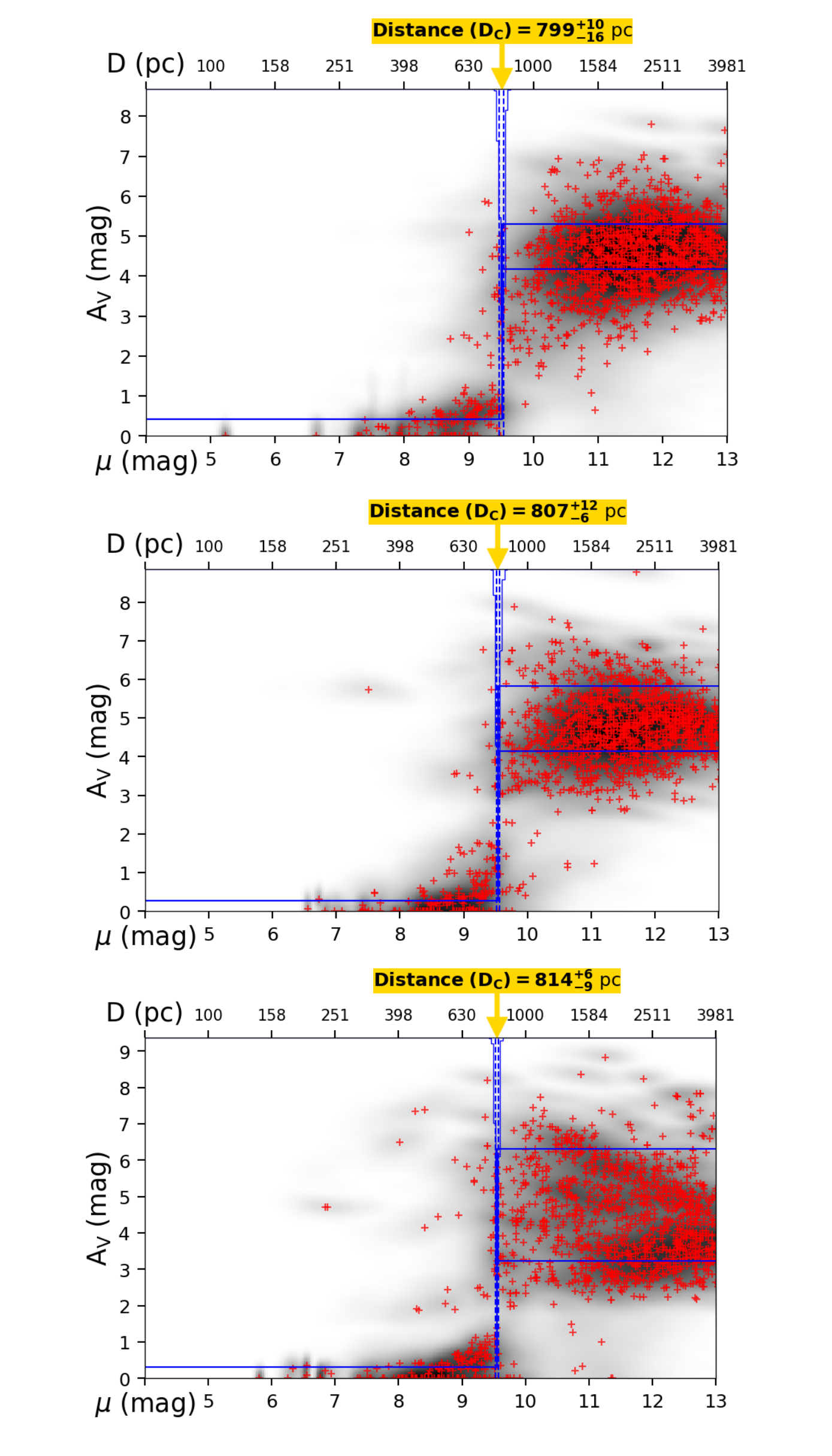}
\vspace{-10pt}
\caption{The distribution of dust extinction $A_V$ as a function of distance modulus $\mu$ (or distance $D$) for three sub-regions of the NAP ($\approx 1 \; \rm deg^{2}$ in area)  centered on ($l=84.0^\circ$, $b=-1.8^\circ$) (top), ($l=84.4^\circ$, $b=0.3^\circ$) (middle), and ($l=85.1^\circ$, $b=-0.3^\circ$) (bottom). The blue line shows the inferred distribution of dust along the line of sight for each sub-region. In each panel, a sharp increase in extinction (marked with a yellow arrow) indicates the most likely distance to that sub-region of the NAP. The red crosses indicate the most likely distance and integrated dust extinction for stars toward each sub-region, which are collectively used to constrain the dust distribution. Based on similar analyses across the full NAP region, we infer an average distance to the cloud of $797 \pm 40$ pc. 
\label{fig:nan3d}}
\end{figure}

We have produced an interactive distance map 
of the NAP based on 3D dust mapping (available \href{https://faun.rc.fas.harvard.edu/czucker/Paper_Figures/NorthAmerica.html}{here}),
following the technique outlined in
\citet{2019ApJ...879..125Z,2020A&A...633A..51Z}.
Figure \ref{fig:nan3d} shows the distribution of dust along the line of sight toward three representative sub-regions ($\approx 1 \; \rm deg^{2}$ in area) of the map, centered at ($l=84.0^\circ$, $b=-1.8^\circ$), ($l=84.4^\circ$, $b=0.3^\circ$), and ($l=85.1^\circ$, $b=-0.3^\circ$), respectively. Similar region-by-region plots over the full cloud are available online.  The distance uncertainties for the nebula do not include any systematic uncertainties, which we estimate to be $5\%$ in distance. 

For the full map, see \url{https://faun.rc.fas.harvard.edu/czucker/Paper_Figures/NorthAmerica.html}.  To produce the map, we grid the NAP region into individual pixels \citep[based on the Healpix system with a resolution parameter ($ N_{side}$)
of 64,][]{Gorski_2005} following
\citet{2019ApJ...879..125Z}, where each pixel covers about $1 \; \rm deg^{2}$ on the Sky. Each pixel is color-coded according to the distance we infer for the dust in that sub-region. We are able to target all area with an extinction of at least 5 mag. Each sub-region is fit independently, with a typical uncertainty of about $5-6\%$ in distance or about 40 pc. 
The distances obtained via 3D dust mapping are complementary to YSO-based distances, obtained either from Gaia DR2 or from VLBI observations 
\citep[e.g., towards masers;][]{2019ApJ...885..131R}. \citet{2020A&A...633A..51Z} find that their 3D dust mapping approach agrees with maser distances to within 10\% out to 2.5 kpc with no systematic offset.

While it is difficult to definitively 
constrain distance gradients within the NAP
given our typical systematic uncertainty ($\approx 40$ pc),
we determine that the entire NAP region traced by dust
is consistent with being at the same distance 
to within a few tens of parsecs. Using distance information across the full cloud, we find an average distance of $797 \pm 40$ pc, in excellent agreement with Gaia DR2 distances towards YSOs in Figure \ref{fig:hist}. Our results are also consistent with the updated distance to Bajamar's star -- the presumed ionizing star -- based on recent Gaia EDR3 results \citep[from][]{Kuhn_2020_RNAAS}, who find $d = 785 \pm 16$ pc.

On a side note, the Gulf cluster distance 
histogram in Figure \ref{fig:hist} shows a 
tentative excess at $\la$800 pc, while the 
Pelican cluster shows a clearer excess at
$\ga$800 pc. This distance difference implies 
that the Gulf cluster has members that are
closer to us, which is consistent
with the Gulf cluster being associated 
with a near-side clump/filament.
Indeed, \citet{2020ApJ...899..128K} 
reported that stars in the Gulf region (their group E)
are $\sim$35 pc closer than stars in the 
Atlantic and Pelican regions
(their groups A to D at $\sim$795 pc).


\section{Discussion}

\subsection{Star Formation and Feedback}\label{subsec:sf}

Out of the three clusters discussed in \S\ref{subsec:gaia}, and show in  Figure \ref{fig:rebull},
the Pelican cluster does not have a 
morphological match with molecular gas structures.  
The Pelican Head region (northwest of the
Pelican cluster from R11 and south of the PelHat cluster from R11) has multiple
MHOs identified in B14,
so the young stars there are
still accreting from the molecular 
gas. However, the rest of the Pelican cluster (outside the Pelican Head region)
has very limited $^{13}$CO(1-0)
or C$^{18}$O(1-0) emission, except for
those young stars near the Atlantic region. 
R11 also noted that the cluster
is in a low-extinction region \citep[based on 
the extinction map from][]{2002AJ....123.2559C}. 

\citet{2020ApJ...899..128K} noticed that 
the Pelican cluster (their Group D) shows
evidence of expansion.
As discussed in \S\ref{subsec:loc}, the Atlantic
clouds are clearly associated with the bright rims, 
and are likely being dispersed by the Bajamar Star.
A possible scenario is that the molecular gas
associated with the Pelican cluster, 
probably the gas in the Atlantic region, 
is being dispersed by the massive star.
The forming cluster was originally confined 
by the cloud gravitational potential, but now it is expanding due
to the (sudden) dispersal of the cloud.
If true, then the Pelican cluster is a 
good example of cluster formation halted
by the feedback from a nearby
massive star. In other words, the cluster 
formation reaches an end, not because its
associated gas is consumed or dispersed
by the cluster itself but by a neighboring
massive star. 

Here we carry out a simple analysis to test
the possibility that the cloud (hereafter the 
pre-Pelican cloud) in which the Pelican
cluster formed can be dispersed by the Bajamar Star.
We note that the PelHat cluster (north of the Pelican cluster) overlaps with relatively high extinction regions (see Figure \ref{fig:rebull}). It is at the edge of the W80 bubble, and has ample molecular gas to continue forming stars and feed the accretion of the existing protostars. Indeed, a recent survey of outflows \citep{2020ApJS..248...15Z} has revealed a few molecular outflows in PelHat, indicating active protostellar accretion. Meanwhile, we know the Pelican cluster has little molecular gas. So the pre-Pelican cloud was probably stripped from the Pelican cluster. Either the pre-Pelican cloud was pushed away by radiation pressure or bubble expansion, or the cloud was ionized by the Bajamar Star. In any case, the Pelican cluster should be closer to the Bajamar Star than PelHat. In other words, the cluster should be inside the bubble by now.

To carry out our basic feedback analysis, we need to estimate and make various assumptions about the HII region and the pre-Pelican cloud. 
The ionizing photon flux from the Bajamar Star is
$S_{\rm UV}\sim$ 10$^{49}$ s$^{-1}$ and the electron 
density $n_e$ of the HII region is $\sim$10 cm$^{-3}$
\citep{1968ZA.....68..368W,2005A&A...430..541C}.
The ionizing flux $S$ from the Bajamar Star
corresponds to a luminosity of  2.2$\times10^{39}$ erg s$^{-1}$
 (assuming 13.6 eV photon energy). 
The W80 bubble radius is $\sim$20 pc, much larger 
than the Str\"omgren radius of $\sim$2.8 pc 
\citep[assuming a mean gas number density $n$ of 100 cm$^{-3}$ before ionization, see equation 7.24 in][]{2017stfo.book.....K}.
Hence, the bubble has likely expanded significantly.
Assuming a temperature of $10^4$ K for the HII region,
the corresponding sound speed is $\sim$11 km s$^{-1}$.
If we assume a mean stellar mass of 0.5 M$_\odot$
\citep{2009ApJS..181..321E}
for the Pelican cluster members,
then the total cluster mass (with 247 members, R11) is $\sim$120 M$_\odot$.
Assuming a star formation efficiency of 3\%
\citep{2012ARA&A..50..531K},
the total pre-Pelican cloud 
mass $M$ would be 4$\times$10$^3$ M$_\odot$, 

First, we consider that the pre-Pelican cloud was pushed by the bubble expansion. Assuming the thermal pressure in the bubble dominates the expansion, then the total expansion time can be computed following equation 7.32 in \citet{2017stfo.book.....K}. 
\begin{equation}
r_i = 9.4S_{49}^{1/7}n_2^{-2/7}T_{i,4}^{2/7}t_6^{4/7}~{\rm pc},
\end{equation}
where $r_i$ is the radius of the ionized sphere, $S_{49}$ is the ionizing photon flux (in units of 10$^{49}$ s$^{-1}$), $n_2$ is the initial number density before ionization (in units of 100 cm$^{-3}$), $T_{i,4}$ is the temperature of the ionized gas (in units of 10$^4$ K), and $t_6$ is the time of expansion (in units of 10$^6$ yr).
Adopting $r_i$ = 20 pc, $S_{49}$ = 1, $n_2$ = 1, and $T_{i,4}$ = 1, the total expansion time $t$ is about 3.7 Myr. Taking the time derivative of the equation, the current bubble expansion velocity is
\begin{equation}
\frac{{\rm d}r_i}{{\rm d}t} = 5.4\times t_6^{-3/7}~{\rm km~s^{-1}},
\end{equation}
which is about 3 km s$^{-1}$. The typical relative velocity between the molecular filaments in the NAN region and the W80 bubble is a few km s$^{-1}$. So the bubble expansion is a possible mechanism of dispersing the pre-Pelican cloud. Considering the geometry, it is possible that the molecular gas to the north-west of the Pelican Head is made in part from gas from the pre-Pelican cloud that has been pushed to that position.

Second, we consider that the Lyman-alpha photons from the HII region exert a radiation pressure on the molecular gas. Assuming the solid angle subtended by the pre-Pelican cloud (as viewed from the Bajamar Star) is $\Omega$, the force caused by the radiation pressure is $S\Omega/4\pi c$, where c is the speed of light. The final velocity of the pre-Pelican cloud after $t$ is $(S\Omega/4\pi c)t/M$. If we assume $\Omega=4\pi/10$, the pre-Pelican cloud reaches about 1.1 km s$^{-1}$,  smaller than the velocity driven by the bubble expansion. 

\begin{figure*}[htbp]
\epsscale{1.15}
\plotone{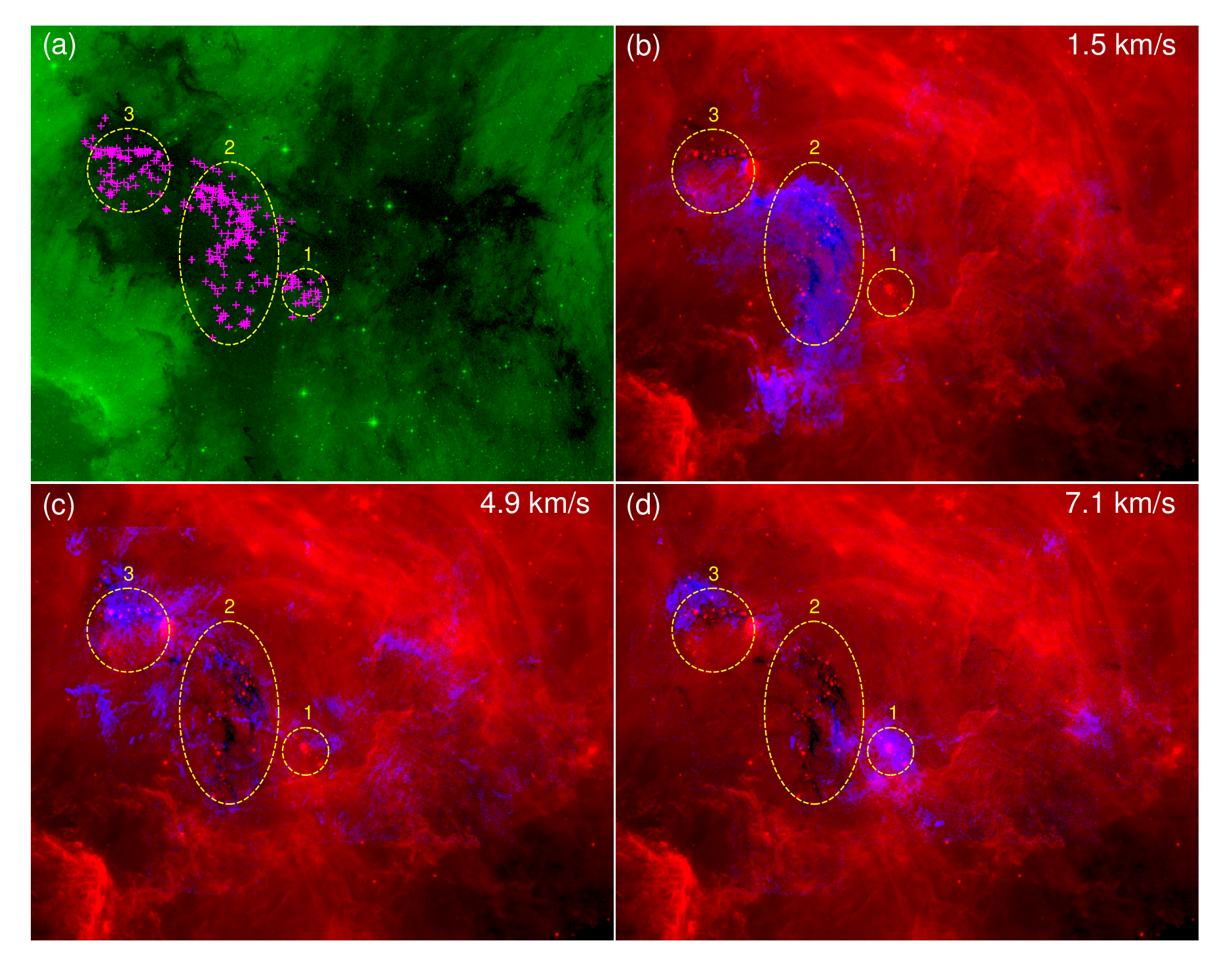}
\vspace{-0.5cm}
\caption{
{\bf (a):} POSS-II red color plate for the Gulf region,
same as Figures \ref{fig:compare1}-\ref{fig:compare7}.
The magenta crosses mark the R11 Gulf cluster members.
The yellow dashed ellipses mark the G09 sub-clusters.
{\bf (b):} The background red color shows the 
Spitzer 24$\mu$m image from R11. The blue color shows
the CARMA+DLH14 combined $^{13}$CO channel map at 
1.5 km s$^{-1}$. The 24$\mu$m dark regions are partially
covered by the blue color. Readers can see them better
in Figure \ref{fig:spitzercover}.
{\bf (c):} Same as panel (b) but showing the
4.9 km s$^{-1}$ channel.
{\bf (d):} Same as panel (b) but showing the
7.1 km s$^{-1}$ channel.
\label{fig:twopanel}}
\end{figure*}

Finally, we consider that the pre-Pelican cloud was
ionized by the Bajamar Star so that we do not
see the molecular gas today.
For an ideal Str\"omgren sphere, the ionized
gas mass is $S_{\rm UV}\mu m_H/n\alpha_B$, 
where $\mu=1.4$ is the mean particle weight, 
$m_H$ is the hydrogen mass, $n$ is the particle 
density which we assumed as 100 cm$^{-3}$ earlier \citep[see equation 7.23 in][]{2017stfo.book.....K},
$\alpha_B=3.5\times10^{-13}$ cm$^3$ s$^{-1}$
is the case B recombination coefficient for hydrogen.
Given the aforementioned flux $S_{\rm UV}$,
the HII region mass is
about $3\times10^3$ M$_\odot$,
comparable to the pre-Pelican cloud
mass $M$. 

The actual history of the cloud dispersal 
may be due to a combination of the bubble expansion, radiative pressure, and ionization. In other words, part of the pre-Pelican 
cloud have been ionized and the rest  could have been pushed away.
In any case, it is possible that the Bajamar Star and the W80 bubble
together could have dispersed the pre-Pelican cloud.

\subsection{The Gulf Cluster}\label{subsec:gulf}

The Gulf cluster, however, needs a closer 
examination. There are probably three 
sub-clusters in this region. From west to east,
\citet[][hereafter G09]{2009ApJ...697..787G} 
named them clusters 1, 2, 3 (see their figure 9).
From our combined molecular line data, we 
find evidence that the three sub-clusters are 
probably separated.

In Figure \ref{fig:twopanel}, we show the $^{13}$CO
gas associated with the three G09 sub-clusters.
In panel (c), the G09 cluster 3 is associated 
with a 24$\mu$m dark region at the northeast end
of the Gulf region (the ``Dark Crab'').
It is embedded in highly extincted molecular
gas at $\sim$4.9 km s$^{-1}$. Meanwhile, the G09
cluster 2 spatially matches the 24$\mu$m dark 
region which is the F-6 filament at a velocity
of $\sim$ 1.5 km s$^{-1}$. Consequently,
the two G09 clusters are associated with two 
distinct molecular gas regions. They are likely
not part of the same cluster. 

More interestingly, the location of the G09 
cluster 1 has no 24$\mu$m dark counterpart, as 
can be seen in Figure \ref{fig:twopanel} (b)(c).
This cluster nicely coincides with the location of 
the background F-7 filament (Figures~\ref{fig:compare}
and \ref{fig:compare7}). This is especially
true at 7.1 km s$^{-1}$ (panel d). 
It is possible that the
G09 cluster 1 is embedded
in F-7, which lies on the far side of W80. There are six members in the cluster 1 with Gaia
DR2 data, and  their distances are (in ascending order) 
1103, 1111, 1611, 1863, 2457, and 3565 pc. These distances are significantly larger than the estimated distance to W80, and thus consistent with the scenario that this is a cluster background to W80.

Overall, we see that at least part of 
the star formation
activities in the NAP complex are strongly
impacted by the feedback from the 
Bajamar Star.

\section{Summary and Conclusion}

We  present high-resolution CARMA
molecular line data toward the North America
and Pelican Nebulae. The new interferometer data is combined with existing single-dish data from the DLH14 telescope to study the structure and kinematics of the molecular gas in this region.
The combined maps show ongoing 
interaction between the molecular gas and 
feedback from the massive Bajamar Star
(O3.5) that is responsible for
the W80 HII region. The molecular gas 
is being dispersed by W80.

The high-resolution data reveal the intricate morphology of the gas.
A substantial fraction of the clouds have a very clumpy structure with sharp edges. Some of these bright edges coincide with the location of
the bright infrared rims which generally 
point toward the Bajamar Star. We argue
that they are heated by the massive star and 
the gas is being dispersed.

Using multi-wavelength data (optical,
infrared, and millimeter), we have identified
a number of distinct structures in the cloud
complex, including a number of dark clouds and globules.
We have cross-matched the structures
in images at different wavelengths, 
and determined the relative  line-of-sight distances
of the molecular filaments and clumps.

We find that there are 
two groups of molecular clouds  between the
W80 bubble and us. The first group (``gray regions'' in the optical image) contains most of the
molecular gas in the region, stretching from the Gulf region in the south to the PelHat region in the north. It is also the
main host of the young stellar objects in the
Nebulae. However, this group of molecular clouds is being impacted by, and will soon be dispersed,
by the massive star feedback.
The second group of molecular gas (``dark regions'' in the optical image) consists of
scattered molecular gas clumps that are less massive.
The clumps appear darker in the optical image, indicative of higher extinction towards these clumps. They show
limited star formation (except for one clump we call the Dark Crab). They are being approached by the expanding W80 bubble. 

To determine the distance to the NAP complex, 
we obtained the Gaia distances to stars in  three young 
stellar object clusters identified by
\citet{2011ApJS..193...25R}.
The results show that the majority of the cluster
members are locate at $\sim$ 800 pc, indicating that
the NAP complex is at this distance. 
We provide an interactive
distance map based on the method developed in
\citet{2019ApJ...879..125Z,2020A&A...633A..51Z}.
The map is consistent with the $\sim$ 800 pc 
NAP distance, and it gives more details on the 
3D structures of the complex, which is broadly
consistent with our conclusions based on matching the morphology of structures seen in optical, infrared,
and millimeter molecular line maps of the region.

The PelHat and Gulf clusters defined by
\citet{2011ApJS..193...25R}
show a clear association with molecular gas.
The third cluster, the Pelican cluster, is associated with substantially less  molecular gas.
We suggest the gas in this cluster has been
dispersed by the O-type Bajamar Star,
not by sources in the Pelican cluster.
It seems, therefore that the star formation in the Pelican cluster
has been interrupted by the massive star.
The pre-Pelican cloud was probably dispersed
by a combination of bubble expansion,
radiative pressure, and gas ionization.
This scenario provides a good 
example of feedback-regulated star formation.
The North America and Pelican Nebulae
thus provide a close-by example for 
studying star formation under the influence
of an isolated massive star.

\acknowledgments 
We thank the anonymous referee for a thorough check on
the paper and helpful comments. 
We thank Luisa Rebull for providing the Spitzer images. We thank Laurent Cambr\'esy for providing the extinction maps. SK acknowledges fruitful discussions with Min Fang and Serena Kim. SS acknowledges support from the European Research Council under the Horizon 2020 Framework Program via the ERC Consolidator Grant CSF-648505. TGSP gratefully acknowledges support by the National Science Foundation under grant No. AST-2009842. SK and HGA were (partially) funded by the National Science Foundation, award AST-1140063, which also provided partial support for CARMA operations. CARMA operations were also supported by the California Institute of Technology, the University of California-Berkeley, the University of Illinois at Urbana-Champaign, the University of Maryland College Park, and the University of Chicago. ASM and VO carried out this research within the Collaborative Research Centre 956 (subprojects A6 and C1), funded by the Deutsche Forschungsgemeinschaft (DFG) - project ID 184018867. Part of this research was carried out at the Jet Propulsion Laboratory, California Institute of Technology, under a contract with the National Aeronautics and Space Administration. SZ acknowledges a funding from the National Natural Science Foundation of China through grants NSF 11803091. RJS acknowledges funding from an STFC ERF (grant ST/N00485X/1). RSK acknowledges support from the Deutsche Forschungsgemeinschaft (DFG) via the Collaborative Research Center (SFB 881, Project-ID 138713538) ``The Milky Way System'' (sub-projects A1, B1, B2 and B8) and from the Heidelberg cluster of excellence (EXC 2181 - 390900948) ``STRUCTURES: A unifying approach to emergent phenomena in the physical world, mathematics, and complex dat'', funded by the German Excellence Strategy. He also thanks for funding form the European Research Council in the ERC Synergy Grant ``ECOGAL -- Understanding our Galactic ecosystem: From the disk of the Milky Way to the formation sites of stars and planets'' (project ID 855130). 

This research made use of the data from the Milky Way Imaging Scroll Painting (MWISP) project, which is a multi-line survey in $^{12}$CO/$^{13}$CO/C$^{18}$O along the northern galactic plane with the Purple Mountain Observatory Delingha 13.7m telescope. MWISP project is supported by National Key R\&D Program of China with grant no. 2017YFA0402700, and the Key Research Program of Frontier Sciences, CAS with grant no. QYZDJ-SSW-SLH047. The Second Palomar Observatory Sky Survey (POSS-II) was made by the California Institute of Technology with funds from the National Science Foundation, the National Aeronautics and Space Administration, the National Geographic Society, the Sloan Foundation, the Samuel Oschin Foundation, and the Eastman Kodak Corporation. The Oschin Schmidt Telescope is operated by the California Institute of Technology and Palomar Observatory. 

\software{Astropy \citep{Astropy-Collaboration13}, Numpy \citep{numpy}, APLpy \citep{Robitaille12}, Matplotlib \citep{matplotlib}}

\facilities{CARMA, DLH:13.7m}

\newpage

\appendix
\restartappendixnumbering 

\section{WISE image}\label{app:ancfig}

\begin{figure*}[htbp]
\epsscale{1.15}
\plotone{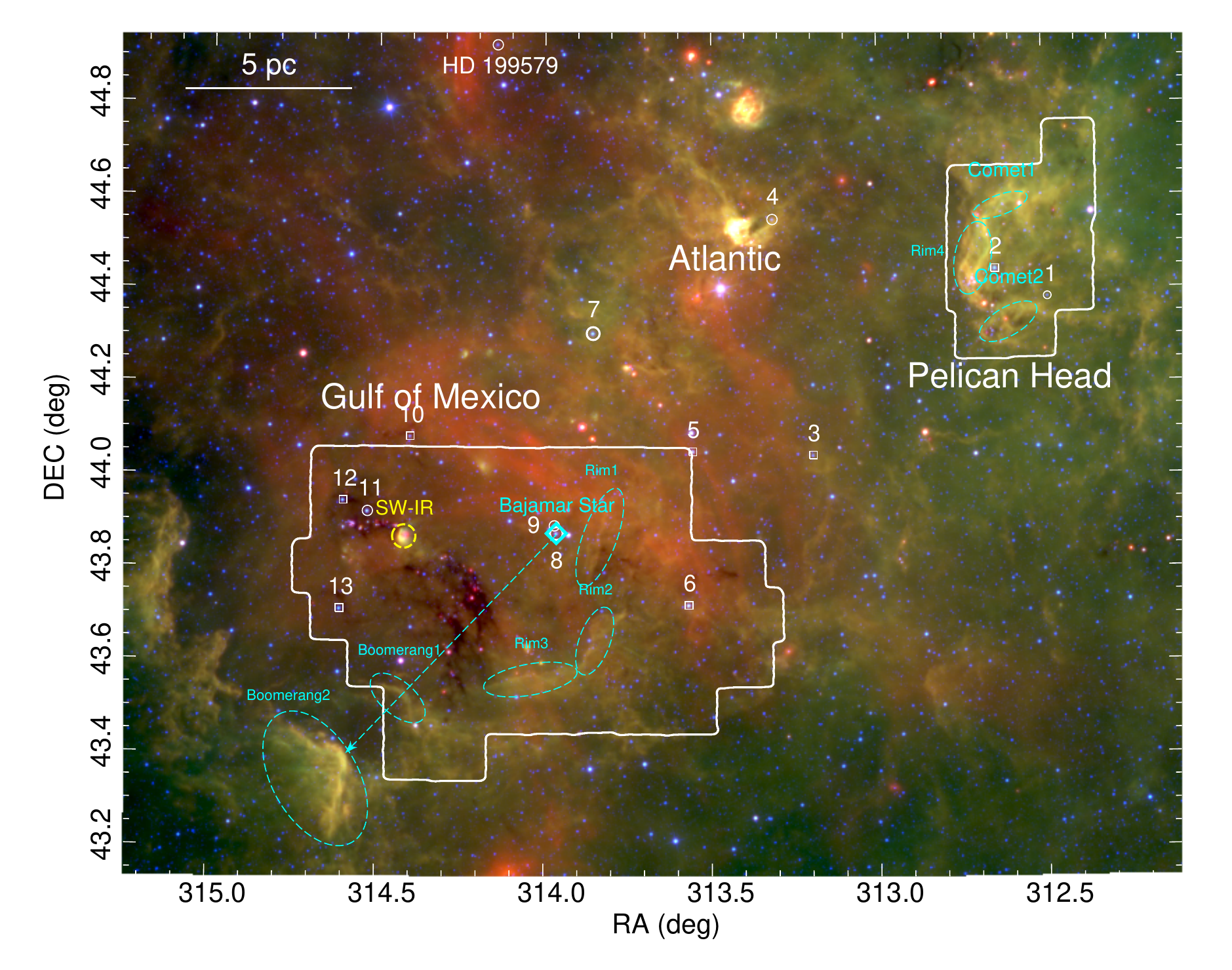}
\vspace{-0.5cm}
\caption{
WISE false color RGB image using the 
22$\mu$m (Red), 
12$\mu$m (Green), and
4.6$\mu$m (Blue) bands.
The cyan diamond marks the position of the
Bajamar Star. The cyan arrow points from the 
Bajamar Star toward the two boomerang features 
at the south-east corner. The two cometary
clouds are seen in molecular gas in
Figure \ref{fig:peakp}.
The stars with numbers 1-13 are from SL08 (as shown in B14).
Circles are stars classified as spectral type OB. 
Squares are possible AGB stars (see SL08 and B14).
The Bajamar Star is \#8.
\label{fig:wisecover}}
\end{figure*}

In Figure \ref{fig:wisecover}, we present a WISE RGB color image as a supplement to Figure \ref{fig:spitzercover}. 

\section{Structure identification}\label{app:struc}

\begin{figure*}[htbp]
\epsscale{1.2}
\plotone{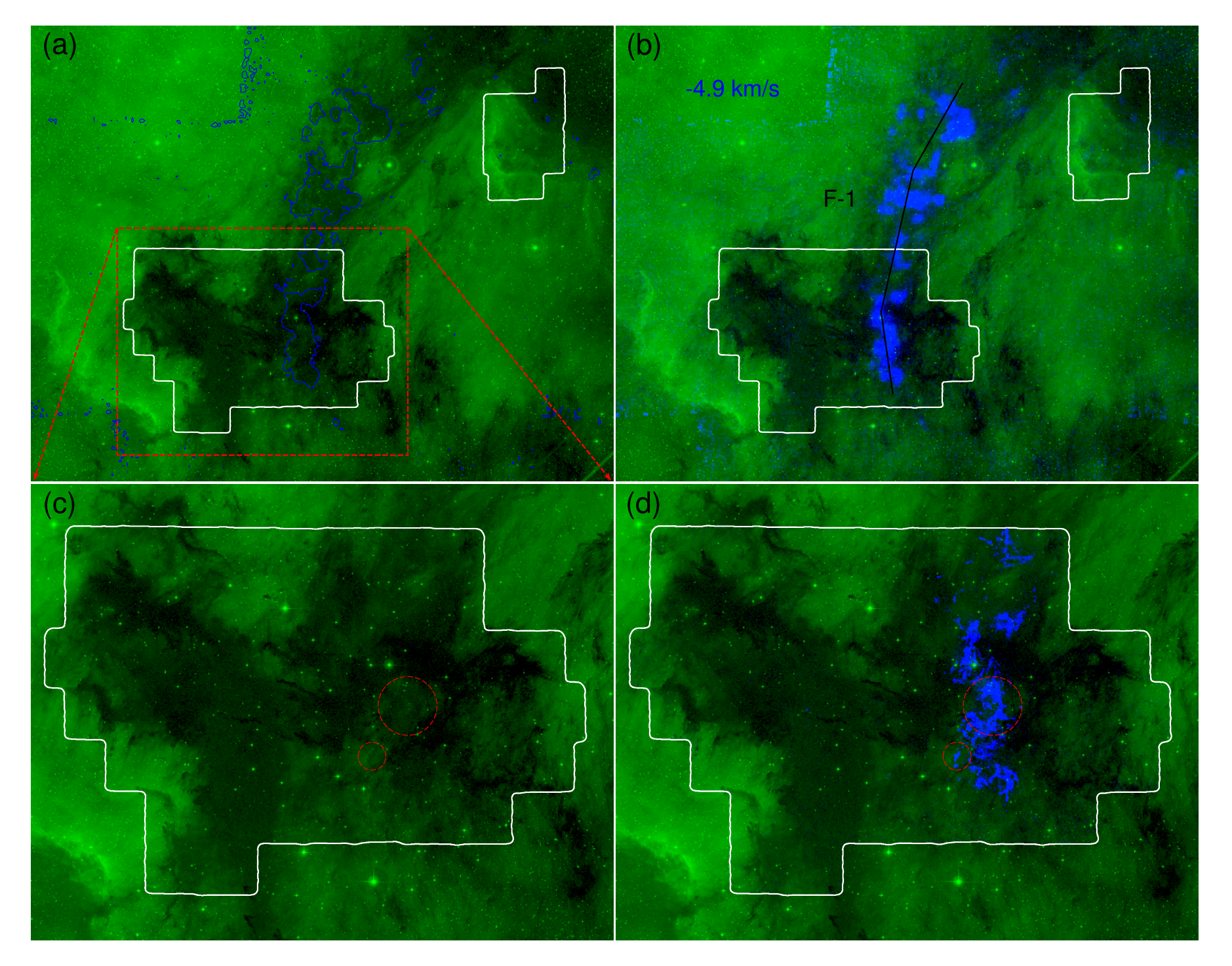}
\caption{
Comparison between the POSS-II red color (0.7 $\mu$m)
and the $^{13}$CO(1-0) cube from Z14 and the combined
$^{13}$CO(1-0) cube from this paper.
{\bf (a):} POSS-II red color plate shown in green color map.
The white boundaries show the CARMA map coverage.
The blue contours show the $^{13}$CO(1-0) emission
from panel (b). The red dashed box shows the zoom-in
region in panel (c). {\bf (b):} The green color and
the white boundaries are the same as panel (a), but the
blue color represents the $^{13}$CO(1-0) emission from Z14.
The black segment marks the F-1 filament defined in Z14.
The blue text shows the  channel velocity.
{\bf (c):} Zoom-in view of the red box in panel (a).
{\bf (d):} The blue color shows 
the $^{13}$CO(1-0) combined (CARMA+DHL14) data  from this paper.
The channel velocity is the same as that in panel (b).
\label{fig:compare1}}
\end{figure*}

\begin{figure*}[htbp]
\epsscale{1.2}
\plotone{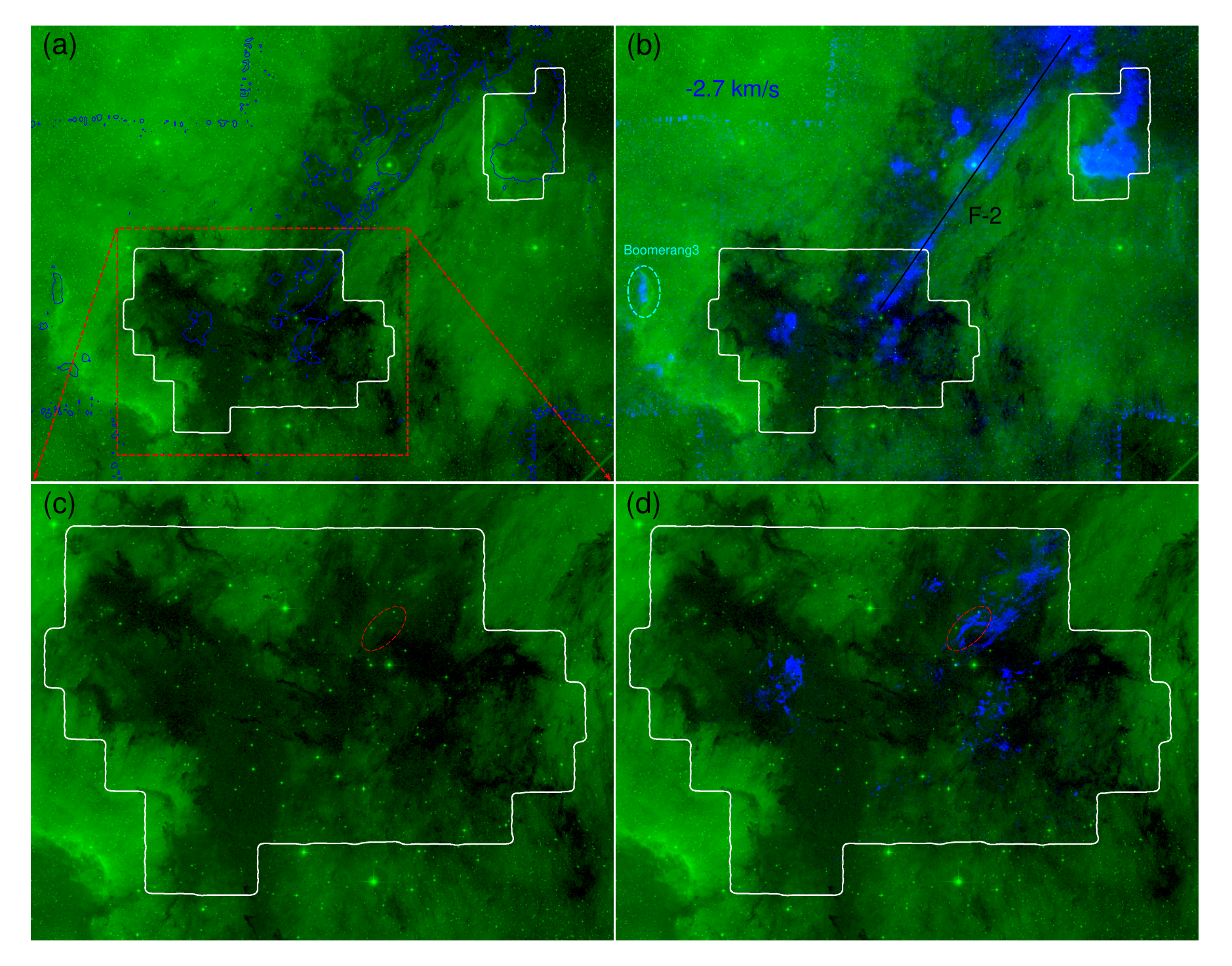}
\caption{
Same as \ref{fig:compare1}, but at a velocity
representative of F-2.
\label{fig:compare2}}
\end{figure*}

\begin{figure*}[htbp]
\epsscale{1.2}
\plotone{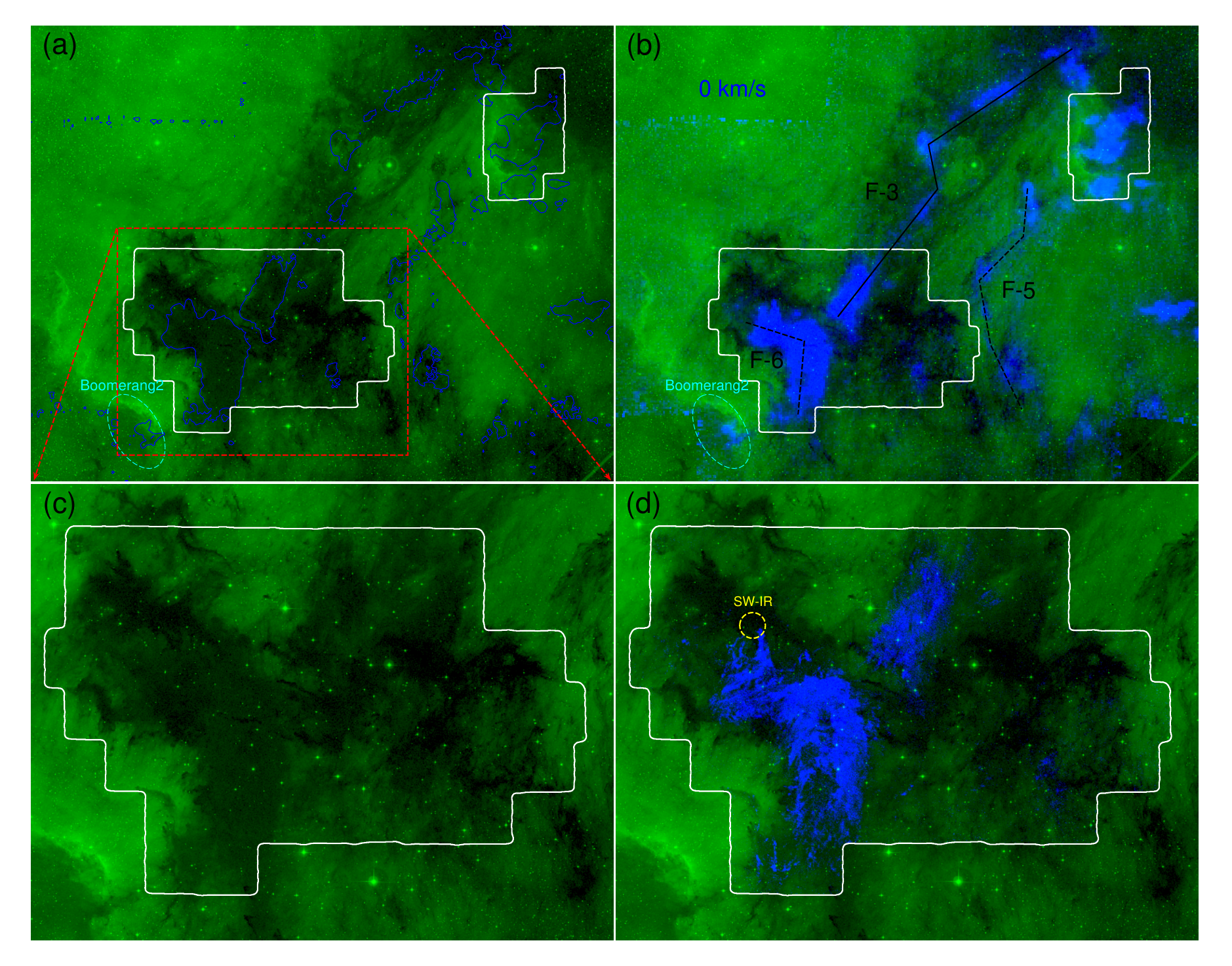}
\caption{
Same as \ref{fig:compare1}, but at a velocity
representative of F-3, F-5, and F-6.
\label{fig:compare3}}
\end{figure*}

\begin{figure*}[htbp]
\epsscale{1.2}
\plotone{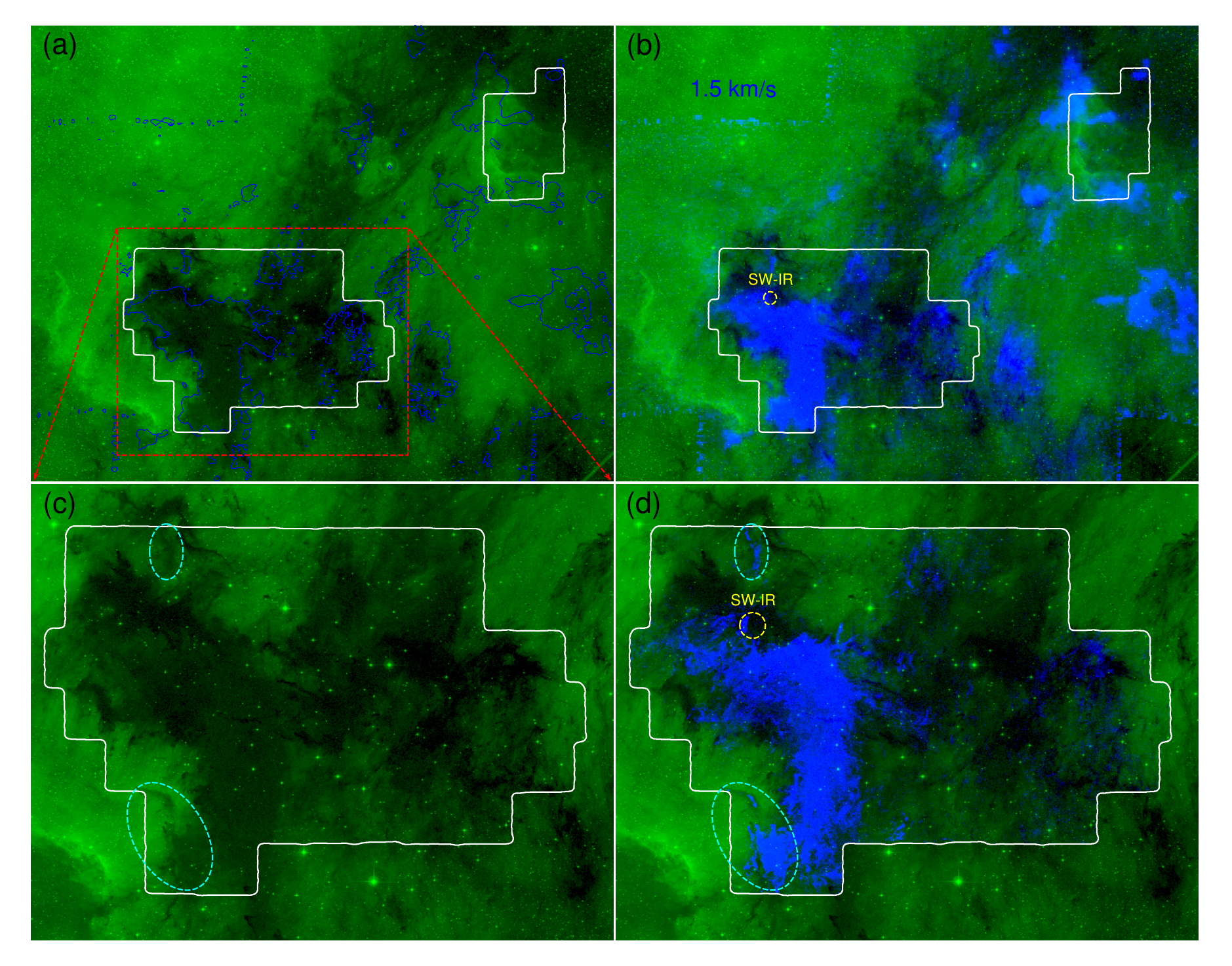}
\caption{
Same as \ref{fig:compare1}, but at a velocity
representative of F-6 and Boomerang1.
\label{fig:compare4}}
\end{figure*}

\begin{figure*}[htbp]
\epsscale{1.2}
\plotone{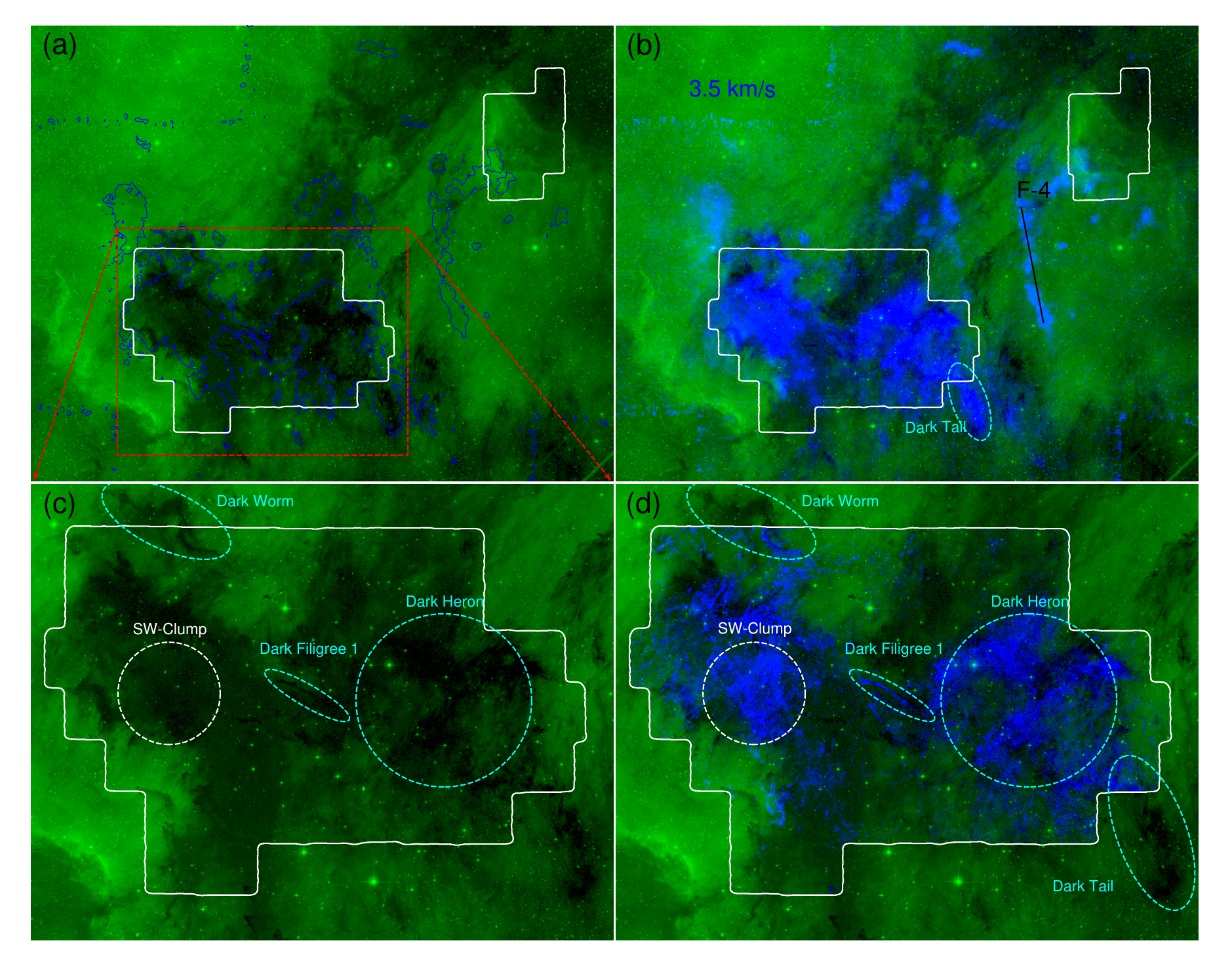}
\caption{
Same as \ref{fig:compare1}, but at a velocity
representative of the dark clumps.
\label{fig:compare5}}
\end{figure*}

\begin{figure*}[htbp]
\epsscale{1.2}
\plotone{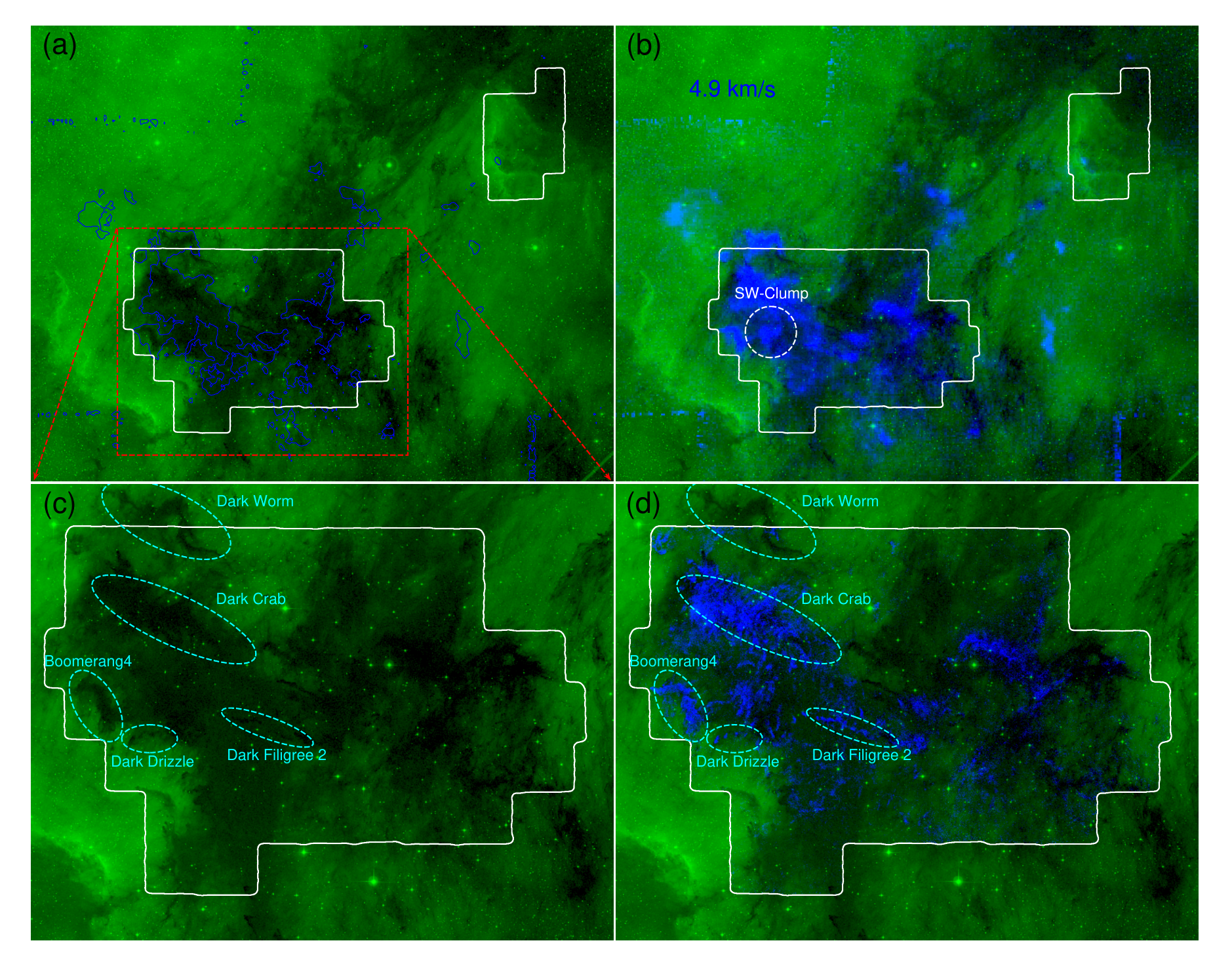}
\caption{
Same as \ref{fig:compare1}, but at a velocity
representative of the dark clumps.
\label{fig:compare6}}
\end{figure*}

\begin{figure*}[htbp]
\epsscale{1.2}
\plotone{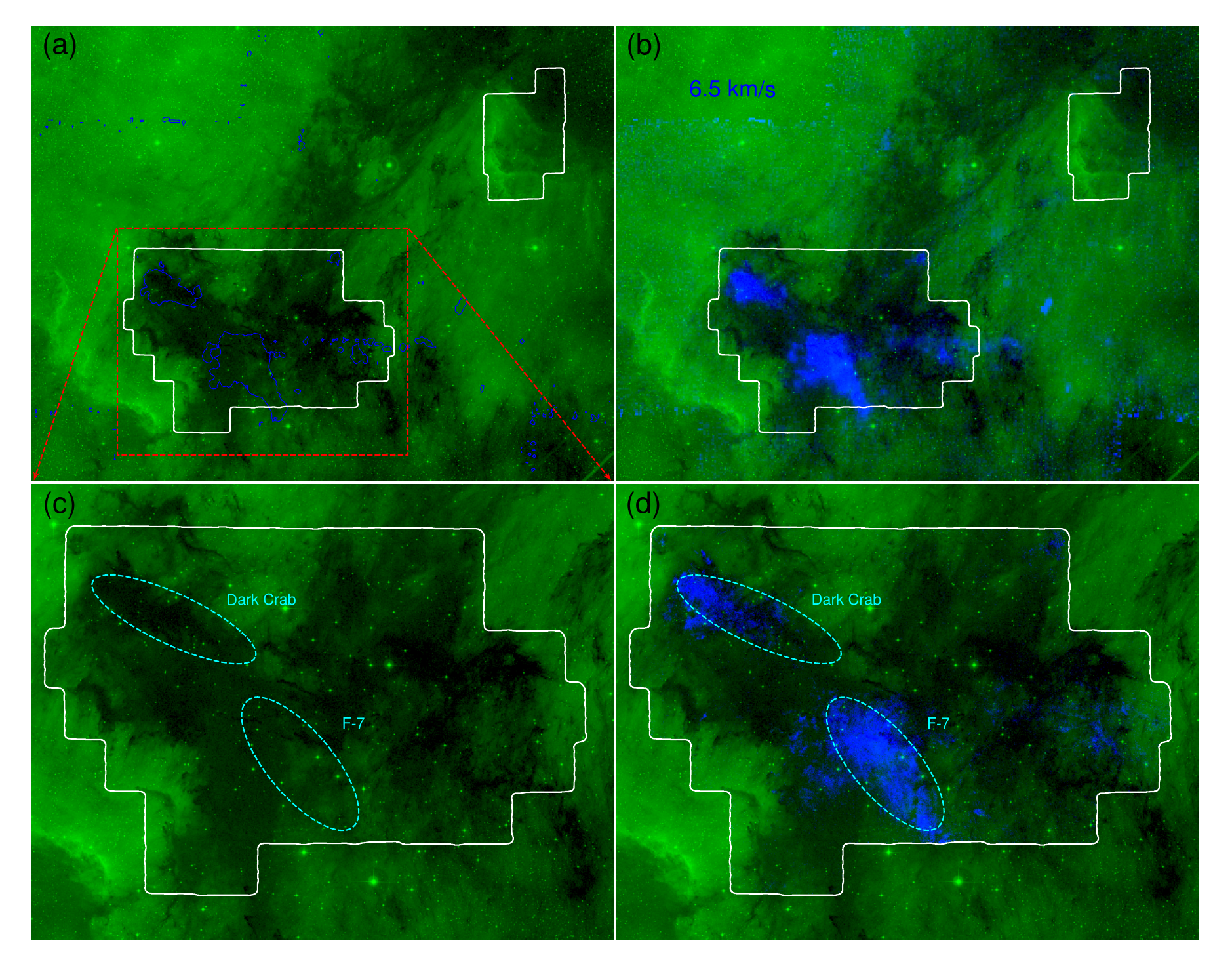}
\caption{
Same as \ref{fig:compare1}, but at a velocity
representative of F-7.
\label{fig:compare7}}
\end{figure*}

Figures \ref{fig:compare1}-\ref{fig:compare7} 
show the comparison
between the POSS-II red plate and the $^{13}$CO(1-0)
emission at different velocities. 
The seven figures have
the same four-panel setup that help compare
the gray/dark features in the optical images 
with the molecular gas distribution at large 
and small scales. In all panels the green colored 
map shows the POSS-II red plate image. 
Panels (a) and (b) show the large-scale NAP area
(similar to the area shown in the optical and
IR images in Figures \ref{fig:DSScoverage} and \ref{fig:spitzercover}),
while panels (c) and (d) zoom in on the Gulf region.  
The blue color in panel (b)
shows the $^{13}$CO(1-0) emission in one channel 
from the DLH14-only maps. This same emission 
is represented with blue contour lines in panel (a).
The blue color in panel (d) shows the $^{13}$CO(1-0) 
emission from the combined (CARMA+DLH14) cube
at the same velocity ($V_{lsr}$) as the emission
shown in panels (a) and (b).

Figure \ref{fig:compare1} shows the distribution
of the $^{13}$CO(1-0) emission 
in the $V_{lsr} = -4.9$ km s$^{-1}$  channel
in comparison with the optical emission. 
At this velocity, all of the
$^{13}$CO emission belongs to the F-1 filament
that was defined by Z14.
In panel (b) we see that F-1 stretches from
the Gulf to the northern Atlantic region, 
following the general structure of the
``gray'' clouds in the optical image.
This is also the case at small scales, 
as can be seen in panel (d) (particularly see
regions enclosed by the two red dashed circles). 
The most straightforward assumption is that the 
gray region is in the foreground of the W80 bubble
as it shields light from the bubble. Therefore,
we conclude that the F-1 filament is on the 
near side of the bubble, moving toward us with
a $V_{lsr}$ of $\sim$ -5 km s$^{-1}$. 

The molecular gas in the small red circle in
Figure \ref{fig:compare1} is part of the Rim2
gas that is heated by the Bajamar Star
(see \S\ref{subsec:midir}).
In fact, both Rim1 and Rim2 are
part of the southern extension of F-1, 
coincident with gray optical regions 
in the Gulf, and most prominently detected at 
velocities around -5 to -3 km s$^{-1}$, 
as seen in the $^{13}$CO 1st-moment map 
(see Figure \ref{fig:king}). Since this gas 
is associated with feedback features produced
by radiation from the Bajamar Star, gas
at these velocities likely resides not far from
the HII region. 
Radio recombination line observations of the W80 HII region show that the $V_{lsr}$ of this source is $5.5 \pm 4.4$ km s$^{-1}$  \citep{1989ApJS...71..469L}.
Given the velocity (and position) of the F-1 filament, 
we argue that the expanding HII region is currently
exerting pressure and pushing on this filament toward us.
The high excitation temperature of the gas associated
with the Rim1 and Rim2 features (see Figure \ref{fig:peakg}(b))
as well as the very clumpy morphology with very sharp and bright edges  is evidence that the 
gas here is heated and sculpted by the UV radiation 
from the nearby high-mass star powering the HII region. Following this idea, we argue that
the ``gray'' regions correspond to molecular gas
very close to the HII region, where the gas is being 
dispersed by the fierce radiation from the
Bajamar Star.

Figure \ref{fig:compare2} shows the distribution 
of the molecular gas (as traced by $^{13}$CO(1-0))
at $V_{lsr}$ = -2.7 km s$^{-1}$.
At this velocity the most prominent molecular gas
feature is the F-2 filament (originally identified 
by Z14), which extends from the northwest edge of 
the map (in the northern Atlantic region),
south to the central part of the Gulf region.
There is clear correspondence in the  morphology
of the F-2 filament and the ``gray'' regions
in the optical image, both at large scales 
(as traced by the DLH14 data) and at small scales
(traced by the CARMA+DLH14 data). The latter  
is highlighted by red dashed ellipses
in panels (c) and (d) of  Figure \ref{fig:compare2}.

Similar to F-1, the F-2 filament is in the 
foreground of, and has a blueshifted velocity 
with respect to, W80.  Given the 
similar low extinction (i.e., being a gray region
instead of a dark region in the optical image) 
and its close position and velocity relative to F-1, 
we argue that F-2 (similar to F-1) is also close to 
the front end of the HII region. However,
since we do not see prominent bright rims or 
clear feedback features in F-2 as in F-1,
we suspect that F-2 is farther from the bubble than F-1.

Other interesting features seen at $V_{lsr} =-2.7$ km s$^{-1}$ 
are the  gas associated with the Pelican Head 
bright rim region (towards the northwest) 
and a bright rim region in the east, which we 
name Boomerang3 (see Figures \ref{fig:compare} 
and \ref{fig:compare2}(b)). These two mark the 
outer regions clearly impacted by the UV 
radiation and the HII region within the DLH14 maps.
Even though these features appear to be at 
(or close to) the edge of the  HII region 
they are most likely to be foreground 
to W80 given there is evident (light gray) 
extinction next to these bright rim regions in
the optical images (see Figures \ref{fig:DSScoverage} 
and \ref{fig:compare}). 

The above analyses of F-1 and F-2 filaments 
have shown that the L935 dark cloud in
the NAP region is the shadow of a combination 
of foreground molecular filaments and clouds. 
The F-3 filament identified by Z14, and seen in
Figure~\ref{fig:compare3} at $V_{lsr} = 0$ km s$^{-1}$, 
appears to be part of the same complex as F-1 and F-2. 
As shown in Figure~\ref{fig:compare3},
filament F-3 winds its way from the Gulf region 
through the Atlantic region and all the way to the
northern end of the Pelican region, coincident with  
gray regions seen in the optical images.   

Two other prominent features are seen at 
velocities similar to those of F-3 (Figure~\ref{fig:compare3}).  
One of these is a filament lying on the west 
side of the L935 dark cloud that follows a
grayish silhouette in the optical image, 
which we name F-5, following the nomenclature 
from Z14. The other feature is the thick
filament with an ``inverted L'' morphology in
the Gulf region. This filament, which we name F-6, 
corresponds to the 24 $\mu$m dark region in Figure 
\ref{fig:spitzercover}, and includes the BGPS dust
continuum clumps SW1, SW2 main, SW2 SE, and SW2 S 
identified by B14 (which we show in Figure 
\ref{fig:peakg}), as well as  many 24 $\mu$m
sources, and 3 YSO clusters identified by  R11. 
Figure \ref{fig:compare3} shows that the
optical counterpart of F-6 is a grayish region,
similar in appearance to the optical 
counterparts of F-1, F-2, F-3. This region is
clearly in the foreground of W80, but has not
(yet) been severely impacted by the massive star. 
Other features seen at $V_{lsr} \sim 0$ km s$^{-1}$
include the CO emission associated with Boomerang2 
in the southeast edge of the NAP complex, 
as well as parts of the Pelican region. 
As with similar features discussed above, 
these structures are most likely foreground to W80.

Figure \ref{fig:compare4} shows the
comparison for molecular gas at 
$V_{lsr} = 1.5$ km s$^{-1}$ and the optical image. 
At this velocity the most prominent feature is
gas associated with the F-6 filament described 
above (also seen at $V_{lsr} = 0$ km s$^{-1}$).
In addition, at this velocity there are two  
interesting structures  which show a good match
between the morphology of the CO and gray extinction 
in the optical image. One is the Boomerang1
feature seen in Figure~\ref{fig:compare} as
a gray feature (see also Figure \ref{fig:spitzercover}),
where  the $^{13}$CO emission nicely
matches the morphology of the feedback structure. 
There is also a good match between the optical and
the molecular gas images southwest of Boomerang1.
The other structure is the silhouette highlighted
by the cyan ellipse in  Figure \ref{fig:compare4},
at the northern edge of the Gulf region covered by
our CARMA observations. As with the other features
described above, the clear correspondence between 
the gas morphology and gray extinction implies
these features are foreground to the HII region.
Since their LSR velocity is lower than the
aforementioned W80 velocity ($\sim 5.5$ km s$^{-1}$), 
they are likely moving away from the bubble.  

The molecular gas counterparts of the dark 
extinction regions begin to appear at higher
LSR velocities. Figure \ref{fig:compare5}
shows the comparison for the $^{13}$CO emission at
$V_{lsr} = 3.5$ km s$^{-1}$ and the optical image. 
The dark region in the Gulf region marked by 
the big dashed (cyan) circle (which we name 
``Dark Heron'') nicely matches the molecular gas
structure. The structure extends beyond the 
southwest border of our CARMA coverage, 
where we see the CO emission 
(from the DLH14-only data) match 
the dark structure we name ``Dark Tail''.
Another intriguing structure is seen 
at the center of the Gulf region, 
highlighted by a dashed ellipse in panels
(c) and (d) of Figure \ref{fig:compare5}.
It is a thin dark filament that is only seen
in the high-resolution combined data 
(hereafter we name it ``Dark Filigree 1'').
Another dashed circle in this figure marks
the structure  we name the ``SW-Clump'', which is
to the east of the SW2 Main and SW2 SE  
dust continuum clumps identified by B14.
This structure appears as a darker region 
on top of, but with a  moderately
distinct velocity component in $^{13}$CO gas
compared to, the gray F-6 filament. 
It is, therefore, not clear if it is part of 
F-6 (see \S\ref{subsec:corr}).
At the northeast corner of the Gulf region
we see the ``Dark Worm'' at the border of
the CARMA mosaic footprint. We see its
bottom half in the combined image, which
 matches well the dark filament in the optical
image.

It is evident that these dark features are 
foreground molecular gas, and given their 
LSR velocity compared to that of W80, 
they are moving away from the HII region.
Unlike many of the structures coincident 
with gray extinction regions, the structures
at this velocity do not show clear evidence 
of being impacted by feedback from the 
high-mass star powering the HII region.
Because of this and the higher extinction
we suggest that the dark regions
are farther from the bubble compared to
the gray regions.

At $V_{lsr}= 3.5$ km s$^{-1}$
we also detect emission from the filament F-4, 
originally defined by Z14 (see Figure 
\ref{fig:compare5}). This CO structure 
does not show any gray or dark counterpart 
in the optical image. Since it shows no optical 
extinction counterparts, the filament is very 
likely on the back side of the bubble. 
There is another clump to the northeast of
the Gulf white boundary. It shows no foreground
extinction. So it should also be on the far side.
Its velocity roughly peaks at 4 km s$^{-1}$.

Figure \ref{fig:compare6} shows the comparison 
between the optical image and $^{13}$CO emission  
at $V_{lsr} = 4.9$ km s$^{-1}$. At this velocity,
a significant fraction of the molecular gas is 
coincident with ``dark'' regions that are
significantly smaller than the filaments 
discussed earlier, and have the appearance of 
scattered clumps and drizzles, shielding 
light from W80 in the background. 
We define four structures based on where the CO 
morphology at this velocity matches that of dark
regions. These are the ``Dark Crab'', ``Boomerang4'', 
``Dark Filigree 2'', and ``Dark Drizzle''.
One interesting feature is that the ``Dark Drizzle''
and the ``Dark Filigree 2'' structures
are broken into small dark globules, each globule
having its own $^{13}$CO counterpart. The typical
angular size of the globules is about 10-15\arcsec.
At a distance of 800 pc (see \S\ref{subsec:gaia}), 
their physical size is $\sim 10^4$ AU,
reminiscent of Bok globules 
\citep[e.g., Barnard 68,][]{2001Natur.409..159A}.
The Dark Crab approximately coincides in position 
with F-6, but it is at a higher velocity.
We therefore argue that the Dark Crab is
in front of F-6, with a velocity similar to that of W80. 

At a $V_{lsr} = 6.5$ km s$^{-1}$, there are
two main structures in $^{13}$CO emission 
(see Figure \ref{fig:compare7}). One of these
structures is associated with the Dark Crab, 
which we first identified when comparing the 
optical image of the Gulf region with the 
$^{13}$CO emission at $V_{lsr} = 4.9$ km s$^{-1}$
(Figure \ref{fig:compare6}). 
At $V_{lsr} = 6.5$ km s$^{-1}$ 
most of the CO emission is concentrated 
towards the northeast half of the Dark Crab,
whereas at lower LSR velocity the emission 
associated with the Dark Crab covers the full 
extent of this dark cloud. The other structure at  
$V_{lsr} = 6.5$ km s$^{-1}$, which we name F-7,
lies in the southern part of the Gulf region 
and is elongated in the northeast-southwest direction. 
Unlike other CO structures we identify above, 
the  morphology of F-7 does not closely match 
the structure of any of gray or dark extinction
features which it overlaps. We  thus argue that 
F-7 is not responsible for any of the extinction 
in the region as it is on the far side of W80.
Given the LSR velocity of F-7, it is likely 
F-7 is moving away from W80
(see \S\ref{subsec:sf} and Figure~\ref{fig:twopanel}).

\bibliographystyle{aasjournal}
\bibliography{ref}


\end{document}